\documentclass[a4paper,12pt]{article}
\usepackage{a4}
\usepackage{graphics}
\usepackage{epsfig}
\usepackage{amssymb}
\usepackage{color}
\usepackage{cite}
\usepackage{epsf}
\usepackage{axodraw}

\definecolor{red1}{rgb}{0.9,0,0}

\newcommand{\la}{\lambda}

\newcommand{\eps}{\epsilon}

\newcommand{\neu}{\tilde{\chi}^0}
\newcommand{\neutralino}{\tilde{\chi}^0}
\newcommand{\slepton}{\tilde{\ell}}

\newcommand{\stau}{\tilde{\tau}}

\def\lsim{\raise0.3ex\hbox{$\;<$\kern-0.75em\raise-1.1ex\hbox{$\sim\;$}}}
\def\gsim{\raise0.3ex\hbox{$\;>$\kern-0.75em\raise-1.1ex\hbox{$\sim\;$}}}
\DeclareMathAlphabet{\scr}{U}{rsfs}{m}{n}
\begin{document}
%{{{ Title
%\hfill \today\\
\hspace*{\fill} BONN-TH-2007-13\\
%\hspace*{\fill} arXiv:yymm.nnnn\\
%-----------------------------------------------------------------------------
\begin{center}
{
\large
 \bf 
Disentangling CP phases in nearly degenerate resonances: \\
neutralino production via Higgs at a muon collider
}                       
\end{center}
\vspace{0.5cm}

\begin{center}
{
\sc Herbi~K.~Dreiner, Olaf~Kittel, and Federico~von~der~Pahlen}
\end{center}
\begin{center}
{\small \it  Physikalisches Institut der Universit\"at Bonn,
                Nussallee 12, D-53115 Bonn, Germany
}
\end{center}
%}}}

%----------------------------------------------------------------------------------------
%{{{ Abstract
\begin{abstract}

In the CP-violating Minimal Supersymmetric Standard Model,
we study the pair production of neu\-tra\-linos 
at center-of-mass energies around the heavy neutral Higgs boson 
resonances. For longitudinally polarized muon beams, we
analyze CP asymmetries which are sensitive to the interference
of the two heavy neutral Higgs bosons.
Due to radiatively induced scalar-pseudoscalar transitions,
the CP asymmetries can be strongly enhanced when the resonances 
are nearly degenerate, as in the Higgs decoupling limit.
The Higgs couplings to the neu\-tra\-lino sector can then be analyzed
in the presence of CP violating phases.
We present a detailed numerical analysis of the cross sections,
neutralino branching ratios, and the CP observables.
We find that radiatively induced CP violation in the Higgs sector
leads to sizable CP-asymmetries, which are accessible 
in future measurements at a muon collider.
However, we expect that our proposed method should be applicable to 
other processes 
with nearly degenerate scalar resonances,
even at hadron colliders.

\end{abstract}
%---------------------------------------------------------------------------------------

\newpage
%}}}

%----------------------------------------------------------------------------------------
%{{{section Introduction

%----------------------------------------------------------------------------------------
\section{Introduction}
%----------------------------------------------------------------------------------------

The CP-conserving Minimal Supersymmetric Standard Model~(MSSM) 
contains three neutral Higgs 
bosons~\cite{HK,GH,Gunion:1989we,Carena:2002es,Djouadi:2005gj},
the lighter and heavier CP-even scalars $h$ and $H$, respectively, 
and the CP-odd pseudoscalar $A$. 
While the MSSM Higgs sector is CP-conserving at Born level 
even in the presence of CP-phases,
loop effects mediated dominantly by third generation squarks 
can generate significant CP-violating scalar-pseudoscalar
transitions, leading to mixing of the
 neutral Higgs states into 
the mass eigenstates $H_1$, $H_2$, $H_3$, with no definite 
CP parities~\cite{Accomando:2006ga,PilaftsisAH,Pilaftsis:1997dr,Choi:2004kq,
GunionetalHA,%Gunion:1997aq,Grzadkowski:1999ye,
HAeffpotential,%Demir:1999hj,Choi:2000wz,Ibrahim:2000qj,
PW.explicitCP,%Pilaftsis:1999qt,Carena:2000yi,Carena:2001fw,
Heinemeyer:2001qd
%,Heinemeyer:2007aq
}.

\medskip

It is well known that
mixing of states with equal conserved quantum numbers 
is strongly enhanced when these states are nearly degenerate,
i.e., when their mass difference is of the order of their 
widths~\cite{Pilaftsis:1997dr,Choi:2004kq}.
This degeneracy occurs naturally in the Higgs decoupling limit of the MSSM,
where the lightest Higgs boson has Standard Model-like couplings and decouples 
from the significantly heavier Higgs bosons~\cite{decoupling}.
In the decoupling limit, a resonance enhanced mixing of the states 
$H$ and $A$ can occur, which may result
in nearly maximal CP-violating 
effects~\cite{PilaftsisAH,Pilaftsis:1997dr,Choi:2004kq}.
The general formalism for mass mixing in extended Higgs sectors
with explicit CP violation is well 
developed~\cite{GunionetalHA,
%Gunion:1997aq,Grzadkowski:1999ye,
HAeffpotential,%Demir:1999hj,Choi:2000wz,Ibrahim:2000qj,
PW.explicitCP,%Pilaftsis:1999qt,Carena:2000yi,Carena:2001fw,
Heinemeyer:2001qd
%,Heinemeyer:2007aq
},
and sophisticated computer codes are available for numerical 
calculations~\cite{Frank:2006yh,FH,CPSuperH}.
Detailed investigations of the fundamental properties of the
Higgs bosons, both phenomenological and experimental,
 will be crucial for the understanding of the mechanism
of electroweak symmetry breaking.

\medskip

In previous studies of the 
CP-conserving~\cite{Grzadkowski:1995rx,mumu90s,%Barger:1996jm,Casalbuoni:1999mm,Berger:2001et,
Dittmaier:2002nd,Fraas:2003cx,Grzadkowski:2000hm,Asakawa:2000uj,Barger:1999tj,Fraas:2004bq,Kittel:2005ma} 
and CP-violating
Higgs sector~\cite{Asakawa:2000es,Atwood:1995uc,Pilaftsis:1996ac,Babu:1998bf,Choi:1999kn,Choi:2001ks,Bernabeu:2006zs,Hioki:2007jc}, it was shown that the
CP-properties and couplings of the heavy neutral Higgs bosons
can be ideally tested in $\mu^+\mu^-$ collisions.
Such a muon collider is a superb machine  for
measuring the neutral Higgs masses, widths, and couplings with high 
precision, since the Higgs bosons are resonantly produced in the
$s$-channel~\cite{Barger:1996vc,hefreports,Blochinger:2002hj%Autin:1999ci,Barger:2001mi,Blochinger:2002hj,Blondel:2004ae
}.
The well controllable beam energy 
allows the study of the center-of-mass energy dependence 
of observables around the Higgs resonances.
In particular, the beam polarization plays an essential role
for analyzing the CP nature of the Higgs sector itself.
Not only backgrounds can be reduced, but 
the CP-even and CP-odd contributions of the 
interfering Higgs resonances to the observables can be
ideally studied, if the beam polarizations are properly 
adjusted~\cite{Grzadkowski:2000hm,Asakawa:2000uj,Barger:1999tj,Fraas:2004bq,Kittel:2005ma,Asakawa:2000es,Atwood:1995uc,Pilaftsis:1996ac,Babu:1998bf,Choi:1999kn,Choi:2001ks}.

\medskip

Besides the initial beam polarization, 
the final fermion polarizations 
are essential to probe the
Higgs interference.
The secondary decays of the final fermions 
enable their spin analysis, and additional final state polarization observables
allow for a complete determination of the CP-properties of the Higgs 
bosons~\cite{Barger:1999tj,Fraas:2004bq,Kittel:2005ma,Asakawa:2000es}.
For final state SM fermions $f\bar f$, with $f=\tau, b,t$,
such polarization observables have been classified
according to their CP and 
CP$\tilde{\rm T}$\footnote{$\tilde{\rm T}$ is 
   the na\"ive time reversal $t\to-t$, which inverts momenta 
   and spins without exchanging initial 
   and final particles.} 
transformation properties\cite{Asakawa:2000es}.
For the production of neu\-tra\-linos~\cite{Fraas:2004bq} and 
charginos~\cite{Kittel:2005ma}
with longitudinally polarized beams, 
it has been shown that asymmetries in 
the energy distributions of their decay products 
are sensitive to the Higgs interference of the CP-even 
and CP-odd production amplitudes.
Thus the couplings of the Higgs boson 
to the neutralino and chargino sector can be analyzed 
in the CP-conserving MSSM.

\medskip
In this work, we extend the study 
of neutralino production at the muon collider~\cite{Fraas:2004bq} 
to the CP-violating case.
For neutralino production
$\mu^+\mu^- \to \tilde\chi_i^0\tilde\chi_j^0,$
with longitudinally polarized muon beams,
we define a CP-odd polarization asymmetry.
We analyze the longitudinal polarizations of the produced neu\-tra\-linos
by their subsequent leptonic two-body decays
$\tilde\chi_j^0 \to \ell\tilde\ell_n,$
$ \ell=e,\mu,\tau,$
with $n=R,L$ for $\ell=e,\mu$, and $n=1,2$ for $\ell=\tau$.
With the energy distribution of the leptons we can define
a CP-even and a CP-odd polarization asymmetry
for the neutralino decay, which probe the neutralino polarization.
First results for neutralino and also chargino production 
in the MSSM with explicit CP violation in the Higgs sector
have been reported in Ref.~\cite{CPNSH}.

\medskip

In Section~\ref{formalism}, 
we give our formalism for neutralino production and decay 
with longitudinally polarized beams.
In an effective Born-improved approach, we
include the leading self energy corrections
into the Higgs couplings. We give analytical formulas for
the production and decay cross sections and distributions,
and show that the energy distribution of the neutralino
decay products depends sensitively on the Higgs interference. 
In Section~\ref{AsymmetriesforPandD}, we define
CP-odd and CP-even asymmetries of the production cross section
and of the energy  distributions. 
These observables are sensitive to the CP-phases in the
Higgs sector, as well as to absorptive contributions
from the Higgs boson propagators.
In Section~\ref{Numerical results}, we present 
a detailed numerical analysis of the cross sections, 
neutralino branching ratios and the CP-observables.
We give special attention to the $\sqrt s$ dependence
of the observables and analyze their dependence
on the CP violating phase $\phi_A$,
of the common trilinear scalar coupling parameter $A_t=A_b=A_\tau\equiv A$,
and on the gaugino and higgsino mass parameters $M_2$ and $\mu$,
which mainly determine the Higgs couplings to the neu\-tra\-linos.
We summarize and conclude in Section~\ref{Summary and conclusions}.

\begin{figure}[t]
\label{fig:neutralinoprod.res}
                \scalebox{1}{
\begin{picture}(10,2)(-3.4,0)
\DashLine(80,50)(140,50){5}
\Vertex(80,50){2}\Vertex(140,50){2}
\Line(140,50)(170,80)
\Line(170,20)(140,50)
\ArrowLine(50,20)(80,50)
\ArrowLine(80,50)(50,80)
\Text(1.6,0.9)[r]{$\mu^+$}
\Text(1.6,2.7)[r]{$\mu^-$}
\Text(6.2,0.9)[l]{$\tilde\chi_i^0$}
\Text(6.2,2.7)[l]{$\tilde\chi_j^0$}
\put(3.35,1.95){$H_2,H_3$}
%% --------------------------------------------
\end{picture}
}
\caption{Resonant Higgs exchange in neutralino pair production.}
\label{Fig:resHiggsMix}
\end{figure}
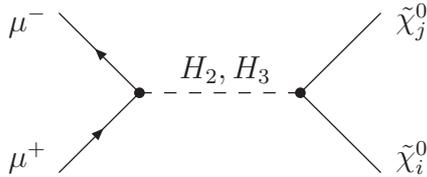

%----------------------------------------------------------------------------------------
\section{Neutralino production and decay formalism
  \label{formalism}}
%----------------------------------------------------------------------------------------

\begin{figure}[t]
\hspace{-2.cm}
\begin{minipage}[t]{6cm}
\begin{center}
{\setlength{\unitlength}{0.6cm}
\begin{picture}(5,5)
\put(-2.5,-8.5){\includegraphics{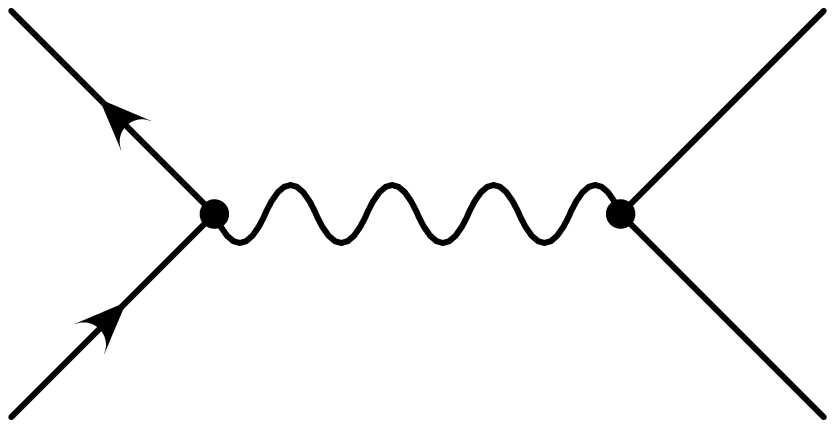}}
\put(1.7,-.4){{\small $\mu^{-}$}}
\put(8.7,-.3){{\small $\tilde\chi^0_j$}}
\put(1.7,3.8){{\small $\mu^{+}$}}
\put(8.7,3.9){{\small $\tilde\chi^0_i$}}
\put(4.7,2.5){{\small $Z, (H_1)$}}
\end{picture}}
\end{center}
\end{minipage}
\hspace{-0.5cm}
\begin{minipage}[t]{5cm}
\begin{center}
{\setlength{\unitlength}{0.6cm}
\begin{picture}(2.5,5)
\put(-4,-9){\includegraphics{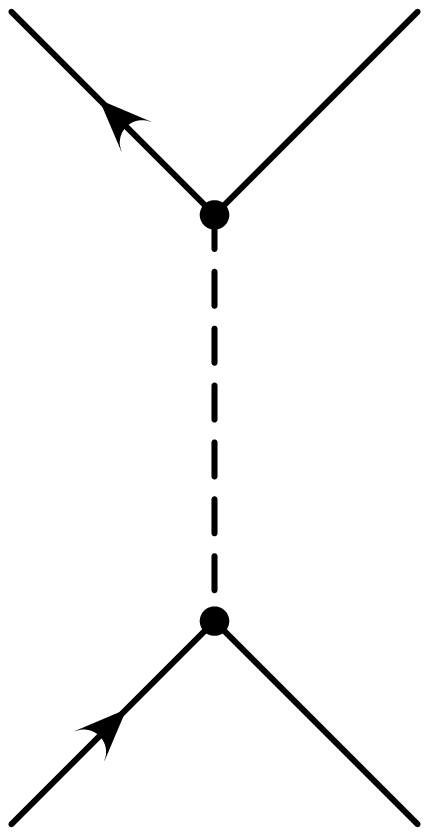}}
\put(1.8,-1.5){{\small $\mu^{-}$}}
\put(1.8,3.8){{\small $\mu^{+}$}}
\put(5.9,-1.4){{\small $\tilde\chi^0_j$}}
\put(5.9,3.9){{\small $\tilde\chi^0_i$}}
\put(4.4,1.5){{\small $\tilde\mu_{L,R}$}}
 \end{picture}}
\end{center}
\end{minipage}
\begin{minipage}[t]{5cm}
\begin{center}
{\setlength{\unitlength}{0.6cm}
\begin{picture}(2.5,5)
\put(-4.5,-9){\includegraphics{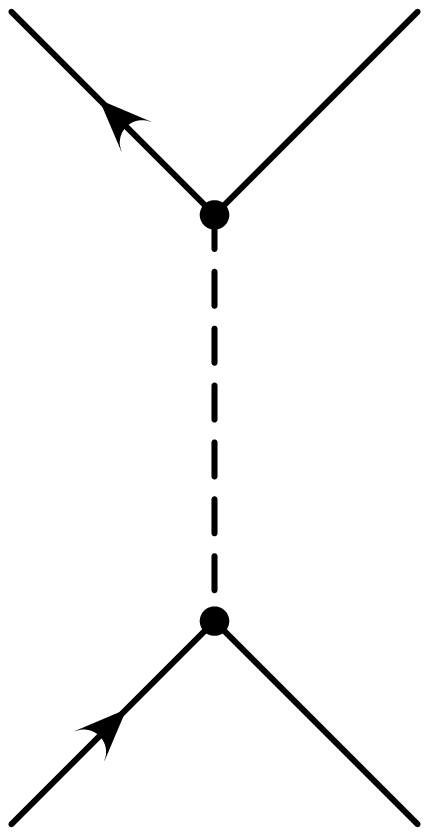}}
\put(1.3,-1.5){{\small $\mu^{-}$}}
\put(1.3,3.8){{\small $\mu^{+}$}}
\put(5.4,-1.4){{\small $\tilde\chi^0_i$}}
\put(5.4,3.9){{\small $\tilde\chi^0_j$}}
\put(3.9,1.5){{\small $\tilde\mu_{L,R}$}}
\end{picture}}
\end{center}
\end{minipage}
\vspace{1.1cm}
\caption{\label{Fig:FeynProd}
Feynman diagrams for non-resonant neu\-tra\-lino production
  $\mu^+ \mu^-\to\tilde\chi^0_i\tilde\chi^0_j$.
}
\end{figure}

We study CP violation in the Higgs sector
in pair production of neu\-tra\-linos 
\begin{equation}
\mu^+  + \mu^- \to \tilde\chi_i^0 + \tilde\chi_j^0,
    \label{production}
\end{equation}
with longitudinally polarized muon beams.
The Feynman diagram for Higgs boson exchange
is shown in Fig.~\ref{Fig:resHiggsMix}.
We will analyze the process at center-of-mass energies 
of the nearly mass degenerate heavy neutral Higgs bosons
$H_2$ and $H_3$.  They will
be resonantly produced in the $s$-channel.
The significantly lighter Higgs boson $H_1$
and the $Z$-boson are also exchanged in the $s$-channel,
however far from their resonances.
Together with the 
smuon exchange $\tilde\mu_{L,R}$ in the $t$- and $u$-channels,
they contribute the
non-resonant continuum to the neutralino production,
see their Feynman diagrams in Fig.~\ref{Fig:FeynProd}.

To analyze the longitudinal polarizations of the produced neu\-tra\-linos,
we consider their subsequent CP-conserving but P-violating 
leptonic two-body decays 
\begin{equation}
        \tilde\chi_j^0 \to \ell^{\pm} + \tilde\ell^{\mp}_n,
\quad \ell=e,\mu,\tau,
   \label{decay}
\end{equation}
with $n=R,L$ for $\ell=e$ or $\ell=\mu$, and $n=1,2$ for $\ell=\tau$.

%

%-----------------------------------------------------------------------------
\subsection{Lagrangians and amplitudes for Higgs exchange
             \label{section:lagrdens}}
%-----------------------------------------------------------------------------

CP violation of the MSSM Higgs sector is induced 
by scalar-pseudoscalar mixing at the loop level.
We will include these mixings effectively
in the interaction Lagrangians
for neu\-tra\-lino production~(\ref{production})        
via Higgs exchange $H_k$, with $k=1,2,3$,
\begin{eqnarray}
\label{mumuphi}
{\scr L}_{\mu^+ \mu^- H} & = &
        \bar\mu \,  
[ c^{H_k\mu\mu}_L P_L +  c^{H_k\mu\mu}_R P_R] 
\, \mu \, H_k,\\
{\scr L}_{\tilde\chi  \tilde\chi H} & = &
\frac{1}{2}
        \bar{\tilde\chi}_i 
[ c^{H_k\chi_i\chi_j}_{L} P_L  +  c^{H_k\chi_i\chi_j}_{R} P_R ] 
                \tilde  \chi_j\,H_k.
\label{chichiphi}
\end{eqnarray}
We obtain the effective Higgs
couplings to the initial muons, $c_{L,R}^{H_k\mu\mu}$,
and the final neu\-tra\-linos, $ c_{L,R}^{H_k\chi_i\chi_j}$,
from their tree level couplings
\begin{eqnarray}
         c_{L,R}^{H_k\mu\mu}  &=&
        C_{k\alpha}   c_{L,R}^{h_\alpha\mu\mu},
\label{eq.Hmu.eff.c}\\
   c_{L,R}^{H_k\chi_i\chi_j}  &=&
  \tilde C_{k\alpha}   c_{L,R}^{h_\alpha\chi_i\chi_j},\quad h_\alpha=h,H,A,
\label{eq.Hchi.eff.c}
\end{eqnarray}
with 
\begin{equation}
\tilde C = {C^{-1} }^T.
\end{equation}
The tree level couplings
$c_{L,R}^{h_\alpha\mu\mu}$ and $ c_{L,R}^{h_\alpha\chi_i\chi_j}$
are defined and discussed, e.g., in Refs.~\cite{Fraas:2004bq,Gunion:1989we}.
The matrix $C$ diagonalizes the Higgs mass matrix ${\rm \bf M}$ at fixed $p^2$.
In the tree-level basis of the CP-eigenstates $h,H,A$,
the symmetric and complex mass matrix at momentum squared $p^2$
is given by~\cite{Frank:2006yh}
\begin{equation}
  {\rm \bf M}(p^2) = 
\left(\begin{array}{ccc}
                        \ m_{h}^2 - \hat\Sigma_{hh}(p^2) \ &
 - \hat\Sigma_{hH}(p^2)  &  - \hat\Sigma_{hA}(p^2)
\\[2mm]
 - \hat\Sigma_{hH}(p^2)  &    \ m_{H}^2 - \hat\Sigma_{HH}(p^2) \ &
- \hat\Sigma_{HA}(p^2)
\\[2mm]
 - \hat\Sigma_{hA}(p^2)  &  - \hat\Sigma_{HA}(p^2) &   \ m_{A}^2 - \hat\Sigma_{AA}(p^2) \
\end{array}\right).
\label{eq:Weisskopf}
\end{equation}
Here 
$\hat\Sigma_{rs}(p^2)$ with $r,s = h,H,A$ are the renormalized self energies 
of the Higgs bosons at one loop,
supplemented with higher-order contributions, see Ref.~\cite{Frank:2006yh}.
When the Higgs bosons are nearly mass degenerate, these
corrections are enhanced by the Higgs mixing.
The propagator matrix 
\begin{equation}
\label{eq:propagatorhHA}
   \Delta_{rs}(p^2) = - i[ p^2 - {\rm \bf M}(p^2) ]^{-1}_{rs},
\end{equation}
 has complex poles at 
 $p^2=\mathcal M_{H_k}^2 \equiv  M_{H_k}^2 - i M_{H_k} \Gamma_{H_k},\ k=1,2,3$,
where $ M_{H_k} $ and $ \Gamma_{H_k} $ are the mass and width of the 
Higgs boson mass eigenstate $H_k$, respectively.
We evaluate the mass matrix ${\rm \bf M}(p^2)$ 
at fixed $p^2=M_{H_2}^2$ 
in its Weisskopf-Wigner form 
with the program {\tt FeynHiggs 2.5.1}~\cite{Frank:2006yh,FH},
in order to obtain the diagonalization matrix $C$, 
as well as the Higgs masses and widths.
%{\color{blue}
Here we neglect the momentum dependence of ${\rm \bf M}(p^2)$,
since, unless Higgs decay thresholds open at this energy,
this dependence is weak in the resonance region ${p^2\approx M_{H_2}^2,M_{H_3}^2}$. 
%}
This approach corresponds to an improved-Born approximation.
It includes the leading-order radiative corrections into
the matrix $C$, but not the specific vertex
and box corrections, as well as the subleading muon and 
neu\-tra\-lino self energy corrections.
We give further details in Appendix~\ref{app:Higgscouplings}. 

With the Born-improved effective couplings 
we write the amplitudes 
for neu\-tra\-lino production via Higgs exchange, see Fig.~\ref{Fig:resHiggsMix},
\begin{eqnarray}
T^P = 
\Delta(H_k)
\left[
\bar{v}(p_{\mu^+})\left(c_L^{H_k\mu\mu}P_L+c_R^{H_k\mu\mu}P_R \right)
u(p_{\mu^-})
\right]\phantom{.}\nonumber\\
\times
\left[
\bar u(p_{\chi^0_j})\left(c_L^{H_k\chi_i\chi_j}P_L
                                   +c_R^{H_k\chi_i\chi_j}P_R \right)
v (p_{\chi^0_i})
\right],
\label{THiggsNoHel}
\end{eqnarray}
with the Breit-Wigner propagator for the Higgs boson
\begin{eqnarray}
   \Delta(H_{k})&=&\frac{i}{s-M^2_{H_{k}}+iM_{H_{k}}\Gamma_{H_{k}}}.
\label{propHiggs}
\end{eqnarray}
The Lagrangians and amplitudes for 
$Z$ and $\tilde\mu_{L,R}$ exchange are given 
in Appendix~\ref{nonresamplitudes},
and the
Lagrangians for the leptonic neu\-tra\-lino decays~(\ref{decay}) are given in
Appendix~\ref{nonreslagrangian}.

%---------------------------------------------------------------------
\subsection{Squared amplitude}
%---------------------------------------------------------------------

In order to calculate the squared amplitude for 
neu\-tra\-lino production~(\ref{production}) and decay~(\ref{decay}),
we use the spin density matrix formalism 
of~\cite{Haber94,Moortgat-Pick:1999di}.
Following the detailed steps in Appendix~\ref{Density matrix formalism}, 
where we also give the production amplitudes,
the squared amplitude in this formalism can be  written as
\begin{eqnarray}
        |{{T}}|^2 &=&  2|\Delta(\tilde\chi_j^0)|^2 
        (P \cdot D + \sum_{a=1}^3 {\Sigma}_P^{a} {\Sigma}_D^{a}),
\label{eq:tsquare}
\end{eqnarray}
with the propagator $\Delta(\tilde\chi_j^0)$ of the decaying neu\-tra\-lino, 
see Eq.~(\ref{eq:neutralinopropagator}).
Here $P$ denotes the unpolarized production of the neutralinos
and $D$ the unpolarized decay.
${\Sigma}_P^{a} $ and $ {\Sigma}_D^{a}$ are the corresponding
polarized terms,
and their product in Eq.~(\ref{eq:tsquare}) describes
the neu\-tra\-lino spin correlations between production and decay. 
With our definition of the spin density production matrix,
Eq.~(\ref{rhoP}),
${\Sigma}_P^{3}/P$ is the 
longitudinal polarization of $\tilde\chi_j^0$, 
${\Sigma}_P^{1}/P$ is the transverse polarization 
in the production plane, and ${\Sigma}_P^{2}/P$ is the polarization
perpendicular to the production plane.
We give explicit expressions for the production terms
$P$ and ${\Sigma}_P^{a}$ in the next Section. The terms 
$D$ and ${\Sigma}_D^{a}$ for neu\-tra\-lino 
decay are given in Appendix~\ref{neutdecay}. 

%--------------------------------------------------------------------
\subsection{Resonant contributions from Higgs exchange}
%-------------------------------------------------------------------

The expansion coefficients of the squared neu\-tra\-lino production 
amplitude~(\ref{eq:tsquare}) subdivide into contributions from the 
Higgs resonances~(${\rm res}$) and the continuum~(${\rm cont}$), 
respectively, 
\begin{equation}
        P=P_{\rm res}+P_{\rm cont}, \qquad {\Sigma}_P^{a} ={\Sigma}_{\rm res}^{a}
        +{\Sigma}_{\rm cont}^{a}, \quad a=1,2,3.
\label{contributions}
\end{equation}
The continuum contributions $P_{\rm cont}$, ${\Sigma}_{\rm cont}^{a}$ are 
those from the non-resonant $Z$ and $\tilde\mu_{L,R}$ 
exchange channels,
with $P_{\rm cont}$ given in Appendix~\ref{nonresamplitudes}
and ${\Sigma}_{\rm cont}^{a}$ 
%can be found 
in \cite{Moortgat-Pick:1999di}.

In order to analyze the dependence
of the resonant contributions
$P_{\rm res}$ and 
$\Sigma_{\rm res}^3$
on the longitudinal $\mu^+$ and $\mu^-$ beam polarizations  
${\mathcal P}_+$ and 
${\mathcal P}_-$, respectively,
we can expand\footnote{
   The resonant contributions 
   $\Sigma_{\rm res}^1$ and $\Sigma_{\rm res}^2$
   to the transverse polarizations of the neutralino vanish
   for scalar Higgs bosons exchange in the $s$-channel.
} 
\begin{eqnarray}
P_{\rm res}&=&
        (1 + {\mathcal{P}}_+{\mathcal{P}}_-)a_0  + 
        ({\mathcal{P}}_+ + {\mathcal{P}}_-)a_1, 
\label{eq.Pr}
\\
\Sigma_{\rm res}^3&=&
         (1 + {\mathcal{P}}_+{\mathcal{P}}_-)b_0 + 
         ({\mathcal{P}}_+ + {\mathcal{P}}_-)b_1.
\label{eq.Sr}
\end{eqnarray}
Such an expansion proofs to be useful 
for discussing CP properties.
The coefficients $a_0$ and $b_1$ are CP-even,
whereas $a_1$ and $b_0$ are CP-odd and vanish
in the case of CP conservation~\cite{Fraas:2004bq}.
They are given by
\begin{eqnarray}
a_n  = \sum_{k\le l} (2-\delta_{kl}) a_n^{kl},\quad
b_n  = \sum_{k\le l} (2-\delta_{kl}) b_n^{kl};
\quad n=0,1; 
\label{eq.anbn}
\end{eqnarray}
with the sum over the contributions from the
Higgs bosons $H_{k}$, $H_{l}$ with $ k,l = 1,2,3$, 
respectively,
and
\begin{eqnarray}
a_0^{kl}~&=&
        \phantom{+}\!   \frac{s}{2}|\Delta_{(kl)}|
\Big[%\phantom{-}
                |c_\mu^+ | |c^+_\chi |
f_{ij}
        \cos(
                {{\delta_\mu^+ }}
                +
                {{\delta_\chi^+ }}
                +
                {{\delta_\Delta }}
        )
\nonumber\\ & & 
\phantom{
%\phantom{+}
\!   \frac{s}{2}|\Delta_{(kl)}|
}
                 -|c_\mu^+ | |c_\chi^{RL} |
                m_i m_j
        \cos(
                {{\delta_\mu^+ }}
                +
                {{\delta_\chi^{RL} }}
                +
                {{\delta_\Delta }}
        )
\Big]_{(kl)},
\label{eq.a0}
\\
a_1^{kl}~&=& \!
        \phantom{+}\!   \frac{s}{2} |\Delta_{(kl)}|
\Big[%\phantom{-}
                |c_\mu^- | |c^+_\chi |
f_{ij}
        \cos(
                {{\delta_\mu^- }}
                +
                {{\delta_\chi^+ }}
                +
                {{\delta_\Delta }}
        )
\nonumber\\ & & 
\phantom{
%\phantom{+}
\!   \frac{s}{2}|\Delta_{(kl)}|
}                -|c_\mu^- | |c_\chi^{RL} |
                m_i m_j
        \cos(
                {{\delta_\mu^- }}
                +
                {{\delta_\chi^{RL} }}
                +
                {{\delta_\Delta }}
        )
\Big]_{(kl)},
\label{eq.a1}
\\
b_0^{kl}~&=&\!
        -\frac{s}{4} |\Delta_{(kl)}|
\Big[
                |c_\mu^+ | |c^-_\chi |
\sqrt{\lambda_{ij}}
        \cos(
                {{\delta_\mu^+ }}
                +
                {{\delta_\chi^- }}
                +
                {{\delta_\Delta }}
        )
\Big]_{(kl)},
\label{eq.b0}
\\
b_1^{kl}~&=&\!
-
\frac{s}{4} |\Delta_{(kl)}|
\Big[
                |c_\mu^- | |c^-_\chi |
\sqrt{\lambda_{ij}}
        \cos(
                {{\delta_\mu^- }}
                +
                {{\delta_\chi^- }}
                +       
                {{\delta_\Delta }}
        )
\Big]_{(kl)},
\label{eq.b1}
\end{eqnarray}
We have defined the products of couplings, 
see Eqs.~(\ref{eq.Hchi.eff.c}) and (\ref{eq.Hmu.eff.c}), 
suppressing the neu\-tra\-lino indices $i$ and~$j$, 
\begin{eqnarray}
c_{\alpha(kl)}^\pm &=&  
c_R^{H_k\alpha\alpha} c_R^{H_l\alpha\alpha\ast}\pm
c_L^{H_k\alpha\alpha} c_L^{H_l\alpha\alpha\ast}
=
\Big[
        | c_\alpha^\pm | \exp({i {\delta_\alpha^\pm }})    
\Big]_{(kl)}  
,\quad \alpha=\mu, \chi,
\label{eq.clambda.mp}
\\
c_{\chi(kl)}^{RL} &=&  
c_R^{H_k\chi\chi} c_L^{H_l\chi\chi\ast}+
c_L^{H_k\chi\chi} c_R^{H_l\chi\chi\ast}
=
\Big[
|c_\chi^{RL}| \exp({i {\delta_\chi^{RL} }})
\Big]_{(kl)}  
,
\label{eq.cchi.RL}
\end{eqnarray}
the product of the Higgs boson propagators~(\ref{propHiggs}),
\begin{eqnarray}
      \Delta_{(kl)}&=&\Delta(H_k)\Delta(H_l)^\ast
=
\Big[
|\Delta|\exp({i {\delta_\Delta }})
\Big]_{(kl)}  
,\label{deltadelta}
\end{eqnarray}
and the kinematical functions 
$f_{ij}=(s-m_{\chi_i^0}^2- m_{\chi_j^0}^2)/2$ and 
${\lambda_{ij}}$, see Eq.~(\ref{triangle}).
Note that the 
coefficients $b_0$ and $b_1$, 
which parametrize the 
neutralino polarization dependence in~Eq.~(\ref{eq.Sr}), 
vanish at threshold as $\sqrt{\lambda_{ij}}$.
The coefficients
$a_{1}$ and $b_{1}$
contribute only for longitudinally polarized muon beams.
These coefficients are 
products of the Higgs boson couplings to the muons and neu\-tra\-linos.
Our aim is to determine these coefficients using polarization asymmetries and the production cross section.
Since a muon collider provides a good beam energy resolution
it is the ideal tool to analyze their strong $\sqrt{s}$ dependence.

We neglect interferences of the chirality violating Higgs 
exchange amplitudes with the chirality conserving continuum amplitudes, 
which are of order $m_\mu/\sqrt{s}$.
Note that contributions from $H_1$ exchange will be small
far from its resonance.

%-----------------------------------------------------------------------------
\subsection{Cross sections}
%-------------------------------------------------------------------

We obtain cross sections and distributions by integrating 
the amplitude squared $|T|^2$~(\ref{eq:tsquare}) over 
the Lorentz invariant phase space element $d{\rm PS}$
\begin{equation}
        d\sigma=\frac{1}{2 s}|T|^2d{\rm PS}.
\label{crossection}
\end{equation}
We use the narrow width approximation for the 
propagator of the decaying neu\-tra\-lino.
Explicit formulas of the phase space for neu\-tra\-lino 
production~(\ref{production}) and decay~(\ref{decay}),
can be found, e.g., in~\cite{Kittel:2005rp}. 
The $\mu^+\mu^-$-spin averaged cross section for 
$\tilde\chi_i^0\tilde\chi_j^0$
neu\-tra\-lino production is
\begin{equation}
\sigma_{ij}= \frac{1}{(1+\delta_{ij})}
             \frac{\sqrt{\lambda_{ij}}}{8\pi s^2}\bar P,
\label{crossectionProd}
\end{equation}
with the triangle function $\lambda_{ij}$~(\ref{triangle}),
and the average over the neu\-tra\-lino production angles
in the center-of-mass system,
\begin{eqnarray}
\bar P  = \frac{1}{4\pi}\int P  d\Omega_{\chi^0_j}.
\qquad
\label{eq:Pbar}
\end{eqnarray}
The integrated cross section for neu\-tra\-lino production~(\ref{production}) 
and subsequent leptonic 
decay $\neu_j\to\ell^\pm\slepton^\mp_n$~(\ref{decay}),
with $n=R,L$ for $\ell=e,\mu$, and $n=1,2$ for $\ell=\tau$, 
is given by
\begin{eqnarray}
 \sigma_{\ell}^{n}= 
                \frac{1}{(1+\delta_{ij})}
                \frac{1 }{64\pi^2}
                \frac{\sqrt{\lambda_{ij}}}{s^2}   \,
                \frac{(m_{\chi_j}^2-m_{\tilde{\ell}}^2)}{m_{\chi_j}^3 {\Gamma_{\chi_j}}}
\,      \bar{P}\cdot D
= \sigma_{ij} \times{\rm BR}(\tilde\chi_j^0\to\ell^\pm\tilde\ell^\mp_n).
\label{eq:sigmatot}
\end{eqnarray}
Explicit expressions for $D$ are given in 
Appendix~\ref{neutdecay}.
Note that the integrated cross section $\sigma_{\ell}^{n}$
is independent of the neutralino polarizations.

%-----------------------------------------------------------------------------
\subsection{Lepton energy distribution}
            \label{energydistr}
%-------------------------------------------------------------------

The differential cross section $d\sigma$~(\ref{crossection}),
and thus the energy distribution of the lepton from
the neu\-tra\-lino decay~(\ref{decay}), depends on the
longitudinal neu\-tra\-lino polarization.
In the center-of-mass system, the kinematical limits of the energy 
of the decay lepton $\ell=e,\mu,\tau$ are
\begin{eqnarray}
E_\ell^{\rm max(min)} &=& \hat{E}_\ell \pm \Delta_\ell,
\label{kinlimits}
\end{eqnarray}
with 
\begin{eqnarray}
\hat{E}_\ell &=& \frac{E_\ell^{\rm max}+E_\ell^{\rm min}}{2} = 
\frac{ m_{\chi^0_j}^2-m_{\tilde\ell}^2}{2 m_{\chi^0_j}^2} E_{\chi^0_j},
\label{ehalf}
\\
\Delta_\ell &=& \frac{E_\ell^{\rm max}-E_\ell^{\rm min}}{2} = 
\frac{ m_{\chi^0_j}^2-m_{\tilde\ell}^2}{2 m_{\chi^0_j}^2}
|\vec{p}_{\chi^0_j}|,
\qquad \ell=e,\mu,\tau.
\label{edif}
\end{eqnarray}
Using the definition of the cross section~(\ref{eq:tsquare}), (\ref{eq:sigmatot}),
and the explicit form of $\Sigma_D^3$~(\ref{etal}),
the energy distribution of the decay lepton $\ell^\pm$ 
is~\cite{Fraas:2004bq,Kittel:2005ma}
\begin{equation}
\frac{d\sigma_{\ell^\pm}^n}{dE_\ell} =
        \frac{\sigma_\ell^n}
        {2\Delta_\ell}\left[ 1 + \,
        \eta_{\ell^\pm}
        \eta_{\ell}^n\frac{\bar{\Sigma}^3_P}{\bar{P}} 
        \frac{(E_\ell- \hat{E_\ell})}{\Delta_\ell} \right],
\label{edist2}
\end{equation}
with $\eta_{\ell^\pm}= \mp1$.
The factor $\eta_\ell^n$ is a measure of 
parity violation in the neu\-tra\-lino decay.
It is maximal
$\eta^R_{e,\mu}=+1$ and $\eta^L_{e,\mu}=-1$ 
for the decay into $\tilde{e}_R,\tilde{\mu}_R$ 
and $\tilde{e}_L,\tilde{\mu}_L$, respectively,
since the sleptons of the first two generations couple 
either purely left or right handed, if mixing is neglected.
For the decay $\tilde\chi_j^0\to\tau^\pm\tilde\tau^\mp_{1,2}$,
the factor 
\begin{equation}
\eta^n_\tau = \frac{|b_{nj}^{\stau}|^2 - |a_{nj}^{\stau}|^2}
                {|b_{nj}^{\stau}|^2 + |a_{nj}^{\stau}|^2},
\label{eq:eta_rl}
\end{equation}
is generally smaller 
$|\eta^n_\tau| <1$
due to stau mixing.
The right and left $\neu_j\stau_n\tau$ couplings $a_{nj}^{\stau}$ and 
$b_{nj}^{\stau}$ are defined in~(\ref{eq:coupl1}).

Further
in Eq.~(\ref{edist2}), 
the coefficients $ P$ and $ \Sigma^3_P$ 
of the squared neu\-tra\-lino production 
amplitude~(\ref{eq:tsquare})
are averaged over the
neu\-tra\-lino production solid angle,
denoted by a bar in our notation~(\ref{eq:Pbar}).
Due to the Majorana character of the neu\-tra\-linos,
the continuum contribution $\Sigma^3_{\rm cont}$~(\ref{contributions})
is forward-backward antisymmetric~\cite{Moortgat-Pick:2002iq},
and vanishes if integrated over the neu\-tra\-lino solid angle.
However, the resonant contribution 
$\Sigma^3_{\rm res}$~(\ref{contributions})
from Higgs exchange is isotropic, thus
\begin{eqnarray}
\bar\Sigma^3_P  = \frac{1}{4\pi}\int \Sigma^3_P  d\Omega_{\chi^0_j}
              =  \Sigma_{\rm res}^3.
\label{eq:sigmabar}
\end{eqnarray}

\medskip

\begin{figure}[t]
\centering
\begin{picture}(14,5.9)
\put(-1.,-17.4){\includegraphics{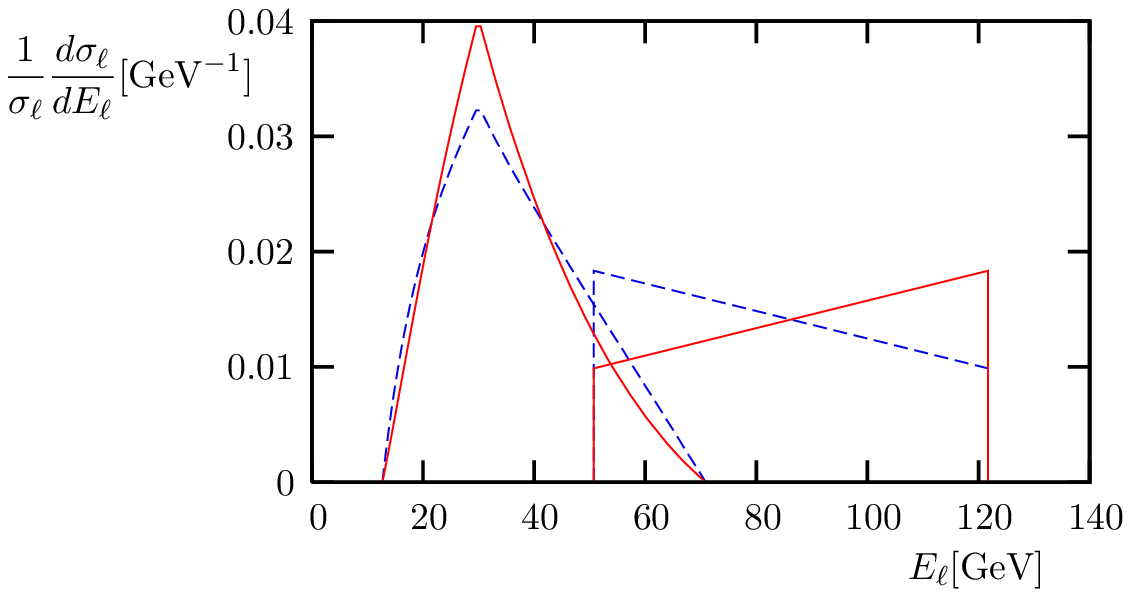}}
\put(6.7,2.95){\color{blue}$  \ell^-$}
\put(8.7,2.95){\color{red1}$  \ell^+$}
\put(4.4,3.7){\color{blue}$  \ell_2^+$}
\put(5.0,4.8){\color{red1}$  \ell_2^-$}
\end{picture}
\caption{\small Normalized energy distributions of the 
primary $\ell^\pm$ and secondary leptons $\ell_2^\mp$ for 
neu\-tra\-lino production $\mu^+\mu^-\to\neu_1\neu_2$  and 
subsequent decay chains
 $ \neu_2\to\ell^-\slepton^+_R $,
$\slepton^+_R\to\neu_1\ell_2^+$ (dashed)
and $\neu_2\to\ell^+\slepton^-_R$, $\slepton^-_R\to\neu_1\ell_2^-$  (solid)
for $ \ell=e$ or $\mu $.
Here  $\mathcal{A}_\ell^R={-15\%} $, 
corresponding to the masses of Table~\ref{scenarioAmasses}, % (scenario~{CP$\chi$}) 
 $\mathcal{P}_{+}=\mathcal{P}_{-}= 0.3$, and $ \sqrt{s}=494$~GeV.
}\label{fig:edist.l1l2}
\end{figure}

In Fig.~\ref{fig:edist.l1l2}, we show the energy distributions~(\ref{edist2}) of the 
leptons $\ell^\pm$ from the decays
$\tilde\chi_j^0\to\ell^+\tilde\ell^-_R$
and $\tilde\chi_j^0\to\ell^-\tilde\ell^+_R$,
for $\ell=e$ or $\mu$.
The cutoffs in the energy distributions of the
primary leptons $\ell^+$ and $\ell^-$ correspond
to their kinematical limits, as given in Eq.~(\ref{kinlimits}).
We see the linear dependence of the distributions 
on the lepton energy. 
The slope of these distributions is proportional 
to the longitudinal neu\-tra\-lino polarization,
and is solely due to the resonant Higgs contributions
$\Sigma_{\rm res}^3$.
In addition, we show in Fig.~\ref{fig:edist.l1l2} the
energy distributions from the secondary leptons $\ell_2^\pm$ from
the subsequent decays $\tilde\ell^+_R\to\ell_2^+\tilde\chi^0_1$ and
$\tilde\ell^-_R\to\ell_2^-\tilde\chi^0_1$. 

\medskip

Note that generally the parity conserving neu\-tra\-lino decays 
into the $Z$ and the lightest Higgs boson,
$\tilde\chi^0_j \to Z\tilde\chi^0_k $, and
$\tilde\chi^0_j \to H_1 \tilde\chi^0_k $, respectively,
cannot be used to analyze the neu\-tra\-lino polarization.
The resulting energy distributions are flat,
due to the Majorana properties
of the neu\-tra\-linos. 
They imply that the left and right couplings obey
$|O_{jk}^{''R}|=|O_{jk}^{''L}|$~(\ref{OLOR}), and
$|c^{H_1\chi_j\chi_k}_R| = |c^{H_1\chi_j\chi_k}_L|$~(\ref{chichiphi}).
The decays are thus parity conserving,
and therefore the parity violating factors analogous to 
$\eta_\tau^n$~(\ref{eq:eta_rl}) vanish.

%%%%%%%%%%%%%%%%%%%%%%%%%%%%%%%%%%%%%%%%%%%%%%%%%%%%%%%%%%%%%%%%%%%%%%%%%%%%%%

%%%%%%%%%%%%%%%%%%%%%%%%%%%%%%%%%%%%%%%%%%%%%%%%%%%%%%%%%%%%%%%%%%%%%%%%%%%%%%

% -----------------------------------------------------------------
%                        P L O T - 1 
% -----------------------------------------------------------------

%}}}
%----------------------------------------------------------------------------------------

%----------------------------------------------------------------------------------------
%{{{ Asymmetries
\section{Asymmetries for neu\-tra\-lino production and decay
             \label{AsymmetriesforPandD}}
%----------------------------------------------------------------------------------------
%{{{ Asymmetries intro
%

In Eqs.~(\ref{eq.Pr}) and (\ref{eq.Sr}), 
we have expressed the resonant contributions $P_{\rm res}$ and $\Sigma^3_{\rm res}$ 
to the spin density matrix elements
for neutralino pair production 
in terms of the longitudinal muon beam  polarizations.
In order to experimentally determine the four different combinations of products of couplings 
$a_0$, $a_1$, $b_0$, and $b_1$, 
we define asymmetries of the neutralino production cross section, 
as well as asymmetries of the energy distributions of the decay leptons.
Together with the neutralino production cross section,
these coefficients can then be experimentally determined, 
and thus the 
Higgs couplings to muons and neutralinos.
\medskip
%}}}

%}}}

%-------------------------------------------------------------------
%{{{Asymmetry of production

\subsection{Asymmetry of the neutralino production cross section}
%-------------------------------------------------------------------

For the cross section of neu\-tra\-lino pair production
$\sigma_{ij}=\sigma(\mu^+\mu^- \to \tilde\chi_i^0\tilde\chi_j^0)$,
Eq.~(\ref{crossectionProd}),
we define, for equal muon beam polarizations 
${\mathcal P}_+ = {\mathcal P}_- \equiv {\mathcal P}$,
the CP-odd asymmetry~\cite{Blochinger:2002hj,CPNSH}
\begin{eqnarray}
\mathcal{A}^{{\rm pol}}_{\rm prod} &=& \frac
        {\sigma_{ij}(\mathcal{P})-\sigma_{ij}(\mathcal{-P})}
        {\sigma_{ij}(\mathcal{P})+\sigma_{ij}(\mathcal{-P})}.
\label{eq:apolprod.ij}
\end{eqnarray}
$\mathcal{A}^{{\rm pol}}_{\rm prod}$ is sensitive to
the CP phases of the Higgs boson 
couplings to the neu\-tra\-linos and to the muons.
Denoting by $\tilde{\rm{T}}$ the na\"ive time reversal $t\to-t$,
which inverts momenta and spins without exchanging initial 
and final particles,
the asymmetry $\mathcal{A}^{{\rm pol}}_{\rm prod}$
is also CP$\tilde{\rm{T}}$-odd.
Thus the asymmetry is due to 
the interference of the CP phases with the absorptive
phases from the transition amplitudes.
The absorptive phases are also called strong phases,
and originate from intermediate particles which go on-shell.
The asymmetry 
$\mathcal{A}^{{\rm pol}}_{\rm prod}$ is therefore sensitive 
to the CP phases of the Higgs boson couplings,
as well as to the phases of the Higgs propagators.

\medskip

In the Higgs decoupling limit~\cite{decoupling},
the heavy neutral Higgs bosons are nearly mass degenerate.
Thus a mixing of the CP-even and CP-odd Higgs states $H$ and $A$ 
can be resonantly enhanced, and large CP-violating Higgs couplings can be
obtained~\cite{PilaftsisAH,Pilaftsis:1997dr}. 
In addition, CP phases in the Higgs sector lead to a larger splitting of
the mass eigenstates $H_2$ and $H_3$. This in general
tends to increase the phase difference between the Higgs propagators,
giving rise to larger absorptive phases in the transition
amplitudes.

\medskip

Using the definitions of the neu\-tra\-lino production
cross section $\sigma_{ij}$~(\ref{crossectionProd}), 
and of the $P$ term~(\ref{eq.Pr}), we obtain
\begin{eqnarray}
\mathcal{A}^{{\rm pol}}_{\rm prod} = 
        \frac{ {2\mathcal P} a_1 }{ (1 + {\mathcal P}^2)a_0 + \bar P_{\rm cont} }.
\label{eq:apolprodadep}
\end{eqnarray}
We can thus employ the asymmetry to determine the CP-odd coefficient $a_1$.
The maximum absolute value of the asymmetry depends on the beam
polarization~${\mathcal P}$
\begin{eqnarray}
\mathcal{A}^{{\rm pol,\, max}}_{\rm prod}= \frac{2\mathcal{P}}{1+\mathcal{P}^2},
\label{eq:amax}
\end{eqnarray}
which follows from Eq.~(\ref{eq:apolprodadep}),
for vanishing continuum contributions $\bar P_{\rm cont}=0$.

Note that the coefficient $a_0$ can be obtained from 
the neu\-tra\-lino production 
cross section $\sigma_{ij}$~(\ref{crossectionProd}).
For example, 
for unpolarized beams, 
${\mathcal P}_+ = {\mathcal P}_- ={\mathcal P}=0$,
\begin{equation}
a_0 = \sigma_{ij}
\frac{8\pi s^2}{\sqrt{\lambda_{ij}}}(1+\delta_{ij}).
\end{equation}
Here we assume that the continuum contributions
$\bar P_{\rm cont}$~(\ref{contributions}) 
to the cross section $\sigma_{ij}$ are already
subtracted, e.g, through an extrapolation of 
$\sigma_{ij}$ around the resonances~\cite{Fraas:2003cx}, and/or by
neu\-tra\-lino cross section measurements at the 
International Linear Collider (ILC)~\cite{TDR,Djouadi:2007ik}.

%}}}

%----------------------------------------------------------------
%{{{Asymmetry of decay

\subsection{Asymmetries of the lepton energy distribution}
\label{sec:asym.edis}
%----------------------------------------------------------------

The longitudinal neu\-tra\-lino polarization
is also sensitive to the Higgs interference
in the production $\mu^+\mu^-\to\tilde\chi^0_i\tilde\chi^0_j$.
The neu\-tra\-lino polarization can be analyzed by the
subsequent decays 
$\neu_j\to\ell^\pm\slepton_{R,L}^\mp$, with $\ell=e,\mu$, 
and $\neu_j\to\tau^\pm\stau_{1,2}^\mp$. 
In Section~\ref{energydistr}, we have shown that the slope of the 
lepton energy distribution, see Fig.~\ref{fig:edist.l1l2},
is proportional to the averaged longitudinal neu\-tra\-lino polarization
$\Sigma^3_{\rm res}/\bar P$.
The polarization can be determined by the
the energy distribution asymmetry~\cite{Fraas:2004bq} 
\begin{eqnarray}
        {\mathcal A}_{\ell^\pm}^n &=& 
        \frac{\Delta \sigma_{\ell^\pm}^n }{\sigma_{\ell^\pm}^n}
 =  \frac{1}{2}
        \eta_\ell^n \eta_{\ell^\pm}
        \frac{\Sigma^3_{\rm res}}{\bar P}
\nonumber\\
 &=& 
 \frac{1}{2}
        \eta_\ell^n \eta_{\ell^\pm}
        \frac{(1 + {\mathcal{P}}_+{\mathcal{P}}_-)b_0 +
 ({\mathcal{P}}_+ + {\mathcal{P}}_-)b_1}{P_{\rm res} +\bar P_{\rm cont}},
\label{asymmetryenergydist}
\end{eqnarray}
with $\Delta \sigma_{\ell^\pm}^n=\sigma_{\ell^\pm}^n(E_\ell>\hat E_\ell)-
\sigma_{\ell^\pm}^n(E_\ell < \hat E_\ell)$,
and $n=R,L$ for $\ell=e,\mu$, and $n=1,2$ for $\ell=\tau$.
Here we have used the explicit formula for the energy distribution
of the decay lepton $\ell^\pm$~(\ref{edist2}),
with $\bar\Sigma^3_P= \Sigma^3_{\rm res}$~(\ref{eq:sigmabar}).
The continuum contributions to the neu\-tra\-lino
polarization $\bar\Sigma^3_{\rm cont}=0$ vanish 
due to the Majorana properties of the neu\-tra\-linos,
see Section~\ref{energydistr}.

The average neu\-tra\-lino 
polarization is thus solely due to the Higgs exchange,
and receives CP-even and CP-odd contributions, 
proportional to $b_1$ and $b_0$,
respectively.
In order to separate these coefficients  
we define the polarization asymmetries
\begin{eqnarray}
% -------------------- A.pol.ell ---------------------------
        {\mathcal{A}}_{{\ell^\pm}}^{{\rm pol}, n}&=&
        \frac{
               \Delta \sigma^n_{\ell^\pm}(\mathcal{P})
               -
               \Delta \sigma^n_{\ell^\pm}(\mathcal{-P})
        }{
                \sigma^n_{\ell^\pm}(\mathcal{P})
               +
                \sigma^n_{\ell^\pm}(\mathcal{-P})
        }
=
        \eta_{\ell}^n\eta_{\ell^\pm}
        \frac{  \mathcal{P} b_1 }{(1+\mathcal{P}^2) a_0 +  \bar{P}_{\rm cont}},
\label{eq.apolijA_a}
\\
% ------------------ Aprime.pol.ell ------------------------
{\mathcal{A}}_{{\ell^\pm}}^{\prime {\rm pol}, n}&=&
        \frac{
               \Delta \sigma^n_{\ell^\pm}(\mathcal{P})
               +
               \Delta \sigma^n_{\ell^\pm}(\mathcal{-P})
        }{
                \sigma^n_{\ell^\pm}(\mathcal{P})
               +
                \sigma^n_{\ell^\pm}(\mathcal{-P})
        }
=
        \frac{1}{2}
        \eta_{\ell}^n\eta_{\ell^\pm}
        \frac{   (1+ \mathcal{P}^2) b_0 }
                {(1+\mathcal{P}^2) a_0 +  \bar{P}_{\rm cont}},
\label{eq.apolijAprime_a}
\end{eqnarray}
for equal muon beam polarizations 
${\mathcal P}_+ = {\mathcal P}_- \equiv {\mathcal P}$.

The slepton from the neu\-tra\-lino decay,
$\tilde\chi_j^0 \to \ell^\pm\tilde\ell^\mp_n,$
subsequently decays into a neu\-tra\-lino and a secondary lepton. 
The primary and secondary leptons have to be distinguished from 
each other, for example, by using 
their different energy distributions, see Fig.~\ref{fig:edist.l1l2}. 
However, the largest part of that irreducible background from the 
secondary lepton cancels in forming the charge conjugated
asymmetries~\cite{Fraas:2004bq}
\begin{eqnarray}
% -------------------- A.pol.ell ---------------------------
{\mathcal{A}}_\ell^{{\rm pol}, n}&=&
 \frac{1}{2}
 ({\mathcal{A}}_{{\ell^+}}^{{\rm pol}, n}-
  {\mathcal{A}}_{{\ell^-}}^{{\rm pol}, n})
=
        \eta_{\ell}^n
        \frac{  \mathcal{P} b_1 }{(1+\mathcal{P}^2) a_0 +  \bar{P}_{\rm cont}},
\label{eq.apolijA}
\\
% ------------------ Aprime.pol.ell ------------------------
{\mathcal{A}}_\ell^{\prime {\rm pol}, n}&=&
\frac{1}{2}
 ({\mathcal{A}}_{\ell^+}^{\prime {\rm pol}, n}-
  {\mathcal{A}}_{\ell^-}^{\prime {\rm pol}, n})
=
        \frac{1}{2}
        \eta_{\ell}^n
        \frac{   (1+ \mathcal{P}^2) b_0 }
                {(1+\mathcal{P}^2) a_0 +  \bar{P}_{\rm cont}}.
\label{eq.apolijAprime}
\end{eqnarray}

Due to the pure left or right coupling structure
of the neu\-tra\-linos to the selectrons and smuons,
the asymmetries for the decay into $\tilde{e}_R,\tilde\mu_R$ 
and $\tilde{e}_L,\tilde\mu_L$, respectively, 
have opposite sign:
\begin{eqnarray}
  {\mathcal A}_{\ell}^{{\rm pol}, R}
=-{\mathcal A}_{\ell}^{{\rm pol}, L}, \quad
{\mathcal A }_{\ell}^{\prime {\rm pol}, R}=
-{\mathcal A }_{\ell}^{\prime {\rm pol}, L},  
\quad \ell =e, \mu,
\end{eqnarray}
which follows from 
$\eta^R_{e,\mu}=+1$ and $\eta^L_{e,\mu}=-1$. 
The asymmetries for the decay into 
$\tilde\tau_1$ and $\tilde\tau_2$ are always
smaller than those for the decays into
a selectron or smuon,
\begin{eqnarray}
\label{eq:realtionAtau}
{\mathcal A}_{\tau}^{{\rm pol}, n} = 
\eta_\tau^n{\mathcal A}_{\ell}^{{\rm pol}, R}, 
\quad
{\mathcal A}_{\tau}^{\prime {\rm pol}, n} = 
\eta_\tau^n{\mathcal A}_{\ell}^{\prime {\rm pol}, R},
\quad n=1,2, \quad \ell =e, \mu,
\end{eqnarray}
with $|\eta_\tau^n|\leq 1$,
due to mixing in the stau sector,
see Eq.~(\ref{eq:eta_rl}). 

%{\color{blue}
The CP-even asymmetry ${\mathcal{A}}_{{\ell}}^{{\rm pol},n}$
is due to the correlation between the longitudinal polarizations 
of the initial muons and final neu\-tra\-linos, 
see Appendix~\ref{polarization.correlation}. % for a more detailed analysis.
%}
Large values of the CP-even asymmetry 
${\mathcal{A}}_\ell^{{\rm pol},n}$ can be obtained 
when both Higgs resonances are nearly degenerate,
and if their amplitudes are of the same magnitude.
However, a scalar-pseudoscalar mixing in the presence
of CP phases will in general
increase the mass splitting of the Higgs bosons,
and the reduced overlap of the Higgs resonances
also reduces the CP-even asymmetry 
${\mathcal{A}}_\ell^{{\rm pol},n}$.

The CP-odd asymmetry ${\mathcal{A}}_\ell^{\prime {\rm pol},n}$
vanishes for CP-conserving Higgs couplings.
Similarly to the CP-odd polarization asymmetry 
$\mathcal{A}^{{\rm pol}}_{\rm prod}$
(\ref{eq:apolprodadep})
for neu\-tra\-lino production, the decay asymmetry
${\mathcal{A}}_\ell^{\prime {\rm pol},n}$ is approximately
maximal if the Higgs mixing is resonantly enhanced.
As pointed out earlier, this can happen naturally
in the Higgs decoupling limit.

%}}}

%}}}
% end of section Asymmetries.
%----------------------------------------------------------------------------------------

%----------------------------------------------------------------------------------------
%{{{ Numerics
\section{Numerical results
  \label{Numerical results}}
%---------------------------------------------------------------------------------
%
%{{{ Numerics introduction
We analyze numerically the 
CP-odd asymmetry $\mathcal{A}^{\rm pol}_{\rm prod}$~(\ref{eq:apolprod.ij}) 
for neu\-tra\-lino production
$\mu^+\mu^-\to\tilde\chi^0_1\tilde\chi^0_2$.
For the subsequent decays,
$\tilde\chi^0_2\to e \tilde e_R$
and
$\tilde\chi^0_2\to\tau\tilde \tau_1$,
we study the CP-even and CP-odd polarization asymmetries
${\mathcal{A}}_e^{{\rm pol}, R}$,
${\mathcal{A}}_\tau^{{\rm pol}, 1}$~(\ref{eq.apolijA}),
and
${\mathcal{A}}_e^{\prime {\rm pol}, R}$~(\ref{eq.apolijAprime}),
respectively.
The feasibility of measuring the asymmetries also depends
on the neu\-tra\-lino production cross section and decay branching ratios, 
which we discuss in detail.

We induce CP violation in the Higgs sector 
by a non-vanishing phase $\phi_A$ of the common 
trilinear scalar coupling parameter
$A_t=A_b=A_\tau\equiv|A|\exp(i\phi_A)$
for the third generation fermions.
This assignment is also compatible with the
bounds on CP-violating phases from experiments
on electric dipole moments 
(EDMs)~\cite{Yao:2006px,Harris:1999jx,Regan:2002ta,Romalis:2000mg}.
We assume a CP-conserving gaugino sector, i.e., we keep the gaugino 
mass parameters $M_1$, $M_2$, and the Higgs mass 
parameter $\mu$ real.
For the calculation of the Higgs masses, widths 
and couplings, we use
the program {\tt FeynHiggs 2.5.1}~\cite{Frank:2006yh,FH}, 
see also Appendix~\ref{app:Higgscouplings}.
For the branching ratios and decay width
of the neu\-tra\-lino, we include the two-body decays~\cite{Kittel:2005rp}
\begin{eqnarray}
\tilde\chi^0_2 &\to& 
\ell+\tilde\ell_n,~\quad
\nu_\ell+\tilde\nu_\ell,~\quad
\tilde\chi^0_1+ Z,~\quad
\tilde\chi^0_1+ H_1,
\label{neutdecaymodes}
\end{eqnarray}
with $n=R,L$ for $\ell=e,\mu$, and $n=1,2$ for $\ell=\tau$.
We neglect three-body decays. 
In order to enable the leptonic neu\-tra\-lino 
decays $\tilde\chi^0_2 \to \ell\tilde\ell_n$,
we need light sleptons. We 
parametrize their masses by $m_0$ and 
$M_2$, which enter in the approximate solutions 
to the renormalization group 
equations, see Appendix~\ref{neutdecay}.
We parametrize the diagonal entries of the
squark mass matrices by the common SUSY scale
parameter $M_{\rm SUSY}=M_{\tilde Q_3}=M_{\tilde U_3}=M_{\tilde D_3}$.
Finally, in order to reduce the number of parameters, we assume  
the GUT relation for the gaugino mass parameters
$M_1=5/3 \, M_2\tan^2\theta_W $.

\medskip

% ------------------------------------------------------------------------------
%{{{ Tables
% ------------------------------------------------------------------------------
%                                       T A B L E  -1-
% ------------------------------------------------------------------------------
\begin{table}
\renewcommand{\arraystretch}{1.2}
\caption{\label{scenarioA} SUSY parameters for the benchmark 
                           scenario~{CP$\chi$}.
The slepton masses are 
parametrized by $m_0$, %
the squark masses by $M_{\rm SUSY}$.
}
\begin{center}
     \begin{tabular}{|c|c|c|c|}
\hline
$ M_{H^\pm} = 500~{\rm GeV}$ & 
$ \tan\beta = 10 $ & 
$       |{A}| = 1~{\rm TeV}$ & 
$    \phi_A = 0.2 \pi$ \\
\hline
$ M_{\rm SUSY} = 500~{\rm GeV}$ &
$          \mu = 400~{\rm GeV}$ & 
$          M_2 = 300~{\rm GeV}$ & 
$          m_0 = 100~{\rm GeV}$ \\
\hline
\end{tabular}
\end{center}
\renewcommand{\arraystretch}{1.0}
\end{table}
% ------------------------------------------------------------------------------

% ------------------------------------------------------------------------------
%                                       T A B L E  -2-
% ------------------------------------------------------------------------------
\begin{table}
\renewcommand{\arraystretch}{1.2}
\caption{SUSY masses, widths, and branching ratios for the benchmark scenario~{CP$\chi$},
        evaluated with  
       {\tt FeynHiggs 2.5.1}~\cite{Frank:2006yh,FH}.
         \label{scenarioAmasses}}
\begin{center}
      \begin{tabular}{|c|c|c|c|}
\hline
$        M_{H_1}  = 126.0~{\rm GeV}$ & 
$   m_{\chi^0_1}  =   147~{\rm GeV}$ &
$ m_{\tilde{e}_R} =   180~{\rm GeV}$ & 
${\rm BR}(\tilde\chi_2^0\to e^+\tilde e_R^-) = 6.4 \%$ \\
\hline
$        M_{H_2}  = 492.8~{\rm GeV}$ & 
$   m_{\chi^0_2}  =   275~{\rm GeV}$ &
$ m_{\tilde{e}_L} =   289~{\rm GeV}$ & 
${\rm BR}(\tilde\chi_2^0\to \tau^+\tilde \tau_1^-) = 23 \%$ \\
\hline
$        M_{H_3}  = 493.6~{\rm GeV}$ & 
$   m_{\chi^0_3}  =   405~{\rm GeV}$ &
$m_{\tilde\tau_1} =   178~{\rm GeV}$ & 
${\rm BR}(\tilde\chi_2^0\to \tilde\chi_1^0 Z) = 9.1 \%$ \\
\hline
$   {\Gamma_{H_2}  =   0.97~{\rm GeV}}$ & 
$ m_{\chi^\pm_1}  =   274~{\rm GeV}$ &
$m_{\tilde\tau_2} =   290~{\rm GeV}$ & 
${\rm BR}(\tilde\chi_2^0\to \tilde\chi_1^0H_1) = 18 \%$ \\
\hline
$   {\Gamma_{H_3}  =   0.93~{\rm GeV}}$ & 
$ m_{\chi^\pm_2}  =   433~{\rm GeV}$ &
$ m_{\tilde\nu}   =   277~{\rm GeV}$ & 
$     \eta_\tau^1 = -0.53$\\
\hline
\end{tabular}
\end{center}
\renewcommand{\arraystretch}{1.0}
\end{table}

%}}}
%
% ------------------------------------------------------------------------------

We center our numerical discussion around 
scenario~{CP$\chi$}, defined in Table~\ref{scenarioA}.
Inspired by the benchmark scenario
CPX~\cite{Carena:2000ks} 
for studying enhanced CP-violating Higgs-mixing phenomena,
we set $|{A}|=2M_{\rm SUSY}=1~{\rm TeV}$, 
$M_3=800~{\rm GeV}$,
and a non-vanishing phase
$\phi_A = 0.2 \pi$.
We thus obtain large contributions from the trilinear 
coupling parameter ${A}$ of the third generation 
to the Higgs sector, both CP-conserving and CP-violating.
%{\color{blue}
In contrast to the CPX scenario, we do not need large values 
of $|\mu|$ to obtain large CP violating effects,
see discussion in Section~\ref{Adependence}.
%} 
We choose 
$\mu = 400~{\rm GeV}$ and $M_2 = 300~{\rm GeV}$ 
of similar size 
to enhance the branching ratios of the Higgs bosons into neu\-tra\-linos,
which are large only for mixed neu\-tra\-linos.
We fix $\tan\beta=10$, since the Higgs boson decays into neu\-tra\-linos 
are most relevant for intermediate values of $\tan\beta$.
Smaller values of $\tan\beta$
favor the $t\bar t$ decay channel, while
larger values enhance decays into $b\bar b$ and $\tau\bar\tau$.
We give the masses of the Higgs bosons, the charginos, neu\-tra\-linos, 
light sleptons, and the 
widths of the Higgs bosons
for scenario~{CP$\chi$}
in Table~\ref{scenarioAmasses}, 
where we also list the branching ratios for the decaying neu\-tra\-lino.
We choose longitudinal muon beam polarizations of
${\mathcal P}_+={\mathcal P}_-={\mathcal P}=\pm0.3$,
which should be feasible at a muon 
collider~\cite{hefreports}.
        %}}}
% ----------------------------------------------------------------------------

% ----------------------------------------------------------------------------
%{{{ sqrts dependence

\subsection{ $\sqrt{s}$ dependence}
\label{sec:sqrts.dependence}
% ----------------------------------------------------------------------------

For the scenario~{CP$\chi$},
we analyze the dependence of the asymmetries 
and the cross section on the center-of-mass energy $\sqrt{s}$.
The CP-even and CP-odd observables exhibit a characteristic 
$\sqrt s$ dependence, mainly given by
the product of Higgs boson propagators $\Delta_{(kl)}$,
see Eq.~(\ref{deltadelta}).
A  muon collider will have a very precise beam energy resolution,
and thus enables detailed line-shape scans.

\medskip

\begin{figure}[t]
\begin{picture}(16,13)
\put(-3,-10.3){\includegraphics{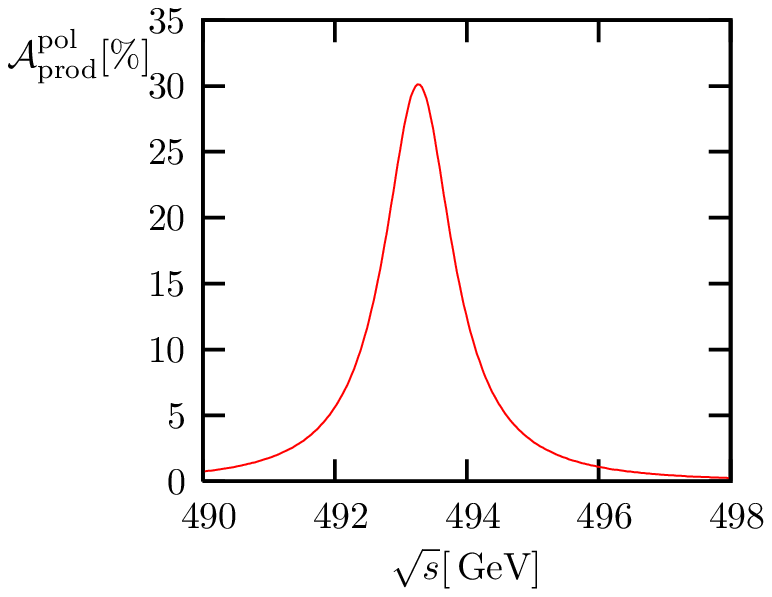}}
\put( 5,-10.3){\includegraphics{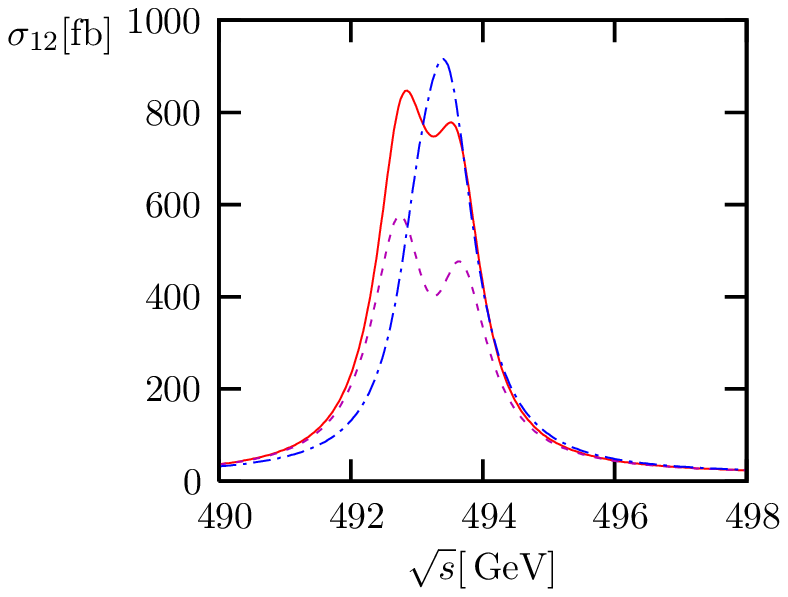}}
\put(0,7.){(a)}
\put(8,7.){(b)}
\put(-3,-17.3){\includegraphics{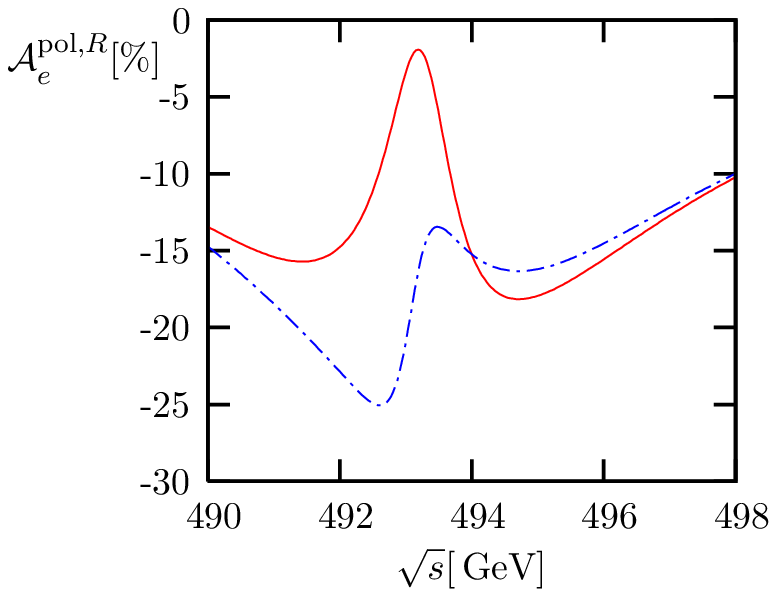}}
\put( 5,-17.3){\includegraphics{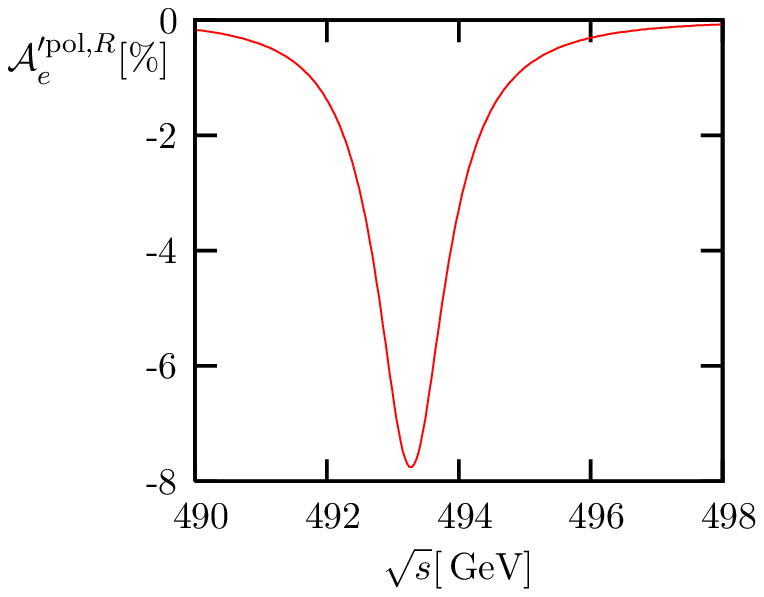}}
\put(0,0.15){(c)}
\put(8,0.15){(d)}
\end{picture}
\caption{\small  
        $\sqrt s$ dependence of 
        {\bf (a)}~the CP-odd production asymmetry 
        $\mathcal{A}^{\rm pol}_{\rm prod}$,
        Eq.~(\ref{eq:apolprod.ij}),
        and {\bf (b)}~the cross section $\sigma_{12}$
        for neu\-tra\-lino production
        $\mu^+\mu^- \to \tilde\chi_1^0\tilde\chi_2^0$.
        For the subsequent decay,
        $\tilde\chi_2^0\to e\tilde e_R$, 
        in {\bf (c)}~the CP-even polarization asymmetry
        $\mathcal{A}^{{\rm pol},R}_{e}$, Eq.~(\ref{eq.apolijA}),
        and in {\bf (d)}~the CP-odd polarization asymmetry
        $\mathcal{A}^{\prime {\rm pol},R}_{e}$, 
        Eq.~(\ref{eq.apolijAprime}),
        for the SUSY parameters as given in Table~\ref{scenarioA}.
        The longitudinal beam polarizations are
        ${\mathcal P}_+={\mathcal P}_-={\mathcal P}=+0.3$~(solid), 
        in {\bf (b)} ${\mathcal P}=-0.3$~(dashed), 
        and in {\bf (b,c)}
        ${\mathcal P}=+0.3$ with $\phi_A=0$~(dot-dashed). 
}
\label{fig:sqrts}
\end{figure}

In Fig.~\ref{fig:sqrts}(a), we show the CP-odd polarization
asymmetry $\mathcal{A}^{\rm pol}_{\rm prod}$
for neu\-tra\-lino production
$\mu^+\mu^- \to \tilde\chi_1^0\tilde\chi_2^0$
as a function of $\sqrt{s}$
around the heavy Higgs resonances $H_2$ and $H_3$.
At the peak value, $\sqrt{s}=(M_{H_2}+M_{H_3})/2\approx 493 $~GeV,
the interference of the two nearly degenerate Higgs bosons
is maximal, leading to an asymmetry of up to 
$\mathcal{A}^{\rm pol}_{\rm prod}=30\%$.
The asymmetry measures the difference
of the neu\-tra\-lino production cross section $\sigma_{12}({\mathcal P})$
for equal positive and negative muon beam polarizations
${\mathcal P}=\pm0.3$,
which we show in Fig.~\ref{fig:sqrts}(b).
We also observe that 
the splitting of the two resonances is increased in the presence of
CP-violating phases. For $\phi_A=0.2\pi$, the two resonances are
clearly visible in the line shape of the  cross section $\sigma_{12}$,
whereas it assumes a single resonance form for $\phi_A=0$,
where the Higgs bosons are extremely degenerate,
see Fig.~\ref{fig:sqrts}(b).

\medskip

The Higgs boson interference in neu\-tra\-lino production also leads
to CP-even and CP-odd contributions to the average longitudinal 
neu\-tra\-lino polarizations. In order to analyze the $\tilde\chi_2^0$
polarization, we discuss the CP-even asymmetry 
$\mathcal{A}^{{\rm pol},R}_e$
and the CP-odd asymmetry
$\mathcal{A}_e^{\prime {\rm pol}, R}$
of the leptonic energy distributions
for the neu\-tra\-lino decay.
For simplicity, we discuss only the decay into a right selectron
$\tilde\chi_2^0\to e\tilde e_R$.
The same asymmetries are obtained for the decay into a right smuon
$\tilde\chi_2^0\to \mu\tilde\mu_R$.
For the neu\-tra\-lino decay into a tau,
$\tilde\chi_2^0\to \tau\tilde\tau_1$,
 the corresponding asymmetries %and significances
are obtained from Eq.~(\ref{eq:realtionAtau}).

\medskip

For the decay
$\tilde\chi_2^0\to e\tilde e_R$,
we show the $\sqrt{s}$ dependence of the
CP- and CP${\tilde{\rm T}}$-even asymmetry
$\mathcal{A}^{{\rm pol},R}_{e}$ in Fig.~\ref{fig:sqrts}(c)
for $\phi_A=0$ and $\phi_A=0.2\pi$.
The CP- and CP${\tilde{\rm T}}$-odd asymmetry 
$\mathcal{A}^{\prime {\rm pol},R}_{e}$ is shown in Fig.~\ref{fig:sqrts}(d).
The phase $\phi_A$ tends to increase the mass splitting of the Higgs resonances. 
Their overlap is now reduced, leading in general to a suppression of the CP-even
asymmetry  $\mathcal{A}^{{\rm pol},R}_{e}$, in particular
at the mean energy of the resonances $\sqrt{s}=(M_{H_2}+M_{H_3})/2$,
see Fig.~\ref{fig:sqrts}(c). On the contrary, the larger Higgs splitting
increases the CP-odd asymmetries $\mathcal{A}^{\prime {\rm pol},R}_{e}$
and $\mathcal{A}^{\rm pol}_{\rm prod}$.

\medskip

All asymmetries for production and decay vanish asymptotically far from the 
resonance region. The continuum contributions from selectron and $Z$ exchange
to the difference of the cross sections and to the average neutralino
polarization cancel in the numerator, but contribute in the denominator
of the corresponding asymmetries, see 
their definitions in Section~\ref{AsymmetriesforPandD}.

\medskip

% ---------------------------------------------------------------------------
%% -- F I G U R E (-x-)      {fig:ACP+}
% ---------------------------------------------------------------------------
% -----------------------------------------------------------------
%                        P L O T - 2 
% -----------------------------------------------------------------
% ----------------------------------------------------------------------------------------

In the following Sections, we analyze 
the dependence of the 
production cross section 
and the asymmetries
on $|{A}|$ and $\phi_A$,
and finally on $M_2$ and $\mu$, 
fixing all remaining parameters to those of 
scenario~{CP$\chi$}. 
We fix the center-of-mass energy to
$\sqrt{s}=(M_{H_2}+M_{H_3})/2$, 
where we expect the largest CP-odd asymmetries
$\mathcal{A}^{\rm pol}_{\rm prod}$ and
$\mathcal{A}^{\prime {\rm pol},R}_{e}$,
see Figs.~\ref{fig:sqrts}(a) and (d), respectively.
For consistency, we also choose 
$\sqrt{s}=(M_{H_2}+M_{H_3})/2$
for the discussion of the CP-even asymmetry
$\mathcal{A}^{{\rm pol},R}_{e}$,
although it is generally suppressed at this value if CP is violated.

        %}}}

% ----------------------------------------------------------------------------
%{{{ A_t dependence

\subsection{$|{A}|$ and $\phi_A$ dependence}
\label{Adependence}
% ----------------------------------------------------------------------------

\begin{figure}[t]
\begin{picture}(16,13.)
\put(-3,-10.3){\includegraphics{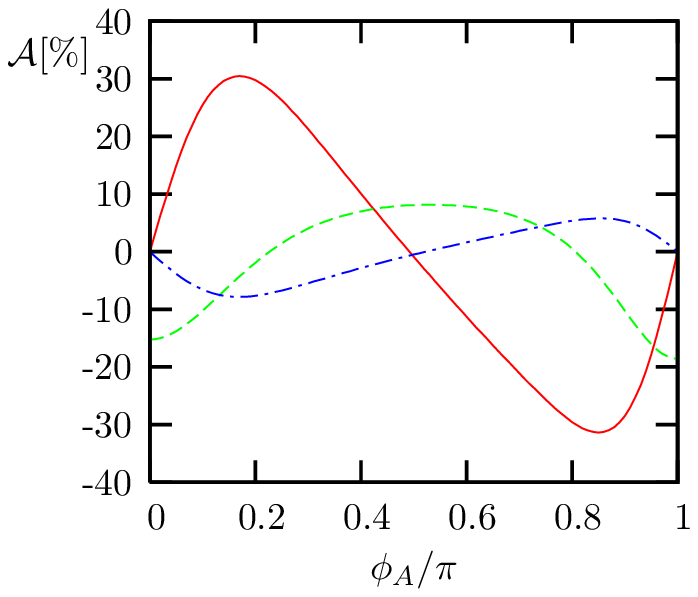}}
\put( 5,-10.3){\includegraphics{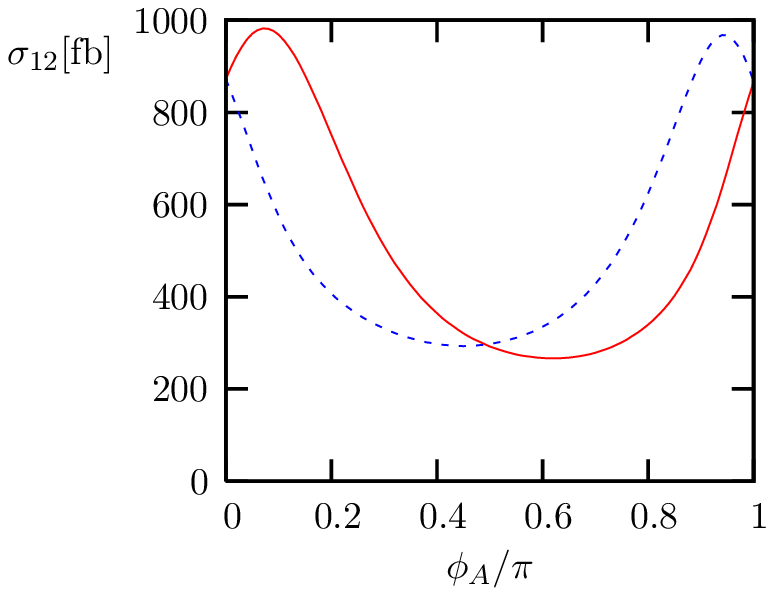}}
\put(.0,7.15){(a)}
\put(8.,7.15){(b)}
\put(-3,-17.3){\includegraphics{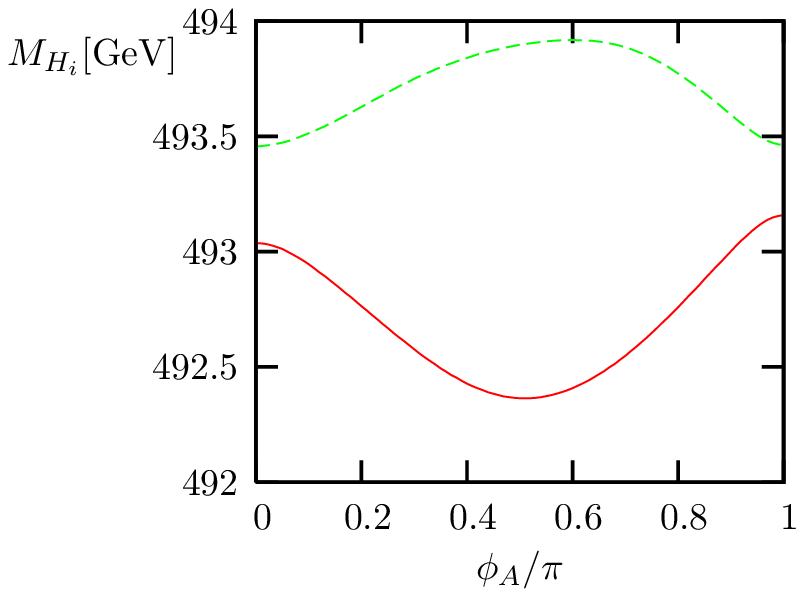}}
\put( 5,-17.3){\includegraphics{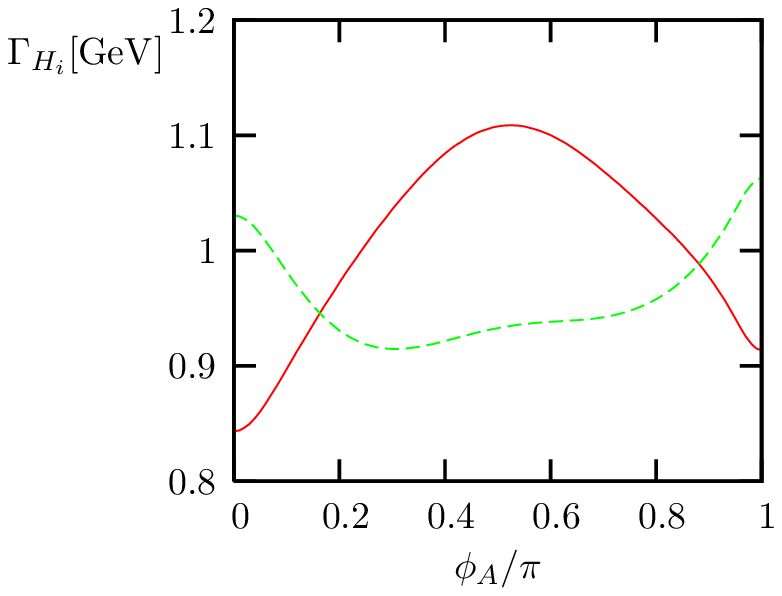}}
\put(.0,0.15){(c)}
\put(8.,0.15){(d)}

\put(2.9,11.8){${\mathcal{A}}^{\rm pol}_{\rm prod}$}
\put(4.9,10.9){${\mathcal{A}}_e^{\prime {\rm pol}, R}$}
\put(1.6,9.0){${\mathcal{A}}_e^{ {\rm pol}, R}$}

\end{picture}
\caption{\small  
      Phase dependence of 
      {\bf (a)}~the CP-odd polarization asymmetry
      $\mathcal{A}^{\rm pol}_{\rm prod}$ 
      (solid), Eq.~(\ref{eq:apolprod.ij}), 
      for neu\-tra\-lino production 
      $\mu^+\mu^- \to \tilde\chi_1^0\tilde\chi_2^0$
      at $\sqrt{s}=(M_{H_2}+M_{H_3})/2$,
      and for the subsequent decay 
      $\tilde\chi_2^0\to e\tilde e_R$
      the CP-even polarization asymmetry 
      ${\mathcal{A}}_e^{ {\rm pol}, R}$ 
      (dashed), Eq.~(\ref{eq.apolijA}),
      and the CP-odd polarization asymmetry
      ${\mathcal{A}}_e^{\prime {\rm pol}, R}$ 
      (dot-dashed), Eq.~(\ref{eq.apolijAprime}).
      In {\bf (b)}~the cross section $\sigma_{12}$
      for neu\-tra\-lino production
      $\mu^+\mu^- \to \tilde\chi_1^0\tilde\chi_2^0$
      with longitudinal beam polarizations
      ${\mathcal P}_+={\mathcal P}_-={\mathcal P}=+0.3$~(solid),
      and ${\mathcal P}=-0.3$~(dotted).
      In {\bf (c)}~the Higgs masses $M_{H_i}$ and
      {\bf (d)}~the Higgs widths $\Gamma_{H_i}$,
      for $i=2$ (solid), and $i=3$ (dashed).
      The SUSY parameters are given in Table~\ref{scenarioA}.
}
\label{fig:phi}
\end{figure}

We analyze the dependence of the CP-asymmetries on the phase $\phi_A$ of
the trilinear coupling ${A}$,
which is the only source of CP violation in our study.
The CP-odd asymmetries, 
$\mathcal{A}^{{\rm pol}\,}_{\rm prod}$ and
${\mathcal{A}}_e^{\prime {\rm pol}, R}$,
see Fig.~\ref{fig:phi}(a),
are approximately maximal, if the 
mixing of the Higgs states is resonantly enhanced.
This is naturally achieved when the diagonal elements 
$m_{H}^2 - \hat\Sigma_{HH}(s)$ and $m_{A}^2 - \hat\Sigma_{AA}(s)$
of the Higgs mass matrix ${\rm \bf M}$, Eq.~(\ref{eq:Weisskopf}), are equal,
provided the corresponding imaginary part is small.
In our scenario, 
where decays into heavy squarks are not kinematically allowed,
this condition is roughly fulfilled for 
$\phi_A\simeq 0.2\pi$. We interpret this condition as a level crossing of the
CP-eigenstates  $H$ and $A$, when 
$m_{H}^2 - m_{A}^2 - \rm{Re}[ \hat\Sigma_{HH}(s)-\hat\Sigma_{AA}(s)]$
changes sign~\cite{Choi:2004kq}. 
The mass difference of the physical Higgs boson masses,
however, is typically increased by the $H$--$A$ mixing, as can be observed
from Fig.~\ref{fig:phi}(c). A splitting of the order of the Higgs
widths $\Gamma_{H_{2,3}}$, shown in Fig.~\ref{fig:phi}(d),
leads to large absorptive phases, which are necessary for the presence of
CP$\tilde{\rm{T}}$-odd observables.
The increased Higgs mass splitting for non-vanishing phases 
leads, however, in general to lower peak cross sections $\sigma_{12}({\mathcal P})$, 
which we show in Fig.~\ref{fig:phi}(b), both for positive and negative
beam polarizations ${\mathcal P}=\pm0.3$.

% ---------------------------------------------------------------------------
%% -- F I G U R E (-x-)      {fig:ACP+}
% ---------------------------------------------------------------------------
% -----------------------------------------------------------------
%                        P L O T - 3 
% -----------------------------------------------------------------
% plots done with fhneut/Plots_n300400/plot_mgama_phi.gpl
% ----------------------------------------------------------------------------------------

\medskip

The asymmetries and cross section for negative $\phi_A$ 
can be obtained from symmetry considerations.
Since the complex trilinear coupling ${A}$ is the only 
source of CP violation in our analysis, the CP-odd asymmetries
$\mathcal{A}^{{\rm pol}\,}_{\rm prod}$ and
${\mathcal{A}}_e^{\prime {\rm pol}, R}$ 
must be odd with respect to 
the transformation $\phi_A \to -\phi_A$,
while the CP-even asymmetry ${\mathcal{A}}_e^{ {\rm pol}, R}$
must be even. Consequently, the cross section transforms as
$\sigma_{12}({\mathcal P}) \to \sigma_{12}({-\mathcal P})$. 

\medskip

\begin{figure}[t]
\centering
\begin{picture}(16,15.5)
\put( 3.5,-6.2){\includegraphics{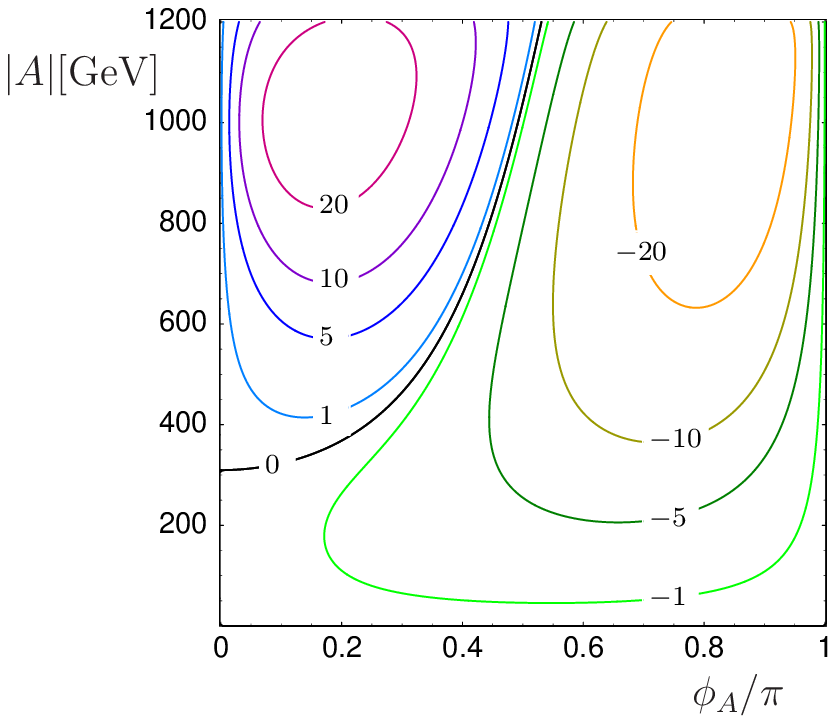}}
\put(-4.5,-6.2){\includegraphics{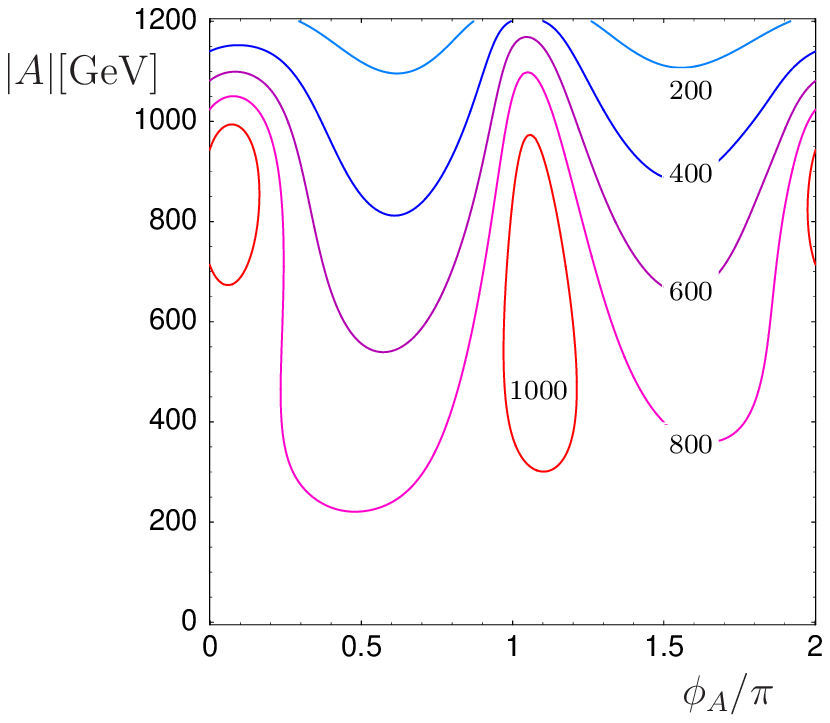}}
\put(1,8.15){(a)}
\put(9,8.15){(b)}
\put(1.6,14.9){ $\sigma(\mu^+\mu^-\to\neutralino_1\neutralino_2)$ 
                         in $\rm{fb}$  }
\put(9.6,14.9){Asymmetry $\mathcal{A}^{\rm pol}_{\rm prod}$ in \%}
\put(-4.5,-14.2){\includegraphics{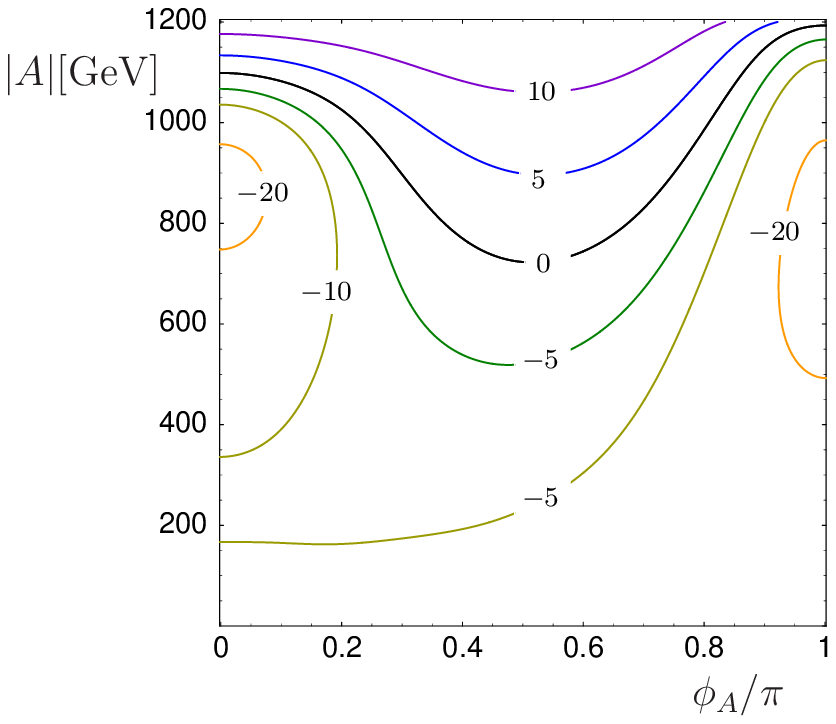}}
\put(3.50,-14.2){\includegraphics{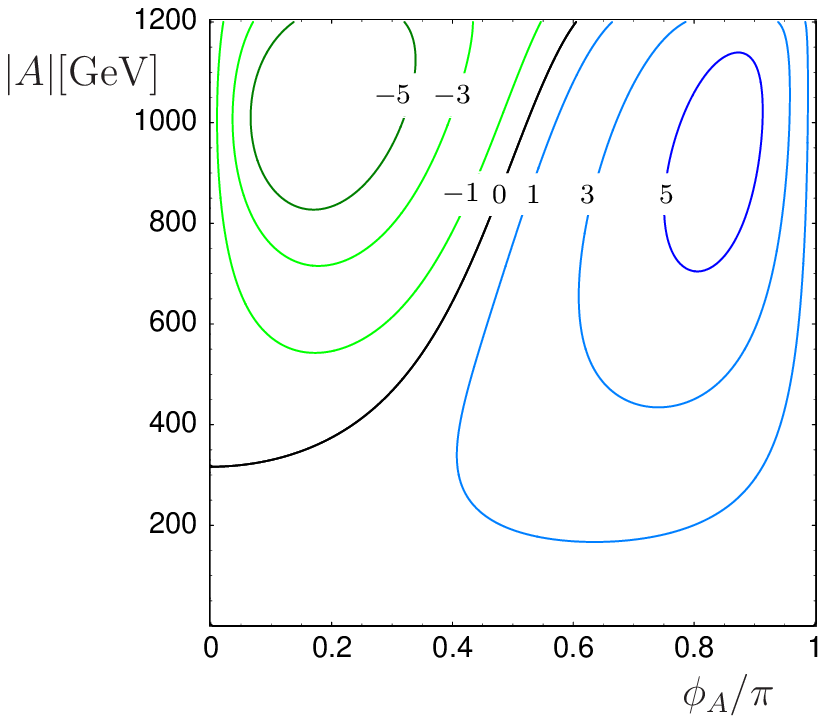}}
\put(1,0.15){(c)}
\put(9,0.15){(d)}
\put(1.4,6.9){  Asymmetry $\mathcal{A}^{{\rm pol},R}_e$ in \%}
\put(9.4,6.9){ Asymmetry ${\mathcal{A}}_e^{\prime {\rm pol}, R}$ in \%}
%%%%%%%%%%%%%%%%%%%%%%%%
\end{picture}
\caption{\small  
      Contour lines in the $\phi_A$--$|{A}|$ plane
      for {\bf (a)} the cross section for neu\-tra\-lino 
      production 
      $\mu^+\mu^- \to \tilde\chi_1^0\tilde\chi_2^0$,
      {\bf (b)} the CP-odd polarization asymmetry
      $\mathcal{A}^{\rm pol}_{\rm prod}$,
      and for the subsequent decay
      $\tilde\chi_2^0\to e\tilde e_R$,
      {\bf (c)} the CP-even polarization asymmetry 
      $\mathcal{A}^{{\rm pol},R}_e$, 
      Eq.~(\ref{eq.apolijA}),
      and {\bf (d)} the CP-odd polarization asymmetry
      ${\mathcal{A}}_e^{\prime {\rm pol}, R}$, 
      Eq.~(\ref{eq.apolijAprime}),
      at $\sqrt{s}=(M_{H_2}+M_{H_3})/2$ with 
      longitudinally polarized beams 
      $\mathcal{P}_{-}=\mathcal{P}_{+}=\pm0.3$.
       The SUSY parameters are given in Table~\ref{scenarioA}.
 }
\label{fig:asymm.AphiA}
\end{figure}

In Fig.~\ref{fig:asymm.AphiA}, we show contour lines of the 
cross section and the asymmetries in the $\phi_A$--$|{A}|$ plane. 
The largest CP-odd asymmetries $\mathcal{A}^{\rm pol}_{\rm prod}$ 
and ${\mathcal{A}}_e^{\prime {\rm pol}, R}$ 
are obtained for $|{A}|\approx 2 M_{\rm SUSY} = 1$~TeV.
For larger values of $|{A}|$,
the lighter stops become kinematically accessible and
$H_2$ decays dominantly into $\tilde{t}_1^+ \tilde{t}_1^-$ pairs,
which leads to a suppression of the neu\-tra\-lino production cross section.
We therefore restrict our discussion to $|{A}| < 1.2$~TeV.

\medskip

As we have observed in Fig.~\ref{fig:phi}(a) of the preceding paragraph,
the CP-even asymmetry $\mathcal{A}^{{\rm pol},R}_e$,
Fig.~\ref{fig:asymm.AphiA}(c), 
is in general larger in the CP-conserving limit. 
The maximum of the asymmetry
is also obtained for ${A}\approx \pm 800~{\rm GeV}$. 
However, this is rather co\"incidental, and is due to the exact 
degeneracy of the Higgs bosons $H$ and $A$.

\medskip
Note that large resonant mixing is possible without requiring large values of $|\mu|$.
The CP-violating scalar-pseudoscalar self energy transitions 
are proportional to the amount of CP violation in the squark sector, described by the quantity
\begin{equation}
\frac{3}{16\pi^2}\frac{{\rm Im}( A_f\mu)}{m^2_{\tilde{f}_2} - m^2_{\tilde{f}_1}},
\end{equation}
with $f=t,b$~\cite{Accomando:2006ga,PilaftsisAH,Asakawa:2000es}.
However we obtain large $H$--$A$ mixing for moderate values of $\mu$, 
since, in the Higgs decoupling limit 
$\rm{Im} \hat\Sigma_{HH}(s)\simeq \rm{Im}\hat\Sigma_{AA}(s)$
for energies below the threshold of heavy squark pair production.
Therefore, the conditions for maximally resonance enhanced mixing discussed in this section may be fulfilled for moderate values of $\hat\Sigma_{HA}(s)$.

%%%%%%%%%%%%%%%%%%%%%%%%%%%%%%%%%%%%%%%%%%%%%%%%%%%%%%%%%%%%%%%%%%%%%%%%%%%%%%
% -----------------------------------------------------------------
%                        P L O T - 4  
% -----------------------------------------------------------------
%%%%%%%%%%%%%%%%%%%%%%%%%%%%%%%%%%%%%%%%%%%%%%%%%%%%%%%%%%%%%%%%%%%%%%%%%%%%%%

%\newpage

%}}}

% ------------------------------------------------------------------------------
%{{{ mu & M2 dependence

\subsection{  $\mu$ and $M_2$ dependence}
\label{sec:mu.and.M_2.dependence}
\begin{figure}[t]
\centering
\begin{picture}(16,7.5)
\put(-4.5,-14.2){\includegraphics{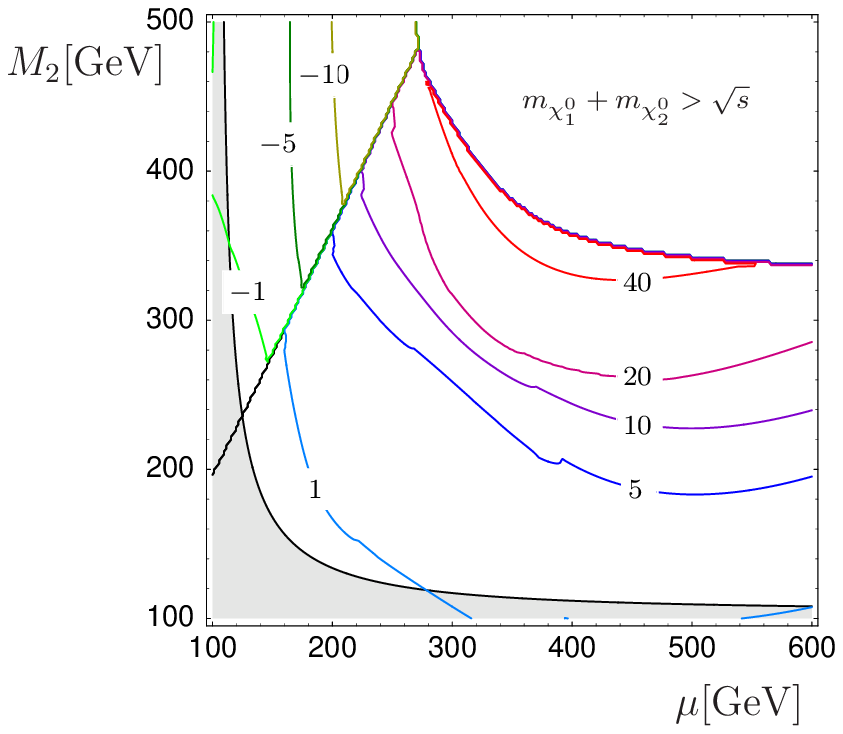}}
\put(3.5,-14.2){\includegraphics{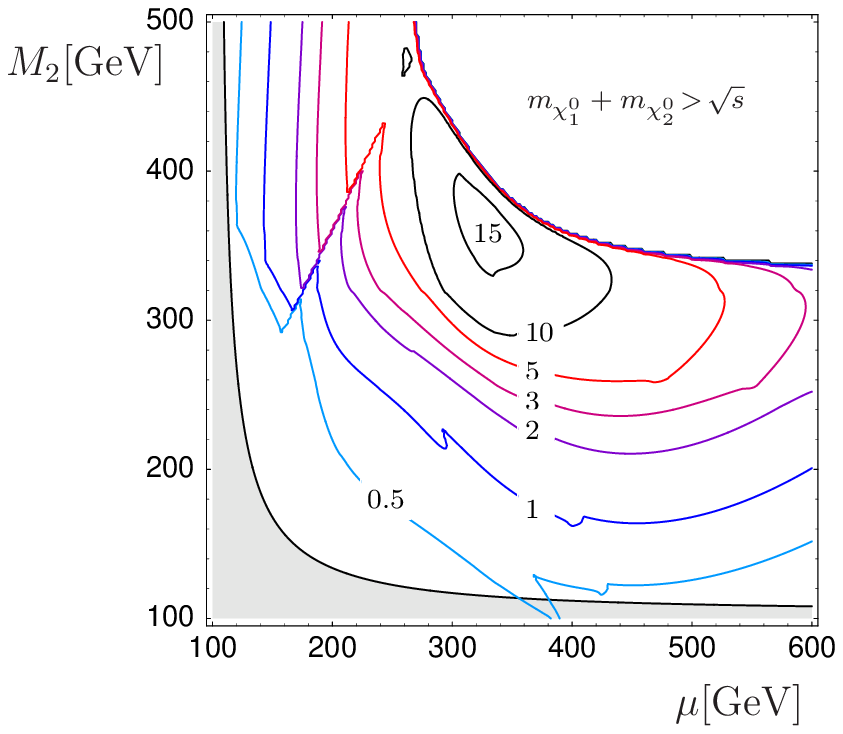}}
\put(1.0,0.15){(a)}
\put(9.0,0.15){(b)}
\put(1.4,6.9){Asymmetry $\mathcal{A}^{\rm pol}_{\rm prod}$ in \%}
\put(10.,6.9){Significance $\mathcal{S}^{\rm pol}_{\rm prod}$}
\end{picture}
\caption{\small 
        Neutralino production
        $\mu^+\mu^-\to\tilde\chi_1^0\tilde\chi_2^0$ at
        $\sqrt{s}=(M_{H_2}+M_{H_3})/2$ with 
        longitudinally polarized beams 
        $\mathcal{P}_{-}=\mathcal{P}_{+}=-0.3$.
        Contour lines in the $\mu$--$M_2$ plane for
        {\bf (a)}~the CP-odd production asymmetry 
        $\mathcal{A}^{\rm pol}_{\rm prod}$~(\ref{eq:apolprod.ij}),
        and {\bf (b)}~the
        significance $\mathcal{S}^{\rm pol}_{\rm prod}$~(\ref{significance1}),
        with ${\mathcal L}=1~{\rm fb}^{-1}$,
        for the SUSY parameters as given in Table~\ref{scenarioA}.
        The corresponding cross section 
        $\sigma(\mu^+\mu^-\to\tilde\chi_1^0\tilde\chi_2^0)$
        is given in Fig.~\ref{fig:sigma.A}.
        The shaded area is excluded by requiring $m_{\chi_1^\pm}>103$~GeV.
}
\label{fig:asymm.prod.A}
\end{figure}

% ------------------------------------------------------------------------------
%
The couplings of the Higgs bosons to the neu\-tra\-linos
strongly depend on the gaugino-higgsino composition of the neu\-tra\-linos,
which are mainly determined by the values of $\mu$ and $M_2$.
For neu\-tra\-lino production $\mu^+\mu^-\to\tilde\chi_1^0\tilde\chi_2^0$,
we show the CP-odd polarization
asymmetry $\mathcal{A}^{\rm pol}_{\rm prod}$~(\ref{eq:apolprod.ij})
in the $\mu$--$M_2$ plane in Fig.~{\ref{fig:asymm.prod.A}}(a).
For ${\mathcal P}=\pm0.3$,
the maximum absolute value of the asymmetry would be
 $\mathcal{A}^{\rm pol, max}_{\rm prod}\approx 55\%$,
as follows from Eq.~(\ref{eq:amax}).
We observe in Fig.~{\ref{fig:asymm.prod.A}}(a)
that the asymmetry reaches $40\%$ near the neu\-tra\-lino production threshold,
where the coefficient $a_1$, see Eq.~(\ref{eq.a1}), 
receives large spin-flip contributions.
For smaller $\mu$ and $M_2$
the Higgs boson widths are increased, since decay channels
into light neu\-tra\-linos and charginos open.
This results in a larger overlap of the Higgs resonances,
which reduces the absorptive phases, and consequently 
suppresses the CP$\tilde{\rm T}$-odd asymmetry
$\mathcal{A}^{\rm pol}_{\rm prod}$.
In the upper left corner of Fig.~\ref{fig:asymm.prod.A}(a), 
the asymmetry changes sign due to a level crossing
of the two neu\-tra\-linos $\tilde\chi_2^0$ and $\tilde\chi_3^0$.

%%%%%%%%%%%%%%%%%%%%%%%%%%%%%%%%%%%%%%%%%%%%%%%%%%%%%%%%%%%%%%%%%%%%%%%%%%%%%%
% contour plots in mu-M2 plane.
%%%%%%%%%%%%%%%%%%%%%%%%%%%%%%%%%%%%%%%%%%%%%%%%%%%%%%%%%%%%%%%%%%%%%%%%%%%%%%
% -----------------------------------------------------------------
%                        P L O T - 5 
% -----------------------------------------------------------------
%%%%%%%%%%%%%%%%%%%%%%%%%%%%%%%%%%%%%%%%%%%%%%%%%%%%%%%%%%%%%%%%%%%%%%%%%%%%%%

\medskip

\begin{figure}[t]
\centering
\begin{picture}(16,7.5)
\put(3.5,-14.2){\includegraphics{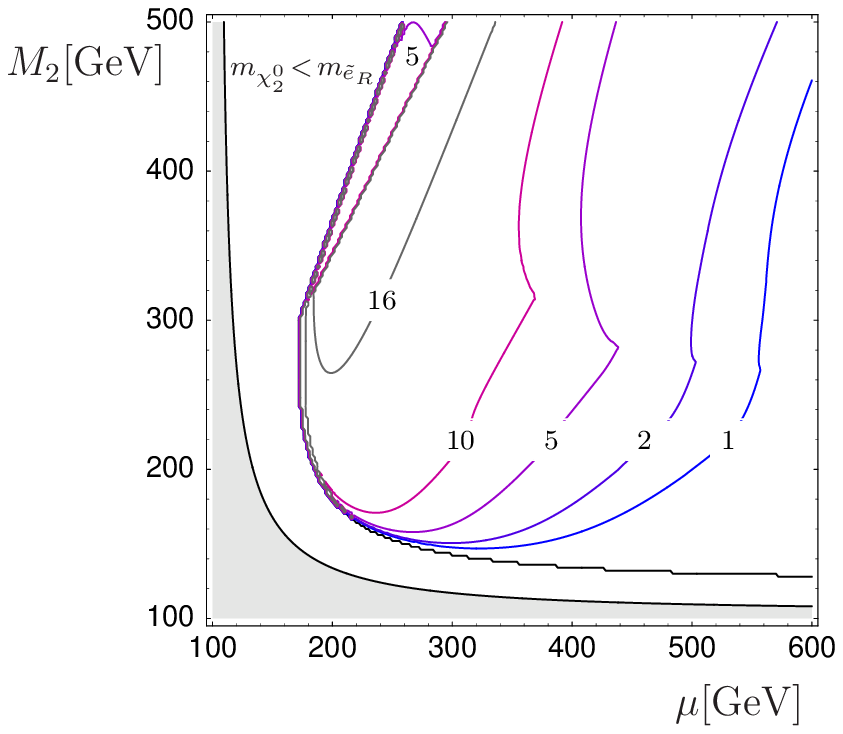}}
\put(-4.5,-14.2){\includegraphics{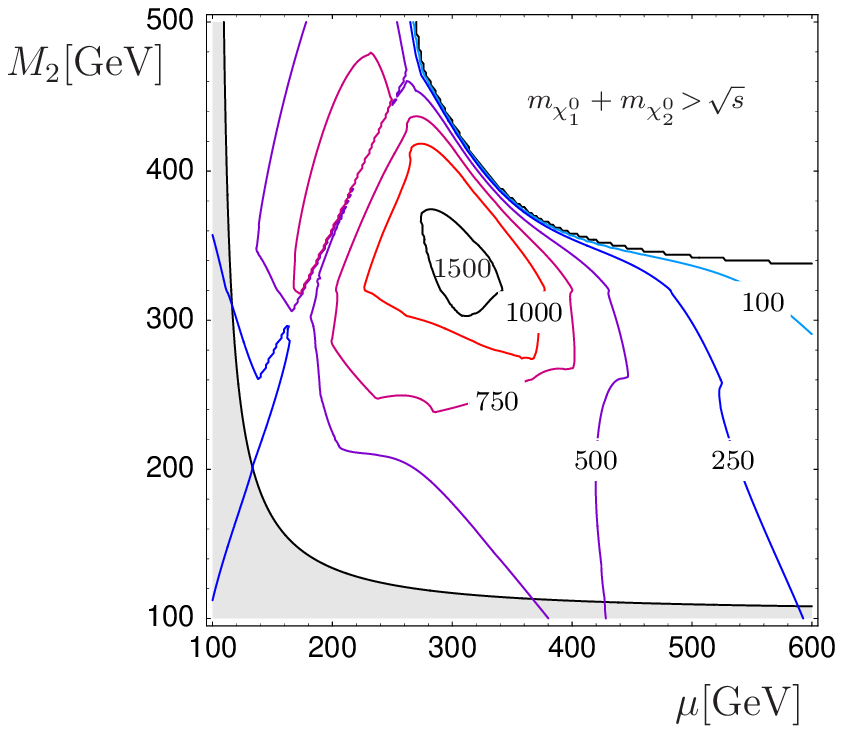}}
\put(1.0,0.15){(a)}
\put(9.0,0.15){(b)}
\put(1.4,6.9){$\sigma(\mu^+\mu^-\to\neutralino_1\neutralino_2)$ 
                         in $\rm{fb}$}
\put(10.,6.9){${\rm BR}(\tilde\chi_2^0\to e^+\tilde e_R^-)$ in \%}
\end{picture}
\caption{\small 
         Contour lines in the $\mu$--$M_2$ of 
        {\bf (a)}~the cross section 
        $\sigma_{12}=\sigma(\mu^+\mu^-\to\tilde\chi_1^0\tilde\chi_2^0)$
        at $\sqrt{s}=(M_{H_2}+M_{H_3})/2$ with 
        longitudinally polarized beams 
        $\mathcal{P}_{-}=\mathcal{P}_{+}=0.3$,
        {\bf (b)}~the branching ratio 
        ${\rm BR}(\tilde\chi_2^0\to e^+\tilde e_R^-)$,
        for the SUSY parameters as given in Table~\ref{scenarioA}.
        The shaded area is excluded by requiring $m_{\chi_1^\pm}>103$~GeV.
}
\label{fig:sigma.A}
\end{figure}

In Fig.~\ref{fig:sigma.A}(a), we show the 
cross section $\sigma_{12}$ for neu\-tra\-lino production
$\mu^+\mu^-\to\tilde\chi_1^0\tilde\chi_2^0$.
In the mixed region $|\mu| \simeq M_2$,
where the Higgs-neu\-tra\-lino couplings are larger,
the cross section reaches up to $\sigma_{12}\approx 1500$~fb.
In addition, since $H_2$ and $H_3$ are mixed CP-eigenstates,
there is no $p$-wave suppression.
However, due to the Majorana nature of the neu\-tra\-linos, 
the continuum contribution from $\tilde\mu$ and $Z$ exchange
to the cross section is $p$-wave suppressed~\cite{Choi:2006mr}.
It is thus negligible near threshold, and reaches $150$~fb 
only for $\mu\lsim150$~GeV.
In Fig.~\ref{fig:sigma.A}(b), we
show the branching ratio for the neu\-tra\-lino decay
$\tilde\chi_2^0\to e^+\tilde e_R^-$.
The decay fraction is reduced  
in the upper right corner, since the channels
$\tilde\chi^0_2 \to \tilde\chi^0_1 Z$, and
$\tilde\chi^0_2 \to \tilde\chi^0_1 H_1$,
open. In particular, the branching ratio
into the lightest Higgs boson can be 
${\rm BR}(\tilde\chi^0_2 \to \tilde\chi^0_1 H_1)> 60\%$,
for $M_2\gsim400$~GeV.

% -----------------------------------------------------------------
%                        P L O T - 6 
% -----------------------------------------------------------------
%%%%%%%%%%%%%%%%%%%%%%%%%%%%%%%%%%%%%%%%%%%%%%%%%%%%%%%%%%%%%%%%%%%%%%%%%%%%%%

\begin{figure}[t]
\centering
\begin{picture}(16,15.5)
\put(-4.5,-6.2){\includegraphics{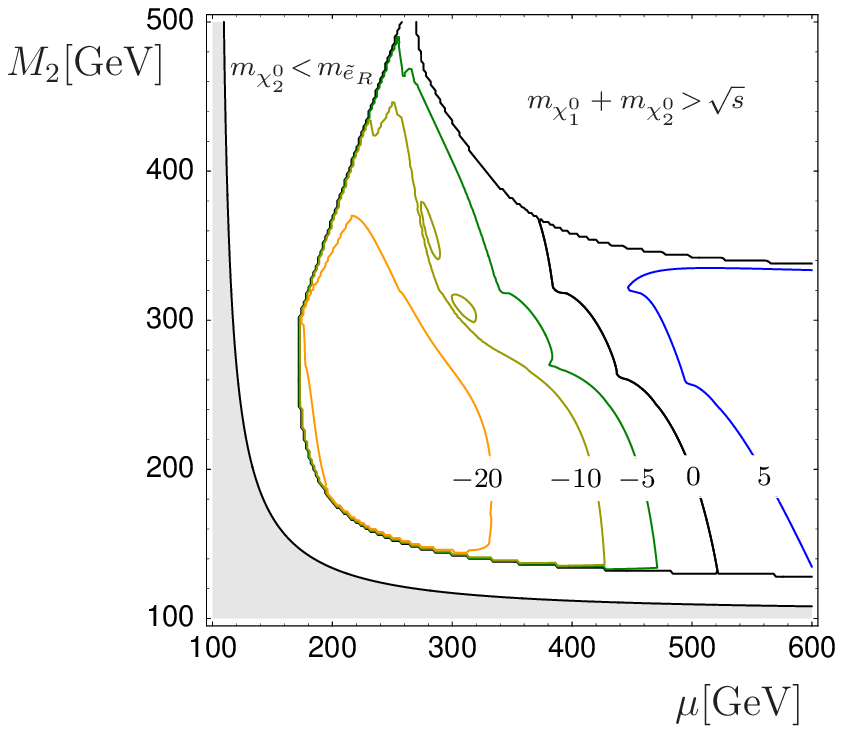}}
\put( 3.5,-6.2){\includegraphics{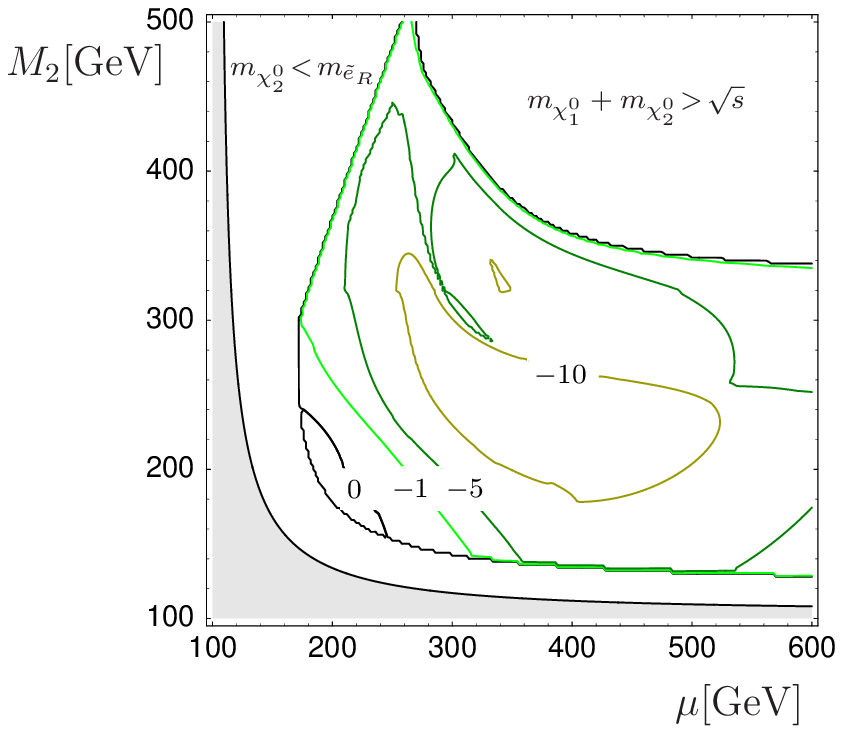}}
\put(1,8.15){(a)}
\put(9,8.15){(b)}
\put(1.4,14.9){ Asymmetry $\mathcal{A}_e^{{\rm pol}, R}$ in \%}
\put(10.,14.9){ Asymmetry $\mathcal{A}_e^{\prime {\rm pol}, R}$ in \%}
\put(-4.5,-14.2){\includegraphics{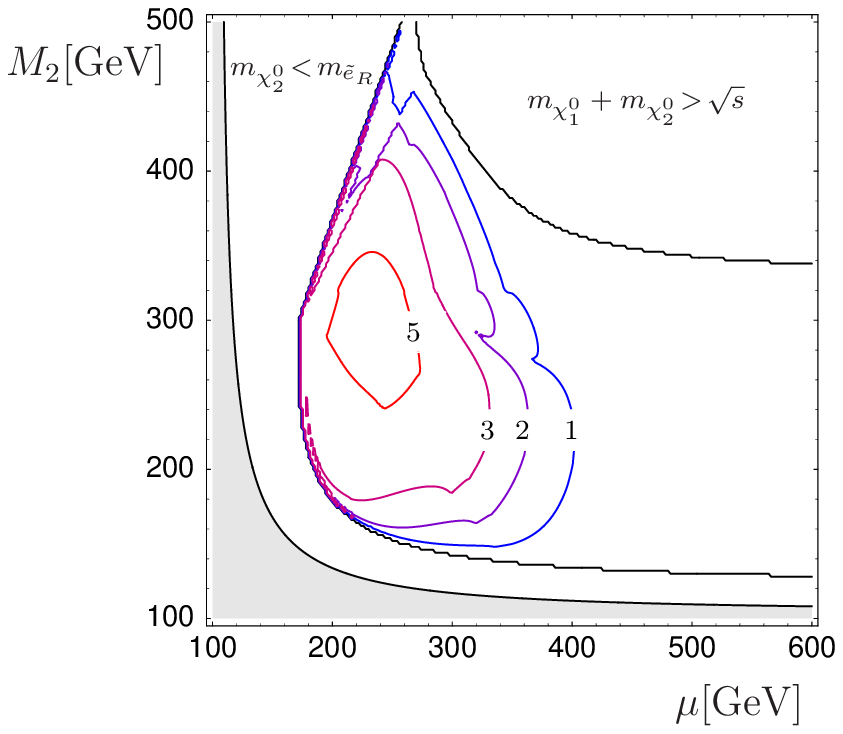}}
\put(3.5,-14.2){\includegraphics{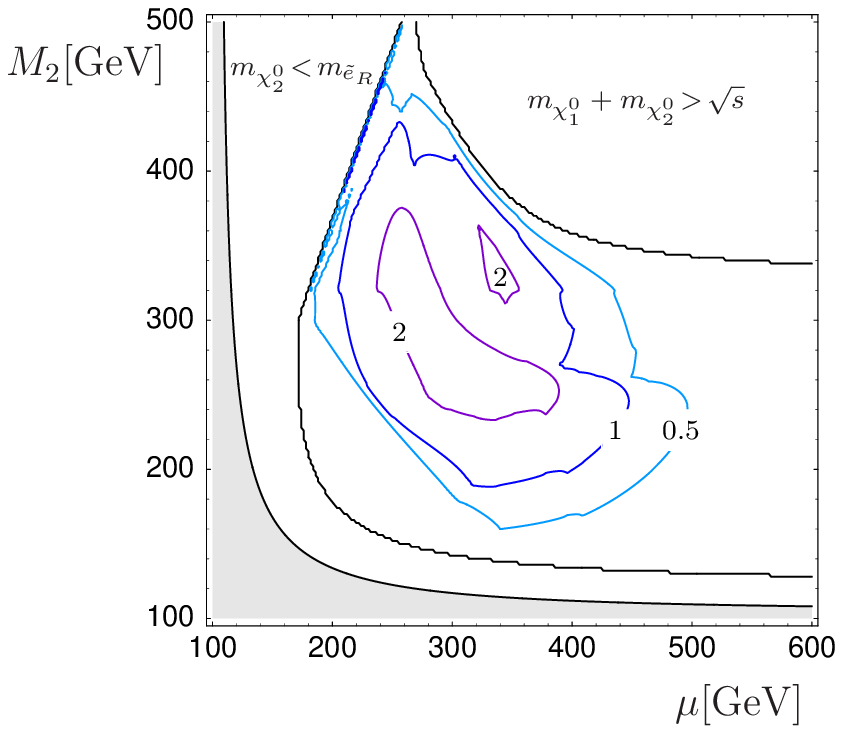}}
\put(1,0.15){(c)}
\put(9,0.15){(d)}
\put(1.4,6.9){Significance $\mathcal{S}_e^{{\rm pol}, R}$}
\put(10.,6.9){Significance $\mathcal{S}_e^{\prime {\rm pol}, R}$}
%%%%%%%%%%%%%%%%%%%%%%%%
\end{picture}
\caption{\small  
       Neutralino production 
       $\mu^+\mu^-\to\tilde\chi_1^0\tilde\chi_2^0$
       and decay 
       $\tilde\chi_2^0\to e\tilde e_R$
       at $\sqrt{s}=(M_{H_2}+M_{H_3})/2$ with 
       longitudinally polarized beams 
       $\mathcal{P}_{-}=\mathcal{P}_{+}=\pm0.3$.
       Contour lines in the $\mu$--$M_2$ plane for
       {\bf (a)}~the CP-even polarization asymmetry
       $\mathcal{A}^{{\rm pol},R}_e$, Eq.~(\ref{eq.apolijA}),
       {\bf (b)}~the CP-odd polarization asymmetry
       ${\mathcal{A}}_e^{\prime {\rm pol}, R}$,
       Eq.~(\ref{eq.apolijAprime}),
       and 
       {\bf (c)}~the significance 
       $\mathcal{S}_e^{{\rm pol}, R}$, 
       Eq.~(\ref{significance3}), 
       and {\bf (d)} the significance
       $\mathcal{S}_e^{\prime {\rm pol}, R}$, 
       Eq.~(\ref{significance4}),  
       with the effective luminosity 
       ${\mathcal L}_{\rm eff}=1~{\rm fb}^{-1}$.
       The SUSY parameters are given in Table~\ref{scenarioA}.
        The neu\-tra\-lino production cross section,
        $\sigma(\mu^+\mu^-\to\tilde\chi_1^0\tilde\chi_2^0)$
        and branching ratio 
        ${\rm BR}(\tilde\chi_2^0\to e^+\tilde e_R^-)$
        are shown in Fig.~\ref{fig:sigma.A}.
        The shaded area is excluded by requiring $m_{\chi_1^\pm}>103$~GeV.
 }
\label{fig:asymm.electron}
\end{figure}

For the neu\-tra\-lino decay
$\tilde\chi_2^0\to e\tilde e_R$,
we show the polarization asymmetries $\mathcal{A}^{{\rm pol},R}_e$
and $\mathcal{A}^{\prime {\rm pol},R}_e$
in Fig.~\ref{fig:asymm.electron}(a) and (b), respectively. 
As discussed before,
the CP-even asymmetry $\mathcal{A}^{{\rm pol},R}_e$ is suppressed 
by CP-violating effects due to the smaller overlap of the resonances 
at $\sqrt{s}=(M_{H_2}+M_{H_3})/2$.
Therefore we only find large values of 
 $\mathcal{A}^{{\rm pol},R}_e$
for light neu\-tra\-linos and charginos in the lower left corner of 
Fig.~\ref{fig:asymm.electron}(a), where the larger Higgs widths counter 
the effect of the larger Higgs mass difference.
On the contrary, in that region the CP-odd asymmetry
${\mathcal{A}}_e^{\prime {\rm pol}, R}$
is reduced due to smaller absorptive phases.
Finally, at threshold the longitudinal polarization of the neu\-tra\-lino 
$\Sigma_{\rm res}^3$~(\ref{eq.Sr}) vanishes, 
and thus also both decay asymmetries, as follows from Eqs.~(\ref{eq.b0}) and~(\ref{eq.b1}).

%%%%%%%%%%%%%%%%%%%%%%%%%%%%%%%%%%%%%%%%%%%%%%%%%%%%%%%%%%%%%%%%%%%%%%%%%%%%%%
% -----------------------------------------------------------------
%                        P L O T - 7 
% -----------------------------------------------------------------
%%%%%%%%%%%%%%%%%%%%%%%%%%%%%%%%%%%%%%%%%%%%%%%%%%%%%%%%%%%%%%%%%%%%%%%%%%%%%%
The significance of the CP-odd polarization asymmetry,
defined in Appendix~\ref{StatSignificances}, reaches 
$\mathcal{S}_e^{\prime {\rm pol}, R}\approx 2$,
see Fig.~\ref{fig:asymm.electron}(c),
and thus the measurement of this asymmetry will be challenging.
Nonetheless, only the asymmetry 
${\mathcal{A}}_\ell^{\prime {\rm pol},n}$
allows to measure the CP-odd contribution $b_0$
to the longitudinal neu\-tra\-lino polarization
$\Sigma_{\rm res}^3$~(\ref{eq.Sr}).
However,
taking 
the other leptonic neu\-tra\-lino decay modes
 into account, 
in particular 
$\tilde\chi_2^0\to \tau\tilde\tau_1$,
to analyze 
$\mathcal{A}^{\prime {\rm pol},1}_\tau$,
results in larger significances\footnote{
          The decay asymmetries
          $\mathcal{A}^{(\prime){\rm pol},L}_e$,
          $\mathcal{A}^{(\prime){\rm pol},L}_\mu$,
          and
          $\mathcal{A}^{(\prime){\rm pol},2}_\tau$
          are only accessible for $\mu\gsim500$~GeV 
          and $M_2\gsim200$~GeV in our scenario.
          They are not relevant for our discussion, since
          the corresponding branching ratios of $\tilde\chi_2^0$ 
          are only as large as a few percent.}.

%}}}

% ------------------------------------------------------------------------------
%{{{ neutralino to tau
\subsubsection{Neutralino decay into a stau-tau pair}
% ------------------------------------------------------------------------------
%
The CP-even and CP-odd neu\-tra\-lino polarization asymmetries
can also be measured for the neu\-tra\-lino decay into a tau,
$\tilde\chi_2^0\to \tau\tilde\tau_1$.
Due to mixing in the stau sector,
the asymmetries for the decay into a tau
are generally smaller than those for the decay into an
electron (or muon),
$\mathcal{A}^{{\rm pol},1}_\tau
=\eta_\tau^1\mathcal{A}^{{\rm pol},1}_e$
and
$\mathcal{A}^{\prime {\rm pol},1}_\tau
=\eta_\tau^1\mathcal{A}^{\prime {\rm pol},R}_e$,
with $|\eta_\tau^1|\leq 1$,
see Eq.~(\ref{eq:realtionAtau}).
%%%%%%%%%%%%%%%%%%%%%%%%%%%%%%%%%%%%%%%%%%%%%%%%%%%%%%%%%%%%%%%%%%%%%%%%%%%%%%
% -----------------------------------------------------------------
%                        P L O T - 8 
% -----------------------------------------------------------------
\begin{figure}[t]
\centering
\begin{picture}(16,15.5)
\put(-4.5,-6.2){\includegraphics{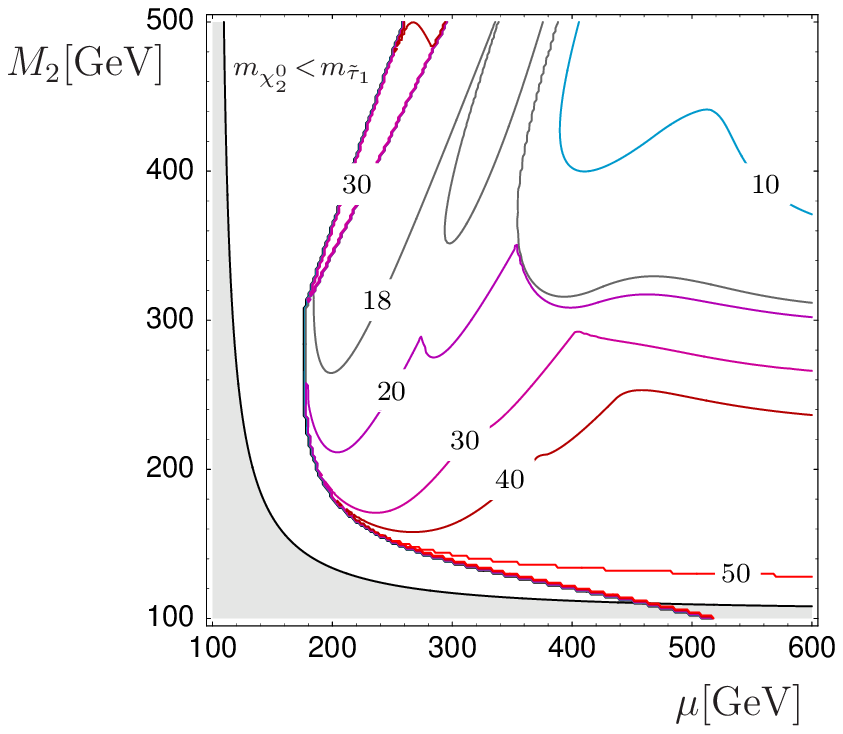}}
\put(3.5,-6.2){\includegraphics{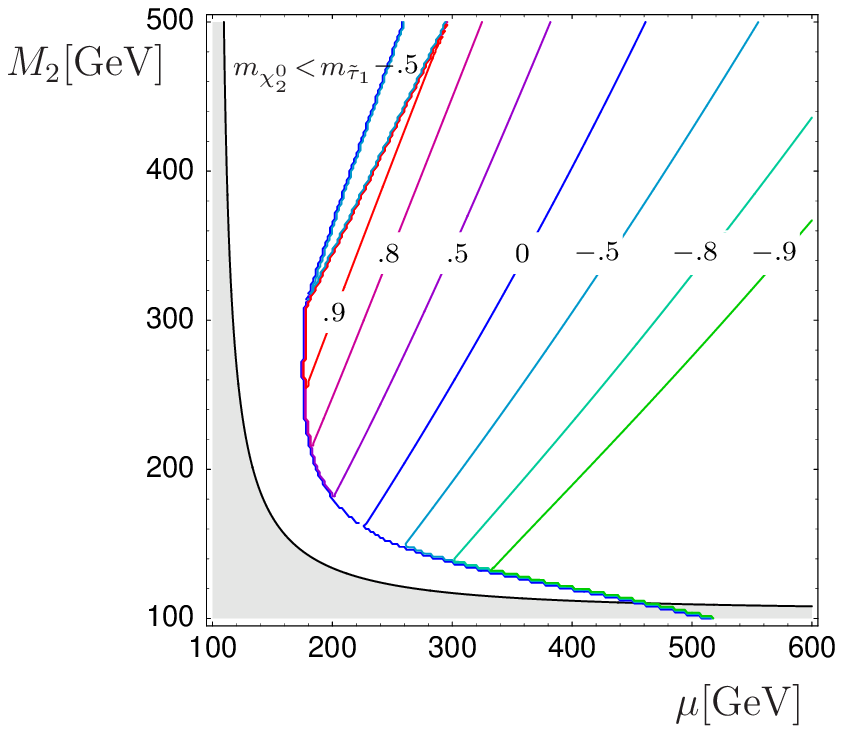}}
\put(1,8.15){(a)}
\put(9,8.15){(b)}
\put(1.4,14.9){${\rm BR}(\tilde\chi_2^0\to \tau^+\tilde \tau_1^-)$ in \%}
\put(10.,14.9){reduction factor $\eta_\tau^1$}
\put(-4.5,-14.2){\includegraphics{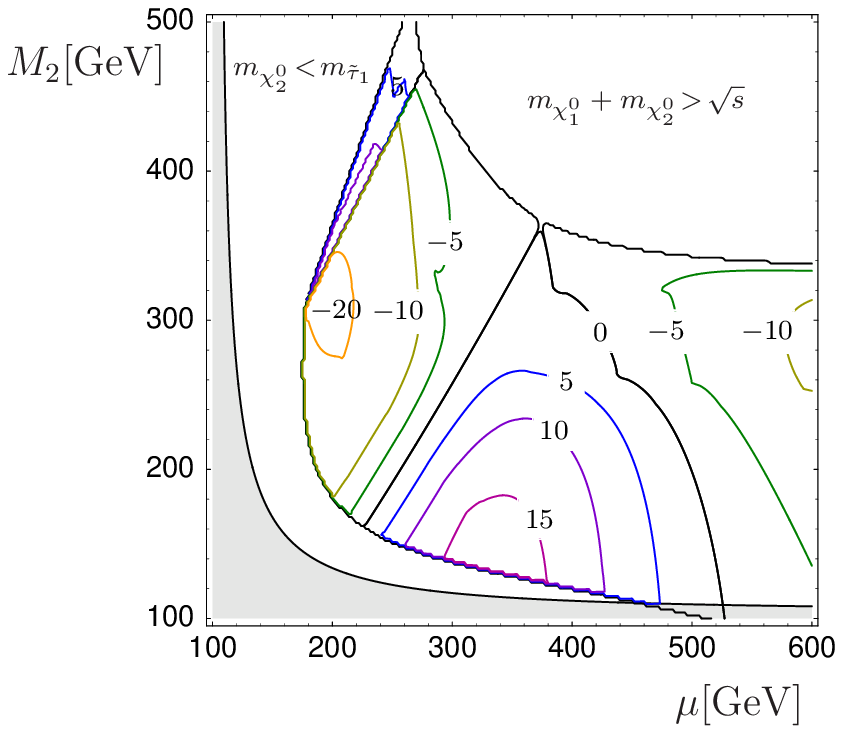}}
\put(3.5,-14.2){\includegraphics{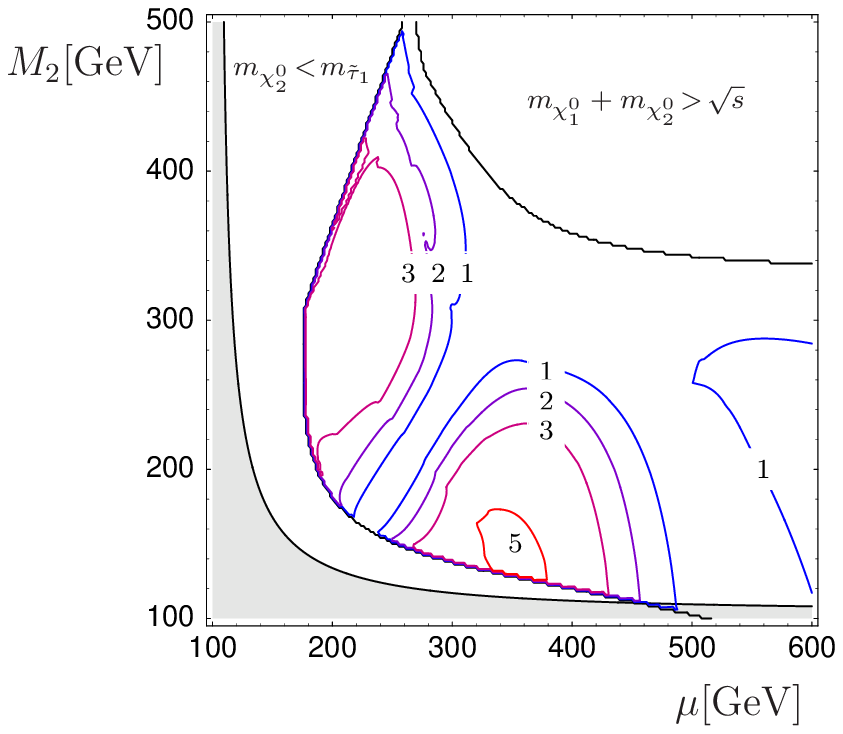}}
\put(1,0.15){(c)}
\put(9,0.15){(d)}
\put(1.4,6.9){Asymmetry $\mathcal{A}^{{\rm pol},1}_\tau$ in \%}
\put(10.,6.9){Significance $\mathcal{S}^{{\rm pol},1}_\tau$}

%%%%%%%%%%%%%%%%%%%%%%%%
\end{picture}
\caption{\small  
       Neutralino production 
       $\mu^+\mu^-\to\tilde\chi_1^0\tilde\chi_2^0$
       and decay into a tau
       $\tilde\chi_2^0\to \tau\tilde \tau_1$,
       at $\sqrt{s}=(M_{H_2}+M_{H_3})/2$ with 
       longitudinally polarized beams 
       $\mathcal{P}_{-}=\mathcal{P}_{+}=\pm0.3$.
       Contour lines in the $\mu$--$M_2$ plane for 
       {\bf (a)}~the branching ratio 
       ${\rm BR}(\tilde\chi_2^0\to \tau^+\tilde \tau_1^-)$, 
       {\bf (b)}~the factor $\eta_\tau^1$~(\ref{eq:eta_rl}),
       {\bf (c)}~the CP-even polarization asymmetry
       $\mathcal{A}^{{\rm pol},1}_\tau$, Eq.~(\ref{eq.apolijA}),
       and {\bf (d)}~the significance 
       $\mathcal{S}_\tau^{{\rm pol}, 1}$, 
       Eq.~(\ref{significance3}), 
       with the effective luminosity 
       ${\mathcal L}_{\rm eff}=1~{\rm fb}^{-1}$, for
       the SUSY parameters as given in Table~\ref{scenarioA}.
        The shaded area is excluded by requiring $m_{\chi_1^\pm}>103$~GeV.
        The cross section 
        $\sigma(\mu^+\mu^-\to\tilde\chi_1^0\tilde\chi_2^0)$
        is shown in Fig.~\ref{fig:sigma.A}.
}
\label{fig:asymm.tau}
\end{figure}
%%%%%%%%%%%%%%%%%%%%%%%%%%%%%%%%%%%%%%%%%%%%%%%%%%%%%%%%%%%%%%%%%%%%%%%%%%%%%%
In Fig.~\ref{fig:asymm.tau}(b), we show the contour lines of the reduction  
factor $\eta_\tau^1$~(\ref{eq:eta_rl}).
In the following, we discuss the CP-even asymmetry 
$\mathcal{A}^{{\rm pol},1}_\tau$ only.
A similar discussion holds however qualitatively also for
the CP-odd asymmetry $\mathcal{A}^{\prime {\rm pol},1}_\tau$.
Note that a measurement of the $\tau$ asymmetries is more involved 
due to $\tau$-reconstruction efficiencies,  
which we however neglect in the following for simplicity.

\medskip

The CP-even asymmetry $\mathcal{A}^{{\rm pol},1}_\tau$
is shown in  Fig.~\ref{fig:asymm.tau}(c).
The significance for measuring the asymmetry
also depends on the cross section for production and
decay, and thus on the different leptonic branching ratios
of the neu\-tra\-lino.
The neu\-tra\-lino decay into a tau dominates
for $M_2\lsim200$~GeV, 
see Fig.~\ref{fig:asymm.tau}(c),
whereas the branching ratio into an electron can attain more than
${\rm BR}(\tilde\chi_2^0\to e^+\tilde e_R^-)= 16\%$,
for $\mu < M_2$, see the contour line in  Fig.~\ref{fig:sigma.A}(b).
We take account of this interplay between the size of the 
branching ratios and the asymmetries
by comparing their statistical significances 
$\mathcal{S}^{{\rm pol},R}_e$ and
$\mathcal{S}^{{\rm pol},1}_\tau$,
which we define in Appendix~\ref{StatSignificances},
Eq.~(\ref{significance3}).
The significances quantify the feasibility of measuring
the asymmetries. 
Both significances can be as large as $5$, however in
different regions of the  $\mu$--$M_2$ plane, 
compare Fig.~\ref{fig:asymm.electron}(c) and
Fig.~\ref{fig:asymm.tau}(d), respectively.

%}}} end of subsection neu\-tra\-lino to tau
%---------------------------------------------------------------------------------

%}}} end of numerics
%----------------------------------------------------------------------------------------
\newpage
{~}
\newpage
%----------------------------------------------------------------------------------------
%{{{section Summary
\section{Summary and conclusions 
  \label{Summary and conclusions}}
%----------------------------------------------------------------------------------------

We have analyzed neu\-tra\-lino production and their
leptonic decays at the muon collider with longitudinally polarized
beams. We have defined polarization asymmetries
to study the interference of the heavy neutral MSSM Higgs
bosons with CP violation, radiatively induced by
the common phase $\phi_A$ of the trilinear scalar coupling
parameter. For nearly degenerate neutral Higgs bosons,
as in the Higgs decoupling limit, the CP violating Higgs mixing can be
resonantly enhanced,  which allows for large CP violating effects.

\medskip

For neu\-tra\-lino production, we have
defined a CP-odd  asymmetry of the cross section
for equal positive and negative muon beam polarizations.
The CP-odd production asymmetry
is sensitive to the CP-phases in the Higgs sector, and 
also receives large contributions from  absorptive phases
of the Higgs propagators. 
In a numerical study, 
we have obtained large values of the production asymmetry
up to $40\%$ for equal beam polarizations of 
$\mathcal P = 0.3$.
For neu\-tra\-linos with mixed gaugino-higgsino character,
the production cross section can be as large as $1500$~fb.
Thus the asymmetry can be measured with high statistical significance.

\medskip

We have shown that also the  neu\-tra\-lino polarization depends 
sensitively on the Higgs interference.
For the subsequent leptonic decays of the neu\-tra\-lino, 
we have analyzed two asymmetries of the
energy distributions of the final leptons 
$e$, $\mu$ and $\tau$.
The decay asymmetries
 probe the CP-odd and the CP-even contributions
to the longitudinal neu\-tra\-lino polarization, respectively.
The decay asymmetries are  complementary to 
the production asymmetry, since they strongly depend
on spin-correlations.
The CP-even asymmetry is due to a correlation between
the longitudinal polarizations of the initial muons and
the final neutralinos. Being CP-even, the asymmetry reaches
$25\%$ for vanishing CP-phases, and is reduced in the
presence of CP-phases.
The CP-odd asymmetry is due to
the spin correlations in the neutralino production and
decay process.
Similarly to the CP-odd asymmetry from the production,
this decay asymmetry is approximately maximal if the 
scalar-pseudoscalar Higgs mixing is resonantly enhanced,
which appears naturally in the Higgs decoupling limit.
The decay asymmetries yield additional information on the
CP nature of the Higgs resonances, 
and complement the production asymmetry.
The asymmetries thus allow a systematic study of the 
interference and mixing effects of CP violating neutral Higgs sector
at the muon collider.

%}}}

% ---------------------------------------------------------------------------------------
%{{{ Aknowledgements
%
\section{Acknowledgments}
 We thank Sven Heinemeyer and Thomas Hahn for helpful discussions, 
 and Hans Fraas for carefully reading the manuscript.
 FP thanks the hospitality of the Institut f\"ur Theoretische Physik 
 und Astrophysik, Universit\"at W\"urzburg.
 OK was supported by the SFB Transregio 33: The Dark Universe.
%
% ----------------------------------------------------------------------------------------
%}}}

\clearpage
%----------------------------------------------------------------------------------------
%{{{ Appendix

\begin{appendix}
\noindent{\Large\bf Appendix}
\setcounter{equation}{0}
\renewcommand{\thesubsection}{\Alph{section}.\arabic{subsection}}
\renewcommand{\theequation}{\Alph{section}.\arabic{equation}}

\setcounter{equation}{0}

%----------------------------------------------------------------------------------------
\section{Note on the effective Higgs couplings 
\label{app:Higgscouplings}}
%----------------------------------------------------------------------------------------
            
In Section~\ref{section:lagrdens}, we have defined the effective 
couplings of the Higgs  bosons to muons and neu\-tra\-linos.
They are obtained by rotating the tree level Higgs couplings
by the matrix $C$. This matrix
includes the leading radiative self-energy corrections,
and is defined by diagonalizing the Higgs propagator matrix 
and the Higgs mass matrix
\begin{equation}
 \Delta_{\rm D}(p^2) =  C  \Delta(p^2) C^{-1} , 
\qquad   {\rm \bf M}_{\rm D}(p^2) =  C   {\rm \bf M}(p^2) C^{-1},
%\label{eq:diagonalization}
\end{equation}
respectively, at fixed momentum squared $p^2=M_{H_2}^2$.
The weak momentum dependence
of the mass matrix ${\rm \bf M}$ is neglected, as in the 
Weisskopf-Wigner approximation~\cite{Weisskopf:1930ps}.
Due to the absorptive parts of the transition amplitude,
the matrix $C$ is in general non-unitary.
Thus the transformed basis of the 
approximate\footnote{We call the Higgs basis $\{ H_1,H_2,H_3 \}$
                     approximate, since it corresponds to 
                     $ {\rm \bf M}_{\rm D}(p^2)$ with fixed momentum  
                     squared $p^2=M_{H_2}^2$,
                     assuming a weak momentum dependence
                     in the resonance region.}
Higgs boson fields $\{ H_1,H_2,H_3 \}$, is 
non-orthonormal~\cite{Choi:2004kq}. 
As a consequence, there exists a dual basis 
$\{ \tilde H_1,\tilde H_2,\tilde H_3 \}$
obtained by the matrix $\tilde C={C^{-1} }^T$.
The corresponding states satisfy the orthogonality relations 
$\langle H_k|\tilde H_l \rangle=\delta_{kl}$.
This leads to different transformations of the tree-level Higgs couplings to 
initial and final state fermions
with $C$ and $\tilde C$, relations~(\ref{eq.Hmu.eff.c}) 
and~(\ref{eq.Hchi.eff.c}), respectively. 
This follows, since the amplitude~(\ref{THiggsNoHel}) 
for neu\-tra\-lino production can also be written in the general form
\begin{eqnarray}
T^P&=&
 \Gamma_{}^{(\chi)} \Delta_{} \Gamma_{}^{(\mu)} \nonumber \\
&=& 
      \Gamma_{}^{(\chi)} C^{-1}_{} C_{} \Delta_{}  C^{-1}_{} C_{}\, \Gamma_{}^{(\mu)} 
\nonumber \\
&=&
     \Gamma_{}^{(\chi)} C^{-1}_{}   \Delta_{D}    C_{}\, \Gamma_{}^{(\mu)} 
     =   \Gamma_{\rm eff}^{(\chi)}   \Delta_{D}  \Gamma_{\rm eff}^{(\mu)},
\label{eq:generalTP}
\end{eqnarray}
where $\Gamma_{}^{(\chi)}$ and $\Gamma_{}^{(\mu)}$ are the one-particle irreducible Higgs vertices
to muons and neutralinos, respectively.
Eq.~(\ref{eq:generalTP}) defines
the effective one-particle irreducible Higgs vertices for initial and
final fermion states, 
\begin{eqnarray}
  \Gamma_{l\, \rm eff}^{(\mu)} &=& C_{lj}\Gamma_{j}^{(\mu)} , \\
  \Gamma_{k\, \rm eff}^{(\chi)} &=& \tilde C_{ki}\Gamma_{i}^{(\chi)} ,
\label{eq:Gammaeffchi}
\end{eqnarray}
 which transform with $C$ and $\tilde C$, respectively.
If the phases of the Higgs boson states are chosen appropriately,
the matrix $C$ can be made complex orthogonal~\cite{orthogonalC}, 
which implies $\tilde C = C$.

%----------------------------------------------------------------------------------------
\section{Correlation between initial and final longitudinal polarizations}
%----------------------------------------------------------------------------------------
\label{polarization.correlation}
In this appendix, we analyze the correlation between initial 
and final longitudinal polarizations
in neutralino pair production in $\mu^+\mu^-$ annihilation via Higgs boson exchange
in the simplified case of a CP conserving Higgs sector.
\medskip

\noindent{\bf Helicity and CP eigenstates}
\\
Out of four possible spin/helicity states of a fermion pair 
only those with $J_z=0$ in the  center-of-mass system (CMS) 
interact with the Higgs bosons.
Here $J$ denotes the total angular momentum and $\hat z$ the direction of the momenta of the fermions.
Since in the CMS the orbital angular momentum of the fermions is orthogonal to their momenta, $L_z=0$, 
their total spin $S$ satisfies  $S_z=J_z-L_z=0$.
The fermion interacting states are thus $|LL\rangle_{f f^\prime}$ and $|RR\rangle_{f f^\prime} $, 
where $L$ and $R$ denote the helicities of the fermions ${f f^\prime}=\mu^+\mu^-,\ \neutralino_i\neutralino_j$.
These states are linear combinations of states $|S,S_z\rangle_{f f^\prime}$ with spin $S=0$ and $S=1$,
\begin{eqnarray}
|LL\rangle_{f f^\prime}
= \frac{1}{\sqrt{2}}(|1,0\rangle - |0,0\rangle)_{f f^\prime},
\label{eq.LLstate}\\
|RR\rangle_{f f^\prime}
= \frac{1}{\sqrt{2}}(|1,0\rangle + |0,0\rangle)_{f f^\prime}.
\label{eq.RRstate}
\end{eqnarray}
For a fermion-antifermion system their spin $S$ is related to their CP quantum number
by ${{\rm CP}}=\eta_{f f^\prime}(-1)^{S+1}$, 
with $\eta_{f \bar f}=1$ for a Dirac fermion-antifermion pair 
and $\eta_{\chi^0_i \chi^0_j }\equiv\eta_{ij}=e^{2i\sigma_{ij}}$ for a pair of neutralinos.
The relative CP phase factor $\eta_{ij}$ is real in our analysis since the neutralino sector is CP-conserving,
with $\sigma_{ij}=0,\pi/2$.
The CP-even and CP-odd muon states 
\begin{eqnarray}
|{\rm CP}+\rangle_{\mu^+\mu^-} = \phantom{i}|1,0 \rangle_{\mu^+\mu^-}, 
\\
|{\rm CP}-\rangle_{\mu^+\mu^-} = i|0,0 \rangle_{\mu^+\mu^-}, 
\label{eq.CPxistates}
\end{eqnarray}
and neutralino states 
\begin{eqnarray}
|{\rm CP}+\rangle_{\chi^0_i\chi^0_j} = (\phantom{-}\cos\sigma_{ij} |1,0 \rangle + i\sin\sigma_{ij}|0,0 \rangle)_{\chi^0_i\chi^0_j},
\label{eq.CPxistate+}
\\
|{\rm CP}-\rangle_{\chi^0_i\chi^0_j} = (-\sin\sigma_{ij} |1,0 \rangle + i \cos\sigma_{ij}|0,0 \rangle)_{\chi^0_i\chi^0_j}
\label{eq.CPxistate-},
\end{eqnarray}
can be expressed as a linear combination of helicity states inverting
Eqs.~(\ref{eq.LLstate}) and (\ref{eq.RRstate}).
Analogously,  the helicity states
are linear combinations of the CP-even and CP-odd states. For the muon-antimuon pairs we obtain
\begin{eqnarray}
&&|LL\rangle_{\mu^+\mu^-}
=
\frac{1}{\sqrt{2}}
( 
|{\rm CP}+\rangle +i |{\rm CP}-\rangle 
)_{\mu^+\mu^-},
\label{eq.LLstateCP}\\[0.1cm]
&&|RR\rangle_{\mu^+\mu^-}
=
\frac{1}{\sqrt{2}}
( 
|{\rm CP}+\rangle -i |{\rm CP}-\rangle 
)_{\mu^+\mu^-}.
\label{eq.RRstateCP}
\end{eqnarray}

% ----------------------------------------------------------------------------------------
\noindent{\bf Transition amplitudes}\\
Assuming a CP conserving Higgs sector implies that, in our Higgs mediated neutralino production process,
a CP-even $\mu^+\mu^-$ state $|{\rm CP}+\rangle_{\mu^+\mu^-}$
can only produce CP-even Higgs bosons, which in turn decay into the CP-even neutralino state
(\ref{eq.CPxistate+}) 
with a real amplitude $\alpha$.
Analogously, a CP-odd $\mu^+\mu^-$ state leads to a CP-odd neutralino state (\ref{eq.CPxistate-}) with an amplitude 
 $\beta$.

An initial state with right handed polarized muon and 
antimuons (\ref{eq.RRstateCP}),
will interact to produce  the neutralino state
\begin{eqnarray}
|RR\rangle_{\mu^+\mu^-}
&\to &
\sqrt{2}\mathcal{N}_R^\prime
[ 
\alpha |{\rm CP}+\rangle  -i \beta  |{\rm CP}-\rangle 
]_{\chi^0_i\chi^0_j},\qquad
\nonumber\\
&=&
\mathcal{N}_R^\prime
[ 
e^{i\sigma_{ij}}(\alpha+\beta) |RR\rangle + 
e^{-i\sigma_{ij}}(\alpha-\beta)  |LL\rangle 
]_{\chi^0_i\chi^0_j},\qquad
\label{eq.RRtochichi}
\end{eqnarray}
where 
$\mathcal{N}_R^\prime$ is a normalization factor.
Here we have used the explicit form of the neutralino states (\ref{eq.CPxistate+}) and (\ref{eq.CPxistate-}) 
and have inverted Eqs.~(\ref{eq.LLstate}) and (\ref{eq.RRstate}).

Similarly,
\begin{eqnarray}
&&|LL\rangle_{\mu^+\mu^-}
\to
\mathcal{N}_L^\prime
[ 
e^{i\sigma_{ij}} (\alpha-\beta)|RR\rangle + 
e^{-i\sigma_{ij}} (\alpha+\beta)|LL\rangle 
]_{\chi^0_i\chi^0_j}.\qquad
\label{eq.LLtochichi}
\end{eqnarray}
From Eqs.~(\ref{eq.RRtochichi}) and (\ref{eq.LLtochichi}) follows that
if either the CP-even or CP-odd amplitudes, $\alpha$ or $\beta$, respectively, vanish, 
then so does the neutralino polarization, 
since in this case the absolute value of the coefficients 
on the r.h.s.\ of Eqs.~(\ref{eq.RRtochichi}) and (\ref{eq.LLtochichi})
are equal.
Note that 
the neutralino polarization depends % in (\ref{eq.RRtochichi}) and (\ref{eq.LLtochichi})	
on the relative signs of the CP-even and CP-odd amplitudes\
since it arises from their interference.
%(for the CP-conserving Higgs sector)

From Eqs.~(\ref{eq.RRtochichi}) and (\ref{eq.LLtochichi}) also follows that
if the muon beams are not longitudinally polarized,
which implies that the initial state has equal proportions of left and right handed $\mu^+\mu^-$ states
(\ref{eq.LLstateCP}) and (\ref{eq.RRstateCP}),
the final neutralinos will also be unpolarized.

Concluding, in the CP conserving Higgs sector
the neutralino polarization 
is sensitive to the relative sign of the transition amplitudes, 
and thus to the product of couplings, 
and can only be non vanishing if the muon beams are longitudinally polarized.
This implies that the correlation between initial and final state longitudinal polarizations
depends on the interference of transition amplitudes 
mediated by Higgs bosons of different CP parities.

In the more general CP violating Higgs sector studied in this paper,
the abovementioned correlation
between initial and final polarizations leads to the CP-even 
asymmetry
${\mathcal{A}}_\ell^{{\rm pol}, n}$~(\ref{eq.apolijA}).
In addition, CP-odd (and CP$\tilde{\rm T}$-odd) effects lead to the 
CP-odd asymmetry
${\mathcal{A}}_\ell^{\prime {\rm pol}, n}$~(\ref{eq.apolijAprime}).
However, since this asymmetry is not due to a correlation
between  initial and final state polarizations, it can be
non-zero even for vanishing beam polarizations.

%-----------------------------------------------------------------------------
\section{Lagrangians for non-resonant neutralino production and leptonic decay}
\label{nonreslagrangian}
%-----------------------------------------------------------------------------

The non-resonant 
neu\-tra\-lino production~(\ref{production}) proceeds via $Z^0$ boson 
exchange in the $s$-channel, and smuon $\tilde\mu_{L,R}$ exchange in the 
$t$- and $u$-channels, see the Feynman diagrams in Fig.~\ref{Fig:FeynProd}.
The interaction Lagrangians for neu\-tra\-lino production
and those for its leptonic  decay
$\tilde\chi_1^0\to \ell\tilde\ell_{L,R}$, with $\ell=e,\mu$
are~\cite{Bartl:1986hp,Moortgat-Pick:1999di}
\begin{eqnarray}
{\scr L}_{Z^0\tilde{\chi}^0_i\tilde{\chi}^0_j} &=&
\frac{1}{2}\frac{g}{\cos\theta_W}
Z^0_{\nu}\bar{\tilde{\chi}}^0_i\gamma^{\nu}
[O_{ij}^{''L} P_L+O_{ij}^{''R} P_R]\tilde{\chi}^0_j~, \quad i, j=1,\dots,4, 
\label{Zchichi}\\[2mm]
{\scr L}_{Z^0 \mu \bar \mu} &=&-\frac{g}{\cos\theta_W}
Z^0_{\nu}\bar \mu\gamma^{\nu}[L_\mu P_L+ R_\mu P_R]\mu,
\label{Zelel}\\[2mm]
        {\scr L}_{\ell \slepton \neu_j} & = & 
                g f_{\ell j}^L \bar{\ell} P_R \neu_j  \slepton_L 
        +       g f_{\ell j}^R \bar{\ell} P_L \neu_j  \slepton_R
        +       \mbox{h.c.}.
\label{slechie}
%\\[2mm]
\end{eqnarray}
with $P_{L, R}=(1\mp \gamma_5)/2$.
In the photino, zino, higgsino basis
($\tilde{\gamma},\tilde{Z}, \tilde{H}^0_a, \tilde{H}^0_b$),
the couplings are~\cite{Bartl:1986hp,Moortgat-Pick:1999di} 
\begin{eqnarray}
        &&O_{ij}^{''L}=-\frac{1}{2} \left[
        (N_{i3}N_{j3}^*-N_{i4}N_{j4}^*)\cos2\beta
  +(N_{i3}N_{j4}^*+N_{i4}N_{j3}^*)\sin2\beta \right],\label{OLOR}\\[2mm]
&&O_{ij}^{''R}=-O_{ij}^{''L*},\\[2mm]
&&L_\mu=-\frac{1}{2}+\sin^2\theta_W, \quad
  R_\mu=\sin^2\theta_W \label{eq_5},\\[2mm]
&&      f_{\ell j}^L = -\sqrt{2}\bigg[\frac{1}{\cos
        \theta_W}(T_{3\ell}-e_{\ell}\sin^2\theta_W)N_{j2}+
        e_{\ell}\sin \theta_W N_{j1}\bigg],
\label{eq:fl}\\[2mm]
&&      f_{\ell j}^R = -\sqrt{2}e_{\ell} \sin \theta_W\Big[\tan 
        \theta_W N_{j2}^*-N_{j1}^*\Big],
\label{eq:fr}
\end{eqnarray}
with $e_{\ell}$ and $T_{3\ell}$ the electric charge 
and third component of the weak isospin of the lepton $\ell$,
the weak mixing angle $\theta_W$,
the weak coupling constant $g=e/\sin\theta_W$, $e>0$,
and the ratio $\tan \beta=v_2/v_1$ of the vacuum expectation values 
of the two neutral Higgs fields.
The neu\-tra\-lino couplings to the $Z$ boson, $O_{ij}^{''L,R}$, 
and to the smuons, $f^{L,R}_{\mu i}$, 
contain the complex mixing elements $N_{ij}$,
which diagonalize the neu\-tra\-lino matrix
 $N_{i \alpha}^*Y_{\alpha\beta}N_{\beta k}^{\dagger}=
 m_{\chi_i}\delta_{ik}$ \cite{HK}, 
with neu\-tra\-lino masses $ m_{\chi_i}>0$.

Mixing can safely be neglected for the scalar leptons of the first 
two generations, $\slepton = \tilde{e},\tilde{\mu}$. 
For the neu\-tra\-lino decay into staus
$\neu_i \to  \stau_n \tau$, $n=1,2$, we take stau mixing into account,
see Appendix~\ref{stau.neutralino}.

%---------------------------------------------------------------------
\section{Density matrix formalism}
\label{Density matrix formalism}
%---------------------------------------------------------------------

We use the spin density matrix formalism 
of~\cite{Haber94,Moortgat-Pick:1999di}
for the calculation of the squared amplitudes for 
neu\-tra\-lino production~(\ref{production}) and decay~(\ref{decay}).  
The amplitude for neu\-tra\-lino production via resonant Higgs exchange, 
Eq.~(\ref{THiggs}),
depends on the helicities $\lambda_\pm$ of the muons $\mu^\pm$ and 
the helicities $\lambda_i,\lambda_j$ of the neu\-tra\-linos 
$\tilde\chi^0_i,\tilde\chi^0_j$
\begin{eqnarray}
T^P_{\lambda_i\lambda_j\lambda_{+}\lambda_{-}} = 
\Delta(H_k)
\left[
\bar{v}(p_{\mu^+},\lambda_{+})\left(c_L^{H_k\mu\mu}P_L+c_R^{H_k\mu\mu}P_R \right)
u(p_{\mu^-},\lambda_{-})
\right]\phantom{.}\nonumber\\
\times
\left[
\bar u(p_{\chi^0_j},\lambda_j)\left(c_L^{H_k\chi_i\chi_j}P_L
                                   +c_R^{H_k\chi_i\chi_j}P_R \right)
v (p_{\chi^0_i},\lambda_i)
\right].
\label{THiggs}
\end{eqnarray}
We include the longitudinal beam polarizations of the muon-beams, ${\mathcal P}_{-}$ and 
${\mathcal P}_{+}$, with $ -1 \le {\mathcal P}_{\pm}\le +1$ in their density matrices
\begin{eqnarray}
\label{eq:density1}
\rho^{-}_{\lambda_{-} \lambda_{-}^\prime}  &=& 
     \frac{1}{2}\left(\delta_{\lambda_{-} \lambda_{-}^\prime} + 
      {\mathcal P}_{-}\tau^3_{\lambda_{-} \lambda_{-}^\prime}\right),\\[2mm]
\label{eq:density2}
\rho^{+}_{\lambda_{+} \lambda_{+}^\prime}  &=& 
     \frac{1}{2}\left(\delta_{\lambda_{+} \lambda_{+}^\prime} + 
   {\mathcal P}_{+}\tau^3_{\lambda_{+} \lambda_{+}^\prime}\right),
\end{eqnarray}
where  $\tau^3$ is the third Pauli matrix. 
The unnormalized spin density matrix 
of  $\tilde\chi_i^0\tilde\chi_j^0$ production 
and $\tilde\chi_j^0$ decay are given by, respectively,
\begin{eqnarray}
\rho^{P}_{\la_j\la^\prime_j} &=& 
\sum_{\la_i,\lambda_{+},\lambda_{+}^\prime,\lambda_{-} \lambda_{-}^\prime} 
\rho^{+}_{\lambda_{+} \lambda_{+}^\prime}
\rho^{-}_{\lambda_{-} \lambda_{-}^\prime}
{T}^{P}_{\la_i\la_j\lambda_{+}\lambda_{-} }
{T}^{P*}_{\la_i\la^\prime_j\lambda_{+}^\prime \lambda_{-}^\prime},
\label{rhop}
\\
        \rho^{D}_{\la^\prime_j\la_j} &=&                
                {{T}}^{D\ast}_{\la^\prime_j}{{T}}^{D}_{\la_j}.
\label{rhod}
\end{eqnarray} 
The amplitude squared for production and decay is then
\begin{eqnarray}
        |{{T}}|^2 &=& |\Delta(\tilde\chi_j^0)|^2 \sum_{\la_j \la_j^\prime} 
        \rho^P_{\la_j \la_j^\prime}\rho^D_{\la_j^\prime\la_j},
\label{tsquare}
\end{eqnarray}
with the neu\-tra\-lino propagator
\begin{eqnarray}
\Delta(\tilde\chi_j^0)=\frac{i}{p^2_{\chi_j^0}-m_{\chi_j^0}^2+
        i m_{\chi_j^0}\Gamma_{\chi_j^0}}.
\label{eq:neutralinopropagator}
\end{eqnarray}
The spin density matrices~(\ref{rhop}) and~(\ref{rhod})
can be expanded in terms of the Pauli matrices $\tau^a$
\begin{eqnarray}
\rho^P_{\la_j \la_j^\prime} &=& 
\delta_{\la_j \la_j^{\prime}}P + \sum_{a=1}^3 \tau_{\la_j \la_j^{\prime}}^a  {\Sigma}_P^{a},
\label{rhoP}
\\
\rho^D_{\la_j^\prime \la_j} &=& 
\delta_{\la_j^\prime \la_j}D + \sum_{a=1}^3 \tau_{\la_j^{\prime} \la_j}^a  {\Sigma}_D^{a},
\label{rhoD}
\end{eqnarray}
where we have defined a set of neu\-tra\-lino spin vectors $s_{\chi_j^0}^a$.
In the center-of-mass system, they are 
\begin{equation}
        s_{\chi^0_j}^{1,\,\mu}= (0;1,0,0), \qquad       s_{\chi^0_j}^{2,\,\mu}= (0;0,1,0), 
        \qquad  s_{\chi^0_j}^{3,\,\mu}= 
        \frac{1}{m_{\chi^0_j}}(|\vec{p}_{\chi^0_j}|;0,0,E_{\chi^0_j}).
        \label{spinvec}
\end{equation}
We have chosen a coordinate frame 
such that the momentum of the neu\-tra\-lino $\tilde\chi^{0}_j$ 
is given by
\begin{equation}
        p_{\chi^0_j}^{\,\mu}=(E_{\chi^0_j};0,0,|\vec{p}_{\chi^0_j}|),
\end{equation}
with
\begin{equation}
        E_{\chi_j^0} =\frac{s+m_{\chi^0_j}^2-m_{\chi^0_i}^2}{2 \sqrt{s}},\quad
    |\vec{p}_{\chi^0_j}|   =\frac{\sqrt{\lambda_{ij}}}{2 \sqrt{s}},
\end{equation}
and the triangle function
\begin{equation}
\lambda_{ij}= \lambda(s,m_{\chi^0_i}^2,m_{\chi^0_j}^2),
\label{triangle}
\end{equation}
with $\lambda(x,y,z)= x^2+y^2+z^2-2(xy+xz+yz)$.

Inserting the density matrices~(\ref{rhoP}) and~(\ref{rhoD}) 
into~(\ref{tsquare}),  gives then the amplitude squared
in the form of Eq.~(\ref{eq:tsquare}).

%-----------------------------------------------------------------------------
\section{Continuum amplitudes and contributions}
\label{nonresamplitudes}
%-----------------------------------------------------------------------------
%

The amplitudes for non-resonant 
$Z$ and $\tilde\mu_{L,R}$ exchange are 
\begin{eqnarray}
%1
T^P_{\lambda_i\lambda_j\lambda_{+}\lambda_{-}}(s,Z) &=&
\frac{g^2}{\cos^2\theta_W}
\Delta^s(Z)\left[\bar{v}(p_{\mu^+},\lambda_{+})\gamma^{\mu}
                (L_{\mu} P_L+R_{\mu}P_R)u(p_{\mu^-},\lambda_{-})\right] 
         \nonumber\\& &
\times\left[\bar u(p_{\chi^0_j},\lambda_j)\gamma_{\mu} 
        (O^{''L}_{ji} P_L+O^{''R}_{ji} P_R) v (p_{\chi^0_i},\lambda_i)\right],
                \label{neut:T1}\\
%2
T^P_{\lambda_i\lambda_j\lambda_{+}\lambda_{-}}(t,\tilde \mu_{L})&=&
                -g^2 f^L_{\mu i}f^{L*}_{\mu j}\Delta^t(\tilde \mu_L)
        \left[\bar v (p_{\mu^+},\lambda_{+}) P_R   v (p_{\chi^0_i},\lambda_i)\right]
 \nonumber\\& & \times
    \left[\bar u (p_{\chi^0_j},\lambda_j)P_L u(p_{\mu^-},\lambda_{-})\right],\label{neut:T2}\\
%3
T^P_{\lambda_i\lambda_j\lambda_{+}\lambda_{-}}(t,\tilde \mu_{R})&=&
                -g^2 f^R_{ \mu i}f^{R*}_{\mu j}\Delta^t(\tilde \mu_R)
                \left[\bar v (p_{\mu^+},\lambda_{+}) P_L  v (p_{\chi^0_i},\lambda_i)\right]
          \nonumber\\&&
           \times\left[\bar u (p_{\chi^0_j},\lambda_j)P_R u(p_{\mu^-},\lambda_{-})\right],\label{neut:T3}\\
%4
T^P_{\lambda_i\lambda_j\lambda_{+}\lambda_{-}}(u,\tilde \mu_L)&=&
                g^2 f^{L*}_{\mu i} f^L_{\mu j}\Delta^u(\tilde \mu_L)
                \left[\bar{v}(p_{\mu^+},\lambda_{+}) P_R v (p_{\chi^0_j},\lambda_j)\right]
                  \nonumber\\& &
                 \times\left[\bar u(p_{\chi^0_i},\lambda_i)P_L u(p_{\mu^-},\lambda_{-})\right],\label{neut:T4}\\
%5
T^P_{\lambda_i\lambda_j\lambda_{+}\lambda_{-}}(u,\tilde \mu_R)&=&
                g^2 f^{R*}_{\mu i} f^R_{\mu j}\Delta^u(\tilde \mu_R)
                \left[\bar v(p_{\mu^+},\lambda_{+}) P_L v(p_{\chi^0_j},\lambda_j)\right]
                \nonumber\\& &
                 \times\left[\bar u(p_{\chi^0_i},\lambda_i)P_R u(p_{\mu^-},\lambda_{-})\right],\label{neut:T5}
\end{eqnarray}
with the propagators 
\begin{equation}\label{neut:propagators}
   \Delta^s(Z)=\frac{i}{s-m^2_Z},\quad
   \Delta^t(\tilde{\mu}_{R,L})=
            \frac{i}{t-m^2_{\tilde{\mu}_{R,L}}},\quad
   \Delta^u (\tilde{\mu}_{R,L})=
             \frac{i}{u-m^2_{\tilde{\mu}_{R,L}}},
         \end{equation}
and 
$t=(p_{\mu^-}-p_{\chi^0_j})^2$ and 
$u=(p_{\mu^-}-p_{\chi^0_i})^2$.
We neglect the $Z$-width in the propagator
$\Delta^s(Z)$ for energies beyond the resonance.
The Feynman diagrams are shown in 
Fig.~\ref{Fig:FeynProd}.
For $e^+e^-$ collisions, the amplitudes are given 
in~\cite{Moortgat-Pick:1999di}.

%%--------------------------------------------------------------------
%{Contributions from the continuum}
%%-------------------------------------------------------------------

The continuum contributions $P_{\rm cont}$ are 
those from the non-resonant $Z$ and $\tilde\mu_{L,R}$ 
exchange channels. 
The coefficient $P_{\rm cont}$ is independent of the neu\-tra\-lino polarization. 
It can be decomposed into contributions from the 
different continuum channels
\begin{equation}
        P_{\rm cont}=P(Z Z)+P(Z \tilde{\mu}_R)+P(\tilde{\mu}_R \tilde{\mu}_R)+
        P(Z \tilde{\mu}_L)+P(\tilde{\mu}_L \tilde{\mu}_L),\label{eq_15}
\end{equation}
with
\begin{eqnarray}
P(Z Z)&=&
        4 \frac{g^4}{\cos^4\theta_W}|\Delta^s(Z)|^2
        (R_{\mu}^2 c_R + L_{\mu}^2 c_L) E_b^2  \nonumber\\ 
        && \times
        \Big\{ |O^{''R}_{ij}|^2 (E_{\chi^0_i} E_{\chi^0_j} + 
        q^2\cos^2\theta)\nonumber\\ &&
        -[(Re O^{''R}_{ij})^2 -(Im O^{''R}_{ij})^2]
        m_{\chi^0_i} m_{\chi^0_j}\Big\}\label{eq_16},\label{P_1}\\
P(Z \tilde{\mu}_R) & = &
        \frac{2g^4}{\cos^2 \theta_W} R_{\mu} c_R E_b^2
        Re\Big\{\Delta^s(Z)
\nonumber\\& & \times
        \Big[ -(\Delta^{t*}(\tilde{\mu}_R) f^{R*}_{\mu i} 
        f^{R}_{\mu j} O^{''R*}_{ij}
        +\Delta^{u*}(\tilde{\mu}_R) f^R_{\mu i} f^{R*}_{\mu j}
        O^{''R}_{ij}) m_{\chi^0_i} m_{\chi^0_j}
\nonumber\\& &
        +(\Delta^{t*}(\tilde{\mu}_R) f^{R*}_{\mu i} f^{R}_{\mu j} O^{''R}_{ij}
        +\Delta^{u*}(\tilde{\mu}_R) f^R_{\mu i} f^{R*}_{\mu j} O^{''R*}_{ij})
(E_{\chi^0_i} E_{\chi^0_j}+q^2\cos^2\theta)
\nonumber\\& & 
        -(\Delta^{t*}(\tilde{\mu}_R) f^{R*}_{\mu i} f^{R}_{\mu j} O^{''R}_{ij}
        -\Delta^{u*}(\tilde{\mu}_R) f^R_{\mu i} f^{R*}_{\mu j} O^{''R*}_{ij})
        2 E_b q \cos\theta
\Big]\Big\},\label{P_2}\\
P(\tilde{\mu}_R \tilde{\mu}_R)&=&
        \frac{g^4}{2} c_R  E_b^2 
        \Big\{|f^R_{\mu i}|^2 |f_{\mu j}^R|^2  \times \nonumber\\& &
                %               \mbox{\hspace*{-.5cm}}
        \Big[ (|\Delta^t(\tilde{\mu}_R)|^2 +|\Delta^u(\tilde{\mu}_R)|^2)
                (E_{\chi^0_i} E_{\chi^0_j}+q^2 \cos^2\theta)
\nonumber\\& &
                        -(|\Delta^t(\tilde{\mu}_R)|^2-|\Delta^u(\tilde{\mu}_R)|^2)
        2 E_b q \cos\theta\Big]
\nonumber\\ &&
        -Re\{(f^{R*}_{\mu i})^2 (f^R_{\mu j})^2
     \Delta^u(\tilde{\mu}_R)\Delta^{t*}(\tilde{\mu}_R)\}
        2 m_{\chi^0_i} m_{\chi^0_j}\Big\}.\label{P_3}
\end{eqnarray}
To obtain the quantities
$P(Z\tilde{\mu}_L),P(\tilde{\mu}_L \tilde{\mu}_L)$ one
has to exchange in (\ref{P_2}) and (\ref{P_3})
\begin{eqnarray}\nonumber
     &&\Delta^{t}(\tilde{\mu}_R)\to\Delta^{t}(\tilde{\mu}_L),\quad
          \Delta^{u}(\tilde{\mu}_R)\to\Delta^{u}(\tilde{\mu}_L),\quad
         c_R \to c_L\\ 
          &&R_{\mu}\to L_{\mu},\quad
         O^{''R}_{ij}\to O^{''L}_{ij},\quad
           f_{\mu i}^R\to f_{\mu i}^L,\quad
                          f_{\mu j}^R\to f_{\mu j}^L. \label{exchange}
\end{eqnarray}
The longitudinal beam polarizations are included in the weighting factors  
\begin{equation}
c_L =(1-{\mathcal P}_-)(1+{\mathcal P}_+), \quad 
c_R= (1+{\mathcal P}_-)(1-{\mathcal P}_+).
\end{equation}
For $e^+e^-$ collisions, the $P$ terms are also given 
in~\cite{Moortgat-Pick:1999di}, however they differ by
a factor of $2$ in our notation~(\ref{eq:tsquare}).
The continuum contributions 
$\Sigma_{\rm cont}^a$~(\ref{contributions}) to the
neu\-tra\-lino polarization can also be found
in~\cite{Moortgat-Pick:1999di}, also differing by
a factor of $2$. However, due to the Majorana character
of the neu\-tra\-linos, the continuum contributions $\Sigma_{\rm cont}^a$
are forward-backward antisymmetric, and vanish if integrated
over the neu\-tra\-lino production angle~\cite{Moortgat-Pick:2002iq},
see Eq.~(\ref{eq:sigmabar}).

%--------------------------------------------------------------------
\section{Neutralino decay into leptons}
\label{neutdecay}
%-------------------------------------------------------------------

The expansion coefficients of the decay matrix~(\ref{rhoD}) 
for the neu\-tra\-lino decay into right sleptons 
$\tilde\chi_j^0 \to \ell^+ \tilde\ell_{R}^-$, 
with  $\ell = e,\mu$, are
\begin{eqnarray}
D &=& \frac{g^2}{2} |f^{R}_{\ell i}|^2 (m_{\chi_i^0}^2 -m_{\tilde\ell}^2 ),\\
\Sigma^a_{D} &=& + g^2 |f^{R}_{\ell i}|^2 
m_{\chi_j^0} (s^a_{\chi_j^0} \cdot p_{\ell_1}).
\label{neutdecay1}
\end{eqnarray}
For the decay into the left sleptons
$\tilde\chi^0_j \to \ell^+ \tilde\ell_L^-$, $\ell=e,\mu$,
the coefficients are
\begin{eqnarray}
D &=& \frac{g^2}{2} |f^{L}_{\ell j}|^2 
                (m_{\chi_j^0}^2 - m_{\tilde\ell}^2 ),\\
\Sigma^a_{D} &=& - g^2 |f^{L}_{\ell i}|^2 m_{\chi_j^0} 
                (s^a_{\chi_j^0} \cdot p_{\ell_1}).
\label{neutdecay2}
\end{eqnarray}
For the decay into the stau 
$\tilde\chi^0_j \to \tau^+ \, \tilde\tau_k^-$, $k=1,2$, one
obtains
\begin{eqnarray}
D &=& \frac{g^2}{2} (
                        |a_{kj}^{\tilde \tau}|^2 +|b_{kj}^{\tilde \tau}|^2 )
                                (m_{\chi_j^0}^2 - m_{\tilde{\tau}_k}^2 ),\\
\Sigma^a_{D} &=&  - g^2 (
                        |a_{kj}^{\tilde \tau}|^2-|b_{kj}^{\tilde \tau}|^2 )
                        m_{\chi_j^0}(s^a_{\chi_j^0} \cdot p_{\ell_1}).
\label{neutdecay3}
\end{eqnarray}
The coefficients $\Sigma^{a}_{D}$ for the charge conjugated processes, 
$\tilde\chi_j^0 \to \ell^- \tilde\ell^+$, 
is obtained by inverting the signs 
of~(\ref{neutdecay1}), (\ref{neutdecay2}) and (\ref{neutdecay3}).

With these definitions we can rewrite the factor 
$\Sigma^{3}_{D}$, that multiplies the 
longitudinal neutralino polarization $\Sigma^{3}_{P}$ in~(\ref{eq:tsquare}), 
\begin{eqnarray}
        \Sigma^{3}_{D} &=& \eta_{\ell\pm} \frac{D}{\Delta_\ell}
        (E_\ell-\bar{E}_\ell),
\label{etal}
 \end{eqnarray}
where we have used 
\begin{eqnarray}
        m_{\chi_j^0} (s^3_{\chi_j^0}\cdot p_\ell)=
        -\frac{m_{\chi_j^0}^2}{|\vec{p}_{\chi^0_j} |}
        (E_\ell-\bar{E}_\ell).
\end{eqnarray}
In order to reduce the free MSSM parameters, 
we parametrize the slepton masses with
their approximate renormalization group equations (RGE)~\cite{Hall:zn}
\begin{eqnarray}
%--------this version including the lepton masses
        m_{\tilde\ell_R}^2 &=& m_0^2 +m_\ell^2+0.23 M_2^2
        -m_Z^2\cos 2 \beta \sin^2 \theta_W,\label{mslr}\\ 
        m_{\tilde\ell_L  }^2 &=& m_0^2 +m_\ell^2+0.79 M_2^2
        +m_Z^2\cos 2 \beta(-\frac{1}{2}+ \sin^2 \theta_W),\label{msll} \\
       m_{\tilde\nu_{\ell}  }^2 &=& m_0^2 +m_\ell^2+0.79 M_2^2 +
       \frac{1}{2}m_Z^2\cos 2 \beta,
\end{eqnarray}
with $m_0$ the common scalar mass parameter the GUT scale.

%-----------------------------------------------------------------------------
\section{Stau-neutralino couplings}
\label{stau.neutralino}
%-----------------------------------------------------------------------------

For the neu\-tra\-lino decay into staus 
$\tilde \chi^0_i \to \tilde \tau_k \tau$, we take stau mixing into
account and write for the Lagrangian \cite{kernreiter:2002}:
\begin{eqnarray}
{\scr L}_{\tau\tilde{\tau} \chi_i }&=&  g\tilde \tau_k \bar \tau
(a^{\tilde \tau}_{ki} P_R+b^{\tilde \tau}_{ki} P_L)\chi^0_i + {\rm h.c.}~,
 \quad k = 1,2; \; i=1,\dots,4, \label{eq:LagStauchi} 
\end{eqnarray}
with
\begin{eqnarray}
a_{kj}^{\tilde \tau}&=&
({\mathcal R}^{\tilde \tau}_{kn})^{\ast}{\mathcal A}^\tau_{jn},\quad 
b_{kj}^{\tilde \tau}=
({\mathcal R}^{\tilde \tau}_{kn})^{\ast}
{\mathcal B}^\tau_{jn},
\quad (n=L,R), 
\label{eq:coupl1}\\
{\mathcal A}^{\tau}_j&=&\left(\begin{array}{ccc}
f^{L}_{\tau j}\\[2mm]
h^{R}_{\tau j} \end{array}\right),\qquad 
{\mathcal B}^{\tau}_j=\left(\begin{array}{ccc}
h^{L}_{\tau j}\\[2mm]
f^{R}_{\tau j} \end{array}\right),
\end{eqnarray}
with ${\mathcal R}^{\tilde \tau}_{kn}$ given in~(\ref{eq:rtau}).
The couplings $f^L_{\tau j}$ and $f^R_{\tau j}$ are defined 
by Eqs.~(\ref{eq:fl}) and (\ref{eq:fr}), respectively, and
\begin{eqnarray}
h^{L}_{\tau j}&=& (h^{R}_{\tau j})^{\ast}
=-Y_{\tau}( N_{j3}^{\ast}\cos\beta+N_{j4}^{\ast}\sin\beta), \\ 
Y_{\tau}&=& \frac{m_{\tau}}{\sqrt{2}\,m_W \cos\beta}, \label{eq:coupl4}
\end{eqnarray}
with $m_W$ the mass of the $W$ boson, $m_{\tau}$ the mass of 
the $\tau$-lepton and $N$ the neu\-tra\-lino mixing matrix in the 
$\tilde{\gamma}, \tilde{Z}, H_1^0, H_2^0$ basis. 
The masses and couplings of the $\tau$-sleptons follow from 
the $\tilde\tau_L - \tilde \tau_R$ mass matrix
\begin{equation}
{\mathcal{L}}_M^{\tilde \tau}= -(\tilde \tau_L^{\ast},\, \tilde \tau_R^{\ast})
\left(\begin{array}{ccc}
m_{\tilde \tau_{L}}^2 & e^{-i\varphi_{\tilde \tau}}m_\tau |\Lambda_{\tilde \tau}|\\[5mm]
e^{i\varphi_{\tilde \tau}} m_\tau |\Lambda_{\tilde \tau}| & m_{\tilde \tau_{R}}^2
\end{array}\right)\left(
\begin{array}{ccc}
\tilde \tau_L\\[5mm]
\tilde \tau_R \end{array}\right),
\label{eq:mm}
\end{equation}
with $m_{\tilde\tau_R}^2$ and $m_{\tilde\tau_L}^2$ given 
by Eqs.~(\ref{mslr}) and (\ref{msll}) replacing $m_\ell^2$ by $m_\tau^2 $, and 
\begin{eqnarray}
\Lambda_{\tilde \tau} & = & A_{\tau}-\mu^\ast\tan\beta, \label{eq:mlr}\\
\varphi_{\tilde \tau}& = & \arg\lbrack \Lambda_{\tilde \tau}\rbrack,
\end{eqnarray}
with $A_{\tau}=A$ the (common) trilinear scalar coupling parameter.
The $\tilde \tau$ mass eigenstates are 
$(\tilde\tau_1, \tilde \tau_2)=(\tilde \tau_L, \tilde \tau_R)
{{\mathcal R}^{\tilde \tau}}^{T}$, 
with 
\begin{equation}
        {\mathcal R}^{\tilde \tau}
         =\left( \begin{array}{ccc}
        e^{i\varphi_{\tilde \tau}}\cos\theta_{\tilde \tau} & 
        \sin\theta_{\tilde \tau}\\[5mm]
        -\sin\theta_{\tilde \tau} & 
        e^{-i\varphi_{\tilde \tau}}\cos\theta_{\tilde \tau}
\end{array}\right).
\label{eq:rtau}
\end{equation}
The mixing angle is 
\begin{equation}
\cos\theta_{\tilde \tau}=
        \frac{-m_\tau |\Lambda_{\tilde \tau}|}{\sqrt{m_\tau^2 |\Lambda_{\tilde \tau}|^2+
        (m_{\tilde \tau_1}^2-m_{\tilde \tau_{L}}^2)^2}},\quad
\sin\theta_{\tilde \tau}=\frac{m_{\tilde \tau_{L}}^2-m_{\tilde \tau_1}^2}
        {\sqrt{m_\tau^2 |\Lambda_{\tilde \tau}|^2+(m_{\tilde \tau_1}^2-
                                m_{\tilde \tau_{L}}^2)^2}},
        \label{eq:thtau}
\end{equation}
and the mass eigenvalues are
\begin{equation}
        m_{\tilde \tau_{\,1,2}}^2 = 
\frac{1}{2}\left[(m_{\tilde \tau_{L}}^2+m_{\tilde \tau_{R}}^2)\mp 
        \sqrt{(m_{\tilde \tau_{L}}^2 - m_{\tilde \tau_{R}}^2)^2
                +4m_\tau^2 |\Lambda_{\tilde \tau}|^2}\;\right].
\label{eq:m12}
\end{equation}

%----------------------------------------------------
\section{Statistical significances}
\label{StatSignificances}
%----------------------------------------------------

We define the statistical significance of the asymmetry
${\mathcal A}^{{\rm pol}}_{\rm prod}$~(\ref{eq:apolprod.ij})
by
\begin{eqnarray}
        {\mathcal S}^{\rm pol}_{\rm prod}
        &=& |{\mathcal A}^{{\rm pol}}_{\rm prod}|
        \sqrt{2 \bar\sigma_{ij} { \mathcal L} },
\label{significance1}
\end{eqnarray}
where ${\mathcal L}$ denotes the integrated luminosity and 
\begin{eqnarray}
\bar\sigma_{ij} =\frac{
|\sigma_{ij}({\mathcal P})-\sigma_{ij}({-\mathcal{P}})|
}{2}
\label{sigmamean}
\end{eqnarray}
is the mean value of the neu\-tra\-lino
production cross section
$\sigma_{ij}$,~(\ref{crossectionProd}),
for both equal muon beam polarizations $\mathcal P$ and $-\mathcal P$.
The significances for the polarization asymmetries
${\mathcal A}_{\ell}^{{\rm pol}, n}$~(\ref{eq.apolijA}) and
${\mathcal A}_{\ell}^{\prime {\rm pol}, n}$~(\ref{eq.apolijAprime})
are defined by
\begin{eqnarray}
        \mathcal{S}_{{\ell}}^{{\rm pol}, n}
        &=& |{\mathcal A}_{\ell}^{{\rm pol}, n}| 
                \sqrt{4\bar\sigma_{ij}
          {\rm BR}(\neu_j\to\ell^+\slepton_n^-){\mathcal{L}_{\rm eff}}},
\label{significance3}
\end{eqnarray}
and
\begin{eqnarray}
\mathcal{S}_\ell^{\prime {\rm pol}, n}
        &=&
|\mathcal{A}_\ell^{\prime {\rm pol}, n}| 
                \sqrt{4\bar\sigma_{ij}
          {\rm BR}(\neu_j\to\ell^+\slepton_n^-){\mathcal{L}_{\rm eff}}},
\label{significance4}
\end{eqnarray}
respectively,
where $\mathcal{L}_{\rm eff} =\eps^n_\ell \mathcal{L}$ is the effective 
integrated luminosity, with $\eps^n_\ell$ the detection efficiency of the leptons 
from the decay $\tilde\chi^0_j\to\ell^\pm\tilde\ell_n^\mp$.
There is a factor $4$ appearing in the significances,
since the asymmetries 
require two sets of equal beam polarizations $\mathcal P$,
as well as two decay modes,
$\tilde\chi_j^0 \to \ell^+\tilde\ell^-_n$,
and the charge conjugated decay
$\tilde\chi_j^0 \to \ell^-\tilde\ell^+_n$.

For an ideal detector, a significance of, e.g., $S = 1$ implies
that the asymmetries  
can be measured at the statistical 68\% confidence level.
In order to predict the absolute values of confidence levels,
clearly detailed Monte Carlo analysis
including detector and background 
simulations with particle identification and reconstruction
efficiencies would be required, which is however beyond the scope
of the present work.

% ----------------------------------------------------------------------------
\end{appendix}
%-----------------------------------------------------------------------------

%}}}

%{{{ Bibliography

%-----------------------------------------------------------------------------

%-----------------------------------------------------------------------------

%}}}
% ---------------------------------------------------------------------------------------

\begin{thebibliography}{99}
%-----------------------------------------------------------------------------

%%%--------MSSM higgs sector basics

\bibitem{HK} H.~Haber, K.~Kane, 
Phys.\ Rep.\ \textbf{117} (1985) 75.
%%CITATION = PRPLC,117,75;%%

\bibitem{GH} J.~Gunion, H.~Haber, 
Nucl.\ Phys.\ \textbf{B 272} (1986) 1.
%%CITATION = NUPHA,B278,449;%%

%cite{Gunion:1989we}
\bibitem{Gunion:1989we}
  J.~F.~Gunion, H.~E.~Haber, G.~L.~Kane and S.~Dawson,
  ``{\it The Higgs Hunter's Guide}'' (Addison-Wesley Publishing Company, Redwood City, CA, 1990).
  %%CITATION = BNL-41644;%%

%\cite{Carena:2002es}
\bibitem{Carena:2002es}
  M.~S.~Carena and H.~E.~Haber,
  %``Higgs boson theory and phenomenology. ((V)),''
  Prog.\ Part.\ Nucl.\ Phys.\  {\bf 50}, 63 (2003)
  [arXiv:hep-ph/0208209].
  %%CITATION = PPNPD,50,63;%%

%\cite{Djouadi:2005gj}
%\bibitem{Djouadi}
\bibitem{Djouadi:2005gj}
  A.~Djouadi,
  %``The anatomy of electro-weak symmetry breaking. II: The Higgs bosons in  the
  %minimal supersymmetric model,''
  arXiv:hep-ph/0503173.
  %%CITATION = HEP-PH 0503173;%%

%\cite{Accomando:2006ga}
\bibitem{Accomando:2006ga}
  E.~Accomando {\it et al.},
  %``Workshop on CP studies and non-standard Higgs physics,''
  arXiv:hep-ph/0608079.
  %%CITATION = HEP-PH/0608079;%%


%\cite{CPNSH}
\bibitem{CPNSH}
  H.~Fraas, O.~Kittel and  F.~von der Pahlen,
  {\it``Higgs boson interference in chargino and neutralino production
   at the muon collider''}, in Ref.~\cite{Accomando:2006ga},
%  E.~Accomando {\it et al.},
%  ``Workshop on CP studies and non-standard Higgs physics,''
%  arXiv:hep-ph/0608079, 
   p. 169.
  %%CITATION = HEP-PH/0608079;%%



%%% -----------Higgs A-H transitions --early works

\bibitem{PilaftsisAH}
%\cite{Pilaftsis:1998pe}
%\bibitem{Pilaftsis:1998pe}
  A.~Pilaftsis,
  %``CP-odd tadpole renormalization of Higgs scalar-pseudoscalar mixing,''
  Phys.\ Rev.\  D {\bf 58}, 096010 (1998)
  [arXiv:hep-ph/9803297];\\
  %%CITATION = PHRVA,D58,096010;%%
%\cite{Pilaftsis:1998dd}
%\bibitem{Pilaftsis:1998dd}
  A.~Pilaftsis,
  %``Higgs scalar-pseudoscalar mixing in the minimal supersymmetric standard
  %model,''
  Phys.\ Lett.\  B {\bf 435}, 88 (1998)
  [arXiv:hep-ph/9805373].
  %%CITATION = PHLTA,B435,88;%%

%%%-----Resonant enahnced H-A mixing


%cite{Pilaftsis:1997dr}
\bibitem{Pilaftsis:1997dr}
  A.~Pilaftsis,
  %``Resonant CP violation induced by particle mixing in transition
  %amplitudes,''
  Nucl.\ Phys.\  B {\bf 504} (1997) 61
  [arXiv:hep-ph/9702393].
  %%CITATION = NUPHA,B504,61;%%

%\cite{Choi:2004kq}
\bibitem{Choi:2004kq}
  S.~Y.~Choi, J.~Kalinowski, Y.~Liao and P.~M.~Zerwas,
  %``H / A Higgs mixing in CP-noninvariant supersymmetric theories,''
  Eur.\ Phys.\ J.\  C {\bf 40} (2005) 555
  [arXiv:hep-ph/0407347].
  %%CITATION = EPHJA,C40,555;%%

%%% ----------- the latest results in  A-H mixings

\bibitem{GunionetalHA}
%\cite{Gunion:1997aq}
%\bibitem{Gunion:1997aq}
  J.~F.~Gunion, B.~Grzadkowski, H.~E.~Haber and J.~Kalinowski,
  %``LEP limits on CP-violating non-minimal Higgs sectors,''
  Phys.\ Rev.\ Lett.\  {\bf 79} (1997) 982
  [arXiv:hep-ph/9704410];\\
  %%CITATION = PRLTA,79,982;%%
%\cite{Grzadkowski:1999ye}
%\bibitem{Grzadkowski:1999ye}
  B.~Grzadkowski, J.~F.~Gunion and J.~Kalinowski,
  %``Finding the CP-violating Higgs bosons at e+ e- colliders,''
  Phys.\ Rev.\  D {\bf 60} (1999) 075011
  [arXiv:hep-ph/9902308].
  %%CITATION = PHRVA,D60,075011;%%


\bibitem{HAeffpotential}
%\cite{Demir:1999hj}
%\bibitem{Demir:1999hj}
  D.~A.~Demir,
  %``Effects of the supersymmetric phases on the neutral Higgs sector,''
  Phys.\ Rev.\  D {\bf 60} (1999) 055006
  [arXiv:hep-ph/9901389];\\
  %%CITATION = PHRVA,D60,055006;%%
%
%\cite{Choi:2000wz}
%\bibitem{Choi:2000wz}
  S.~Y.~Choi, M.~Drees and J.~S.~Lee,
  %``Loop corrections to the neutral Higgs boson sector of the MSSM with
  %explicit CP violation,''
  Phys.\ Lett.\  B {\bf 481} (2000) 57
  [arXiv:hep-ph/0002287];\\
  %%CITATION = PHLTA,B481,57;%%
%
%\cite{Ibrahim:2000qj}
%\bibitem{Ibrahim:2000qj}
  T.~Ibrahim and P.~Nath,
  %``Corrections to the Higgs boson masses and mixings from chargino, W and
  %charged Higgs exchange loops and large CP phases,''
  Phys.\ Rev.\  D {\bf 63}, 035009 (2001)
  [arXiv:hep-ph/0008237];
  %%CITATION = PHRVA,D63,035009;%%
%%%\cite{Ibrahim:2002zk}
%%%\bibitem{Ibrahim:2002zk}
%%%  T.~Ibrahim and P.~Nath,
  %``Neutralino exchange corrections to the Higgs boson mixings with  explicit
  %CP violation,''
  Phys.\ Rev.\  D {\bf 66}, 015005 (2002)
  [arXiv:hep-ph/0204092].
  %%CITATION = PHRVA,D66,015005;%%

\bibitem{PW.explicitCP}
%\cite{Pilaftsis:1999qt}
%\bibitem{Pilaftsis:1999qt}
  A.~Pilaftsis and C.~E.~M.~Wagner,
  %``Higgs bosons in the minimal supersymmetric standard model with explicit CP
  %violation,''
  Nucl.\ Phys.\  B {\bf 553} (1999) 3
  [arXiv:hep-ph/9902371];\\
  %%CITATION = NUPHA,B553,3;%%
%
%\cite{Carena:2000yi}
%\bibitem{Carena:2000yi}
  M.~S.~Carena, J.~R.~Ellis, A.~Pilaftsis and C.~E.~M.~Wagner,
  %``Renormalization-group-improved effective potential for the MSSM Higgs
  %sector with explicit CP violation,''
  Nucl.\ Phys.\  B {\bf 586} (2000) 92
  [arXiv:hep-ph/0003180];\\
  %%CITATION = NUPHA,B586,92;%%
%
%\cite{Carena:2001fw}
%\bibitem{Carena:2001fw}
  M.~S.~Carena, J.~R.~Ellis, A.~Pilaftsis and C.~E.~M.~Wagner,
  %``Higgs-boson pole masses in the MSSM with explicit CP violation,''
  Nucl.\ Phys.\  B {\bf 625}, 345 (2002)
  [arXiv:hep-ph/0111245].
  %%CITATION = NUPHA,B625,345;%%


%\cite{Heinemeyer:2001qd}
\bibitem{Heinemeyer:2001qd}
  S.~Heinemeyer,
  %``The Higgs boson sector of the complex MSSM in the Feynman-diagrammatic
  %approach,''
  Eur.\ Phys.\ J.\  C {\bf 22} (2001) 521
  [arXiv:hep-ph/0108059].
  %%CITATION = EPHJA,C22,521;%%

%%%----------- Decoupling limit in MSSM ------------------------------


%~\cite{decoupling}
\bibitem{decoupling}
%\cite{Dobado:2000pw}
%\bibitem{Dobado:2000pw}
  A.~Dobado, M.~J.~Herrero and S.~Penaranda,
  %``The Higgs sector of the MSSM in the decoupling limit,''
  Eur.\ Phys.\ J.\  C {\bf 17}, 487 (2000)
  [arXiv:hep-ph/0002134];
  %%CITATION = EPHJA,C17,487;%%
%\cite{Gunion:2002zf}
%\bibitem{Gunion:2002zf}
  J.~F.~Gunion and H.~E.~Haber,
  %``The CP-conserving two-Higgs-doublet model: The approach to the  decoupling
  %limit,''
  Phys.\ Rev.\  D {\bf 67}, 075019 (2003)
  [arXiv:hep-ph/0207010];
  %%CITATION = PHRVA,D67,075019;%%
%\cite{Haber:1993jr}
%\bibitem{Haber:1993jr}
  H.~E.~Haber and Y.~Nir,
  %``The Decay Z $\to$ A0 A0 neutrino anti-neutrino and e+ e- $\to$ A0 A0 z in
  %two Higgs doublet models,''
  Phys.\ Lett.\  B {\bf 306}, 327 (1993)
  [arXiv:hep-ph/9302228];
  %%CITATION = PHLTA,B306,327;%%
%\cite{Haber:1995be}
%\bibitem{Haber:1995be}
  H.~E.~Haber,
  %``Challenges for nonminimal Higgs searches at future colliders,''
% CERN-TH/95-109
  arXiv:hep-ph/9505240.
  %%CITATION = HEP-PH/9505240;%%


%%%%%%---------computer-codes------------------


% FeynHiggs
%hep-ph/0611326, hep-ph/0212020, hep-ph/9812472, hep-ph/9812320. 
%\cite{Frank:2006yh}
\bibitem{Frank:2006yh}
  M.~Frank, T.~Hahn, S.~Heinemeyer, W.~Hollik, H.~Rzehak and G.~Weiglein,
  %``The Higgs boson masses and mixings of the complex MSSM in the
  %Feynman-diagrammatic approach,''
  JHEP {\bf 0702} (2007) 047
  [arXiv:hep-ph/0611326].
  %%CITATION = JHEPA,0702,047;%%


\bibitem{FH}
%\cite{Heinemeyer:1998np}
%\bibitem{Heinemeyer:1998np}
  S.~Heinemeyer, W.~Hollik and G.~Weiglein,
  %``The masses of the neutral CP-even Higgs bosons in the MSSM: Accurate
  %analysis at the two-loop level,''
  Eur.\ Phys.\ J.\  C {\bf 9} (1999) 343
  [arXiv:hep-ph/9812472];\\
  %%CITATION = EPHJA,C9,343;%%
%\cite{Heinemeyer:1998yj}
%\bibitem{Heinemeyer:1998yj}
  S.~Heinemeyer, W.~Hollik and G.~Weiglein,
  %``FeynHiggs: A program for the calculation of the masses of the neutral
  %CP-even Higgs bosons in the MSSM,''
  Comput.\ Phys.\ Commun.\  {\bf 124} (2000) 76
  [arXiv:hep-ph/9812320];\\
  %%CITATION = CPHCB,124,76;%%
%\cite{Degrassi:2002fi}
%\bibitem{Degrassi:2002fi}
  G.~Degrassi, S.~Heinemeyer, W.~Hollik, P.~Slavich and G.~Weiglein,
  %``Towards high-precision predictions for the MSSM Higgs sector,''
  Eur.\ Phys.\ J.\  C {\bf 28} (2003) 133
  [arXiv:hep-ph/0212020];\\
  %%CITATION = EPHJA,C28,133;%%
%
%\cite{Heinemeyer:2007aq}
%\bibitem{Heinemeyer:2007aq}
  S.~Heinemeyer, W.~Hollik, H.~Rzehak and G.~Weiglein,
  %``The Higgs sector of the complex MSSM at two-loop order: QCD
  %contributions,''
  arXiv:0705.0746 [hep-ph].
  %%CITATION = ARXIV:0705.0746;%%



%% CPsuperH------

\bibitem{CPSuperH}
%\cite{Lee:2003nta}
%\bibitem{Lee:2003nta}
  J.~S.~Lee, A.~Pilaftsis, M.~S.~Carena, S.~Y.~Choi, M.~Drees, 
  J.~R.~Ellis and C.~E.~M.~Wagner,
  %``CPsuperH: A computational tool for Higgs phenomenology in the minimal
  %supersymmetric standard model with explicit CP violation,''
  Comput.\ Phys.\ Commun.\  {\bf 156}, 283 (2004)
  [arXiv:hep-ph/0307377];
  %%CITATION = CPHCB,156,283;%%
%\cite{Ellis:2006eh}
%\bibitem{Ellis:2006eh}
  J.~R.~Ellis, J.~S.~Lee and A.~Pilaftsis,
  %``Higgs phenomenology with CPsuperH,''
  Mod.\ Phys.\ Lett.\  A {\bf 21}, 1405 (2006)
  [arXiv:hep-ph/0605288].
  %%CITATION = MPLAE,A21,1405;%%



%%%---------determining higgs masses etc in s-resonances etc
%%%------------and CP-conserving studies in \mu+\mu-

%\cite{Grzadkowski:1995rx}
\bibitem{Grzadkowski:1995rx}
  B.~Grzadkowski and J.~F.~Gunion,
  %``Using decay angle correlations to detect CP violation in the neutral Higgs
  %sector,''
  Phys.\ Lett.\  B {\bf 350}, 218 (1995)
  [arXiv:hep-ph/9501339].
  %%CITATION = PHLTA,B350,218;%%

\bibitem{mumu90s}
%\cite{Barger:1996jm}
%\bibitem{Barger:1996jm}
  V.~D.~Barger, M.~S.~Berger, J.~F.~Gunion and T.~Han,
  %``Higgs boson physics in the s-channel at mu+ mu- colliders,''
  Phys.\ Rept.\  {\bf 286} (1997) 1
  [arXiv:hep-ph/9602415];\\
  %%CITATION = PRPLC,286,1;%%
%
%\cite{Casalbuoni:1999mm}
%\bibitem{Casalbuoni:1999mm}
  R.~Casalbuoni, A.~Deandrea, S.~De Curtis, D.~Dominici, R.~Gatto and J.~F.~Gunion,
  %``Analysis of narrow s-channel resonances at lepton colliders,''
  JHEP {\bf 9908} (1999) 011
  [arXiv:hep-ph/9904268];\\
  %%CITATION = JHEPA,9908,011;%%
%
%\cite{Berger:2001et}
%\bibitem{Berger:2001et}
  M.~S.~Berger,
  %``Higgs sector radiative corrections and s-channel production,''
  Phys.\ Rev.\ Lett.\  {\bf 87}, 131801 (2001)
  [arXiv:hep-ph/0105128].
  %%CITATION = PRLTA,87,131801;%%

%\cite{Dittmaier:2002nd}
\bibitem{Dittmaier:2002nd}
  S.~Dittmaier and A.~Kaiser,
  %``Photonic and QCD radiative corrections to Higgs boson production in mu+
  %mu- --> f anti-f,''
  Phys.\ Rev.\  D {\bf 65}, 113003 (2002)
  [arXiv:hep-ph/0203120].
  %%CITATION = PHRVA,D65,113003;%%

%\cite{Fraas:2003cx}
\bibitem{Fraas:2003cx}
H.~Fraas, F.~Franke, G.~Moortgat-Pick, F.~von der Pahlen and A.~Wagner,
%``Precision measurements of Higgs chargino couplings in chargino pair
%production at a muon collider,''
Eur.\ Phys.\ J.\ C {\bf 29} (2003) 587
[arXiv:hep-ph/0303044].
%%CITATION = HEP-PH 0303044;%%

%\cite{Grzadkowski:2000hm}
\bibitem{Grzadkowski:2000hm}
  B.~Grzadkowski, J.~F.~Gunion and J.~Pliszka,
  %``How valuable is polarization at a muon collider? A test case:  Determining
  %the CP nature of a Higgs boson,''
  Nucl.\ Phys.\  B {\bf 583}, 49 (2000)
  [arXiv:hep-ph/0003091].
  %%CITATION = NUPHA,B583,49;%%


%\cite{Asakawa:2000uj}
\bibitem{Asakawa:2000uj}
  E.~Asakawa, A.~Sugamoto and I.~Watanabe,
  %``Production of CP even and CP odd Higgs bosons at muon colliders,''
  Eur.\ Phys.\ J.\  C {\bf 17}, 279 (2000).
  %%CITATION = EPHJA,C17,279;%

%\cite{Barger:1999tj}
\bibitem{Barger:1999tj}
  V.~D.~Barger, T.~Han and C.~G.~Zhou,
  %``Higgs boson decays to tau pairs in the s-channel at a muon collider,''
  Phys.\ Lett.\  B {\bf 480}, 140 (2000)
  [arXiv:hep-ph/0002042].
  %%CITATION = PHLTA,B480,140;%%

%\cite{Fraas:2004bq}
\bibitem{Fraas:2004bq}
H.~Fraas, F.~von der Pahlen and C.~Sachse,
 %``Interference of Higgs boson resonances in mu+ mu- $\to$ neutralino(i)
%neutralino(j) with longitudinal beam polarization,''
Eur.\ Phys.\ J.\ {\bf C37} (2004) 495 [arXiv:hep-ph/0407057].
%%CITATION = HEP-PH 0407057;%%

\bibitem{Kittel:2005ma}
  O.~Kittel and F.~von der Pahlen,
  %``Higgs boson interference in mu+ mu- --> chargino(i)+ chargino(j)- with
  %longitudinally polarized beams,''
  Phys.\ Rev.\  D {\bf 72} (2005) 095004
  [arXiv:hep-ph/0508267].
  %%CITATION = PHRVA,D72,095004;%%


%%%--------- CP violation studies at muon colliders

%\cite{Asakawa:2000es}
\bibitem{Asakawa:2000es}
E.~Asakawa, S.~Y.~Choi and J.~S.~Lee,
 %``Probing the MSSM Higgs boson sector with explicit CP violation through third generation fermion pair production at muon colliders,''
Phys.\ Rev.\ D {\bf 63} (2001) 015012
[arXiv:hep-ph/0005118].
%%CITATION = HEP-PH 0005118;%%


%\cite{Atwood:1995uc}
\bibitem{Atwood:1995uc}
  D.~Atwood and A.~Soni,
  %``Neutral Higgs CP violation at mu+ mu- colliders,''
  Phys.\ Rev.\  D {\bf 52}, 6271 (1995)
  [arXiv:hep-ph/9505233].
  %%CITATION = PHRVA,D52,6271;%%

%\cite{Pilaftsis:1996ac}
\bibitem{Pilaftsis:1996ac}
  A.~Pilaftsis,
  %``Higgs-Z Mixing and Resonant CP Violation at mu+ mu- Colliders,''
  Phys.\ Rev.\ Lett.\  {\bf 77} (1996) 4996
  [arXiv:hep-ph/9603328].
  %%CITATION = PRLTA,77,4996;%%



%\cite{Babu:1998bf}
\bibitem{Babu:1998bf}
  K.~S.~Babu, C.~F.~Kolda, J.~March-Russell and F.~Wilczek,
  %``CP violation, Higgs couplings, and supersymmetry,''
  Phys.\ Rev.\  D {\bf 59} (1999) 016004
  [arXiv:hep-ph/9804355].
  %%CITATION = PHRVA,D59,016004;%%


%\cite{Choi:1999kn}
\bibitem{Choi:1999kn}
  S.~Y.~Choi and J.~S.~Lee,
  %``s-channel production of MSSM Higgs bosons at a muon collider with  explicit
  %CP violation,''
  Phys.\ Rev.\  D {\bf 61} (2000) 111702
  [arXiv:hep-ph/9909315].
  %%CITATION = PHRVA,D61,111702;%%

%cite{Choi:2001ks}
\bibitem{Choi:2001ks}
  S.~Y.~Choi, M.~Drees, B.~Gaissmaier and J.~S.~Lee,
  %``CP violation in tau slepton pair production at muon colliders,''
  Phys.\ Rev.\  D {\bf 64} (2001) 095009
  [arXiv:hep-ph/0103284].
  %%CITATION = PHRVA,D64,095009;%%

%\cite{Bernabeu:2006zs}
\bibitem{Bernabeu:2006zs}
  J.~Bernabeu, D.~Binosi and J.~Papavassiliou,
  %``CP violation through particle mixing and the H - A lineshape,''
  JHEP {\bf 0609}, 023 (2006)
  [arXiv:hep-ph/0604046].
  %%CITATION = JHEPA,0609,023;%%

%\cite{Hioki:2007jc}
\bibitem{Hioki:2007jc}
  Z.~Hioki, T.~Konishi and K.~Ohkuma,
  %``Studying possible CP-violating Higgs couplings through top-quark pair
  %productions at muon colliders,''
  JHEP {\bf 0707}, 082 (2007)
  [arXiv:0706.4346 [hep-ph]].
  %%CITATION = JHEPA,0707,082;%%




%%% ---------------Reports on the muon collider

%\cite{Barger:1996vc}
\bibitem{Barger:1996vc}
  V.~D.~Barger, M.~S.~Berger, J.~F.~Gunion and T.~Han,
  %``Particle physics opportunities at mu+ mu- colliders,''
  Nucl.\ Phys.\ Proc.\ Suppl.\  {\bf 51A} (1996) 13
  [arXiv:hep-ph/9604334].
  %%CITATION = NUPHZ,51A,13;%%

\bibitem{hefreports}
%\bibitem{Autin:1999ci}
Proceedings of {\it Prospective Study of Muon Storage Rings at CERN}, 
Eds. B.~Autin, A.~Blondel, J.~Ellis, CERN yellow report,
CERN 99-02, ECFA 99-197, April 30 (1999);\\
%\href{http://www.slac.stanford.edu/spires/find/hep/www?r=cern-99-02}{SPIRES entry}
%
%\cite{Barger:2001mi}
%\bibitem{Barger:2001mi}
  V.~D.~Barger, M.~Berger, J.~F.~Gunion and T.~Han,
  %``Physics of Higgs factories,''
  %in {\it Proc. of the APS/DPF/DPB Summer Study on the 
  % Future of Particle Physics (Snowmass 2001) } ed. N.~Graf,
  {\it In the Proceedings of APS / DPF / DPB Summer Study 
     on the Future of Particle Physics (Snowmass 2001), 
     Snowmass, Colorado, 30 Jun - 21 Jul 2001, pp E110}
  [arXiv:hep-ph/0110340;\\
  %%CITATION = ECONF,C010630,E110;%%
%\cite{Barger:2001mi}
%
%\bibitem{Blondel:2004ae}
A.~Blondel {\it et al.},
``{\it ECFA/CERN studies of a European neutrino factory complex},''
CERN-2004-002.
%\href{http://www.slac.stanford.edu/spires/find/hep/www?r=cern-2004-002}{SPIRES entry}

\bibitem{Blochinger:2002hj}
C.~Bl\"ochinger et al., Higgs working group of the ECFA-CERN study on Neutrino Factory \& Muon Storage Rings at CERN, {\it Physics Opportunities at $\mu^+\mu^-$ Higgs Factories}
CERN-TH/2002-028, [arXiv:hep-ph/0202199];
%%CITATION = HEP-PH 0202199;%%
%


%%%%----weisskopf-wigner approximation

%\cite{Weisskopf:1930ps}
\bibitem{Weisskopf:1930ps}
  V.~Weisskopf and E.~Wigner,
  %``Over the natural line width in the radiation of the harmonius oscillator,''
  Z.\ Phys.\  {\bf 65} (1930) 18.
  %%CITATION = ZEPYA,65,18;%%




%%%%%------------Spin-density formalism, and continuum-refs, Appendix refs


\bibitem{Haber94}
H.~E.~Haber, 
Proceedings of the 21st SLAC Summer Institute on Particle Physics: {\it Spin Structure in High Energy Processes}, SLAC, Stanford, CA 1993 
%``Spin formalism and applications to new physics searches,''
[arXiv:hep-ph/9405376].
%%CITATION = HEP-PH 9405376;%%

%\cite{Moortgat-Pick:1999di}
\bibitem{Moortgat-Pick:1999di}
  G.~A.~Moortgat-Pick, H.~Fraas, A.~Bartl and W.~Majerotto,
  %``Polarization and spin effects in neutralino production and decay,''
  Eur.\ Phys.\ J.\  C {\bf 9}, 521 (1999)
  [Erratum-ibid.\  C {\bf 9}, 549 (1999)]
  [arXiv:hep-ph/9903220].
  %%CITATION = EPHJA,C9,521;%%

%\cite{Kittel:2005rp}
\bibitem{Kittel:2005rp}
  O.~Kittel,
  %``CP violation in production and decay of supersymmetric particles,''
  arXiv:hep-ph/0504183.
  %%CITATION = HEP-PH 0504183;%%

%\cite{Moortgat-Pick:2002iq}
\bibitem{Moortgat-Pick:2002iq}
  G.~A.~Moortgat-Pick and H.~Fraas,
  %``Influence of CP and CPT on production and decay of Dirac and Majorana
  %fermions,''
  Eur.\ Phys.\ J.\  C {\bf 25}, 189 (2002)
  [arXiv:hep-ph/0204333].
  %%CITATION = EPHJA,C25,189;%%







%%%------------------ILC TDR und RDR---------------------

\bibitem{TDR}
%J.~A.~Aguilar-Saavedra {\it et al.}  [ECFA/DESY LC Physics Working Group
%                 Collaboration],
``TESLA Technical Design Report Part III: 
{\it Physics at an e+e- Linear Collider},''
arXiv:hep-ph/0106315.
%%CITATION = HEP-PH 0106315;%%

%\cite{Djouadi:2007ik}
\bibitem{Djouadi:2007ik}
  A.~Djouadi, J.~Lykken, K.~Monig, Y.~Okada, M.~J.~Oreglia and S.~Yamashita,
  ``International Linear Collider Reference Design Report Volume 2: {\it Physics at the ILC},''
  arXiv:0709.1893 [hep-ph].
  %%CITATION = ARXIV:0709.1893;%%




%%%----------- EDMs experimental-------------------------------

%\cite{Yao:2006px}
\bibitem{Yao:2006px}
  W.~M.~Yao {\it et al.}  [Particle Data Group],
  %``Review of particle physics,''
  J.\ Phys.\ G {\bf 33} (2006) 1.

%\cite{Harris:1999jx}
\bibitem{Harris:1999jx}
  P.~G.~Harris {\it et al.},
  %``New experimental limit on the electric dipole moment of the neutron,''
  Phys.\ Rev.\ Lett.\  {\bf 82} (1999) 904.
  %%CITATION = PRLTA,82,904;%%

%\cite{Regan:2002ta}
\bibitem{Regan:2002ta}
  B.~C.~Regan, E.~D.~Commins, C.~J.~Schmidt and D.~DeMille,
  %``New limit on the electron electric dipole moment,''
  Phys.\ Rev.\ Lett.\  {\bf 88} (2002) 071805.
  %%CITATION = PRLTA,88,071805;%%
 
%\cite{Romalis:2000mg}
\bibitem{Romalis:2000mg}
  M.~V.~Romalis, W.~C.~Griffith and E.~N.~Fortson,
  %``A new limit on the permanent electric dipole moment of Hg-199,''
  Phys.\ Rev.\ Lett.\  {\bf 86} (2001) 2505
  [arXiv:hep-ex/0012001].
  %%CITATION = PRLTA,86,2505;%%




%%%-------------Benchmark scenarios CPX etc

%\cite{Carena:2000ks}
\bibitem{Carena:2000ks}
  M.~Carena, J.~R.~Ellis, A.~Pilaftsis and C.~E.~M.~Wagner,
  %``CP-violating MSSM Higgs bosons in the light of LEP 2,''
  Phys.\ Lett.\  B {\bf 495} (2000) 155
  [arXiv:hep-ph/0009212].
  %%CITATION = PHLTA,B495,155;%%



%%%----------------- 	spin analysis 	------------------------------
%\cite{Choi:2006mr}
\bibitem{Choi:2006mr}
  S.~Y.~Choi, K.~Hagiwara, H.~U.~Martyn, K.~Mawatari and P.~M.~Zerwas,
  %``Spin analysis of supersymmetric particles,''
  Eur.\ Phys.\ J.\  C {\bf 51}, 753 (2007)
  [arXiv:hep-ph/0612301].
  %%CITATION = EPHJA,C51,753;%%







%%%%%------------continuum-refs for Appendix 

\bibitem{orthogonalC}
R.A.~Horn, C.A.~Johnson, Matrix Analysis, Cambridge University Press 1990%(ISBN 0521386322)
;\\
%\cite{Hahn:2006hr}
%\bibitem{Hahn:2006hr}
  T.~Hahn,
  %``Routines for the diagonalization of complex matrices,''
  arXiv:physics/0607103.
  %%CITATION = PHYSICS/0607103;%%

%\cite{Bartl:1986hp}
\bibitem{Bartl:1986hp}
  A.~Bartl, H.~Fraas and W.~Majerotto,
  %``PRODUCTION AND DECAY OF NEUTRALINOS IN e+ e- ANNIHILATION,''
  Nucl.\ Phys.\ B {\bf 278} (1986) 1.
  %%CITATION = NUPHA,B278,1;%%

%\cite{Hall:zn}
\bibitem{Hall:zn}
L.~J.~Hall and J.~Polchinski,
%``Implications Of Supersymmetric Origins For Monojets,''
Phys.\ Lett.\ B {\bf 152} (1985) 335.
%%CITATION = PHLTA,B152,335;%%

\bibitem{kernreiter:2002}
 A.~Bartl, K.~Hidaka, T.~Kernreiter and W.~Porod,
 %``tau-sleptons and tau-sneutrino in the MSSM with complex parameters,''
 Phys.\ Rev.\ D {\bf 66} (2002) 115009
 [arXiv:hep-ph/0207186].
 %%CITATION = HEP-PH 0207186;%%


%-----------------------------------------------------------------------------
\end{thebibliography}
\end{document}